\documentclass[fleqn,usenatbib]{mnras}

\usepackage{amssymb}

\usepackage{newtxtext,newtxmath}
\usepackage[flushleft]{threeparttable}

\usepackage[T1]{fontenc}

\DeclareRobustCommand{\VAN}[3]{#2}
\let\VANthebibliography\thebibliography
\def\thebibliography{\DeclareRobustCommand{\VAN}[3]{##3}\VANthebibliography}

\usepackage{graphicx}	% Including figure files
\usepackage{amsmath}	% Advanced maths commands
\usepackage{amssymb}	% Extra maths symbols
\usepackage{xcolor}     % includes coloured text
\usepackage[normalem]{ulem} % add the ability to strike-out pieces of text

\title[Flexible climate model for exoplanets]{EOS-ESTM: A flexible climate model for habitable exoplanets}

\author[L. Biasiotti et al.]{
L. Biasiotti,$^{1,2}$\thanks{E-mail: lorenzo.biasiotti@inaf.it}
P. Simonetti$^{1,2}$,
G. Vladilo$^{1}$,
L. Silva$^{1,3}$, 
G. Murante$^{1}$,
S. Ivanovski$^{1}$,
\newauthor
M. Maris$^{1}$,
S. Monai$^{1}$,
E. Bisesi$^{1}$,
J. von Hardenberg$^{4}$,
A. Provenzale$^{5}$
\\
$^{1}$INAF - Trieste Astronomical Observatory,
Via G. B. Tiepolo 11,
34143 Trieste, Italy\\
$^{2}$University of Trieste - Dep. of Physics, 
Via G. B. Tiepolo 11, 
34143 Trieste, Italy\\
$^{3}$IFPU - Institute for Fundamental Physics of the Universe, Via Beirut 2, 34014 Trieste, Italy\\
$^{4}$Politecnico di Torino - DIATI, Corso Duca degli Abruzzi 24,
10129 Torino, Italy\\
$^{5}$Institute of Geosciences and Earth Resources - IGG,
National Research Council of Italy, 56127 Pisa, Italy\\
}

\pubyear{2021}

\begin{document}
\label{firstpage}
\pagerange{\pageref{firstpage}--\pageref{lastpage}}
\maketitle

\begin{abstract}
Rocky planets with temperate conditions provide the best chance for discovering habitable worlds and life outside the Solar System. In the last decades, new instrumental facilities and large observational campaigns have been driven by the quest for habitable worlds. Climate models aimed at studying the habitability of rocky planets are essential tools to pay off these technological and observational endeavours. In this context, we present EOS-ESTM, a fast and flexible model aimed at exploring the impact on habitability of multiple climate factors, including those unconstrained by observations. EOS-ESTM is built on ESTM, a seasonal-latitudinal energy balance model featuring an advanced treatment of the meridional and vertical transport. The novel features of EOS-ESTM include: (1) parameterizations for simulating the climate impact of oceans, land, ice, and clouds as a function of temperature and stellar zenith distance; (2) a procedure (EOS) for calculating the radiative transfer in atmospheres with terrestrial and non-terrestrial compositions illuminated by solar- and non-solar-type stars. By feeding EOS-ESTM with Earth’s stellar, orbital and planetary parameters we derive a reference model that satisfies a large number of observational constraints of the Earth’s climate system.  
Validation tests of non-terrestrial conditions yield predictions that are in line with comparable results obtained with a hierarchy of climate models. The application of EOS-ESTM to planetary atmospheres in maximum greenhouse conditions demonstrates the possibility of tracking the snowball transition at the outer edge of the HZ  for a variety of planetary parameters, paving the road for multi-parametric studies of the HZ.
\end{abstract}


%%%%%%%%%%%%%%%%%%%%%%%%%%%%%%%%%%%%%%%%%%%%%%%%%%

%%%%%%%%%%%%%%%%% BODY OF PAPER %%%%%%%%%%%%%%%%%%

\section{Introduction}

\label{sec:introduction}

Over the past two decades, ground- and space-based observations have unveiled thousands exoplanets and planetary systems around other stars in our Galaxy.   
About $4900$ exoplanets are currently confirmed\footnote{e.g. https://exoplanets.nasa.gov/; https://exoplanetarchive.ipac.caltech.edu/},  
in large part detected as transits by the {\it Kepler}\footnote{https://www.nasa.gov/mission\textunderscore pages/kepler/main/index.html} mission \citep{Borucki2010}. It's successor TESS\footnote{https://tess.mit.edu/} \citep[\textit{Transit Exoplanet Survey Satellite},][]{Ricker2015} is expected to boost the detection number, while CHEOPS\footnote{https://www.esa.int/Science\textunderscore Exploration/Space\textunderscore Science/Cheops} \citep[\textit{CHaracterizing ExOPlanet Satellite},][]{Broeg2018} will help to characterize the structural properties of already selected planets. In the short term, PLATO\footnote{https://sci.esa.int/web/plato/} \citep[\textit{PLAnetary Transits and Oscillations of stars},][]{Rauer2014} will search transiting Earth-analogues around bright stars.

The statistically relevant numbers of detected planets are allowing to investigate on all aspects of planetary structure and formation in different size ranges, as a function of stellar spectral type, composition, and even stellar multiplicity. A diversity of planetary systems architectures and a large range of planetary masses and/or radii have been observed, showing that the Solar System is just one possible outcome of the planetary formation process \citep[e.g.][]{UdrySantos2007,Howard2012,WinnFabrycky2015,Kaltenegger2017}.
Observations are necessarily biased toward giant gaseous planets around late-type stars, but the ever increasing statistics has allowed to infer that virtually any star in our Galaxy hosts at least one planet, with the planetary size distribution suggesting a steep increase towards small rocky Earth-like planets with thin atmospheres.
In fact, while it has been found that planetary masses offer a loose  
constraint on composition, currently in all cases it has been found that at small radii, R$_p \leq 1.5-2$ R$_{\oplus}$, all planets are rocky \citep[e.g.][]{Rogers2015} with a gap, i.e. an almost sudden transition, between Earth-like volatile poor and Neptune-like volatile rich planets \citep[e.g.][]{Fulton2017}.

These studies are shifting the current research from detection and statistics to full characterization of planetary properties, with one of the main goals of exoplanetary science being the quest for life outside the Solar System. 
This endeavour can only be tackled through remote atmospheric spectroscopy
\citep[transit, reflection, emission and their time variations, e.g.][]{Kreidberg2018} of potentially habitable rocky planets, in order to identify spectral features of biological origin. This possibility rests on the notion that the metabolic activity by-products for a well developed surface life may impact the atmospheric chemistry to a measurable amount \citep[e.g.][]{Lovelock1965,Kasting2014}. 

This observational challenge \citep[e.g.][for a review]{Fujii2018} should be partly within reach of the recently launched JWST \citep[\textit{James Webb Space Telescope},][]{Gardner2006,Kalirai2018}, probably limited to nearby M-type stars \citep[e.g.][]{Koll2019}, and, within the next decade, of the approved spatial mission ARIEL\footnote{https://arielmission.space/; https://sci.esa.int/web/ariel/} \citep[\textit{Atmospheric Remote-sensing Infrared Exoplanet Large-survey},][]{Tinetti2018}, although mainly for objects with warm H-dominated atmospheres. Nearby terrestrial analogues are expected to be detected with the ground-based E-ELT\footnote{https://elt.eso.org/} \citep{Snellen2015,Morley2017} equipped with the spectrograph HIRES \citep{Maiolino2013}. 
In the longer term, further space-based projects currently under assessment will be selected, that specifically aim to directly detect and characterize nearby temperate terrestrial analogues, e.g. HabEX\footnote{https://www.jpl.nasa.gov/habex/} \citep[\textit{Habitable Exoplanet Observatory},][]{Gaudi2020}, LUVOIR\footnote{https://asd.gsfc.nasa.gov/luvoir/} 
\citep[\textit{Large UV/Optical/IR Surveyor},][]{TheLUVOIRTeam2019}, OST\footnote{https://origins.ipac.caltech.edu/} \citep[\textit{Origins Space Telescope},][]{Wiedner2021}, LIFE\footnote{https://www.life-space-mission.com/} \citep[\textit{Large Interferometer For Exoplanets},][]{Quanz2021}.
The recent decadal survey for astronomy and astrophysics (Astro2020) report\footnote{https://nap.nationalacademies.org/catalog/26141/pathways-to-discovery-in-astronomy-and-astrophysics-for-the-2020s} endorsed 
recommendations for a single UV/Optical/IR flagship mission, that picks a compromise concept between LUVOIR and HabEX.

To accomplish the demanding task of searching for and  deciphering spectral signatures, a thorough and holistic observational and theoretical characterization of carefully selected rocky exoplanets is required.
The selection, among the observationally reachable targets for high-resolution spectroscopy of thin atmospheres, requires habitability studies with climate models. 
These simulations will enable the identification of those exoplanets with the largest chance of potentially hosting a surface diffuse life, i.e.\ with the largest habitability, that must be evaluated over a wide range of mostly unknown conditions. 
Moreover, the interpretation of any detected atmospheric features in terms of physical status of the atmosphere, and of their biotic or abiotic origin, will be unavoidably subjected to huge uncertainties and degeneracies, including false positives even for oxygen \citep[e.g.][]{Schwieterman2018,Meadows2018}. A considerable effort of modelization that exploits all available observations will be needed in order to assess the global physical characterization of the selected exoplanets, and in particular precisely of their potential surface climate and habitability.\\

Habitability studies for exoplanets rely on the concept of the habitable zone (HZ), classically defined as the range of stellar insolation, the main driver of climate, that allows surface temperatures compatible with a long-term presence of surface liquid water for a planet with an N$_2$–CO$_2$–H$_2$O atmosphere and a climate system stabilized by the carbonate-silicate feedback \citep[][]{Walker1981,Kasting1993,Kopparapu2013a,Kopparapu2014}. 
In these reference works, the inner and outer edges of the HZ are defined respectively for a H$_2$O- and a CO$_2$-dominated atmosphere, for an otherwise Earth-like planet orbiting stars of different spectral types. 
The HZ is considered as the prerequisite for potentially inhabited planets with exchange of gases between the biosphere and atmosphere \citep[][]{Kasting2014,Schwieterman2018} and therefore for spectroscopic biosignature searches.
Actually, in addition to insolation and spectral type, a large range of (mostly unknown) climate forcing factors affects planetary surface temperature and habitability, e.g. atmospheric mass and composition, surface gravity, radius, rotation period, obliquity, geography \citep[e.g.][]{Ramirez2019}, in addition to the observable orbital parameters. 
Also, different definitions of habitability could be envisaged and calculated for an optimal selection of exoplanets.

The large variety of planetary situations, expected and already uncovered by observations hints that a large range of non-Earth conditions should be accounted for. The majority of these parameters can currently only be explored with climate simulations.   
Currently $\sim 60$ observed rocky exoplanets are considered potentially habitable\footnote{ https://phl.upr.edu/projects/habitable-exoplanets-catalog, where the reported number refer to the empirical liquid water HZ, as defined by the insolation range received by Venus and Mars respectively $\sim 1$ and 4 Gyrs ago, when they could have hosted surface liquid water \citep[][]{Kasting1993}.}, but their number may change should a multi-parametric analysis of the huge possible parameter space of surface temperature be performed.\\   

The climate and the surface habitability of exoplanets can be explored, as for the Earth, using a hierarchy of models, depending on the aim and problem to be addressed \citep[see e.g.][for a review]{Shields2019}. Climate models should be able to account for, even at different levels of simplification, the complexity of the climate system due to the interplay of different components and processes, giving rise to feedbacks leading to multiple equilibria or even runaway conditions \citep[][]{Provenzale2013}. For instance, the water-vapour and the ice/albedo feedbacks set the spatial and temporal limits of the liquid water HZ. The accounting of these complexities is particularly important to simulate conditions not treated in Earth-tailored climate models.

Fully-coupled ocean-atmosphere General Circulation Models (GCM) are the most detailed and computing resources consuming models, in principle requiring a large amount of information to obtain meaningful results (e.g. detailed geography and orography).
In fact GCM are often applied for exoplanetary studies by adopting an Earth or simplified configurations, such as an aquaplanet \citep[e.g.][]{Leconte2013b,WolfToon2013,WolfToon2015,Shields2014,KS2015,Wolf2022}. They are fundamental tools to compute the coupled atmospheric-ocean dynamics on the long term and to study atmospheric dynamics in particularly complex configurations. These include rocky planets in the HZ of M-type stars, the most numerous and easiest targets for spectroscopy follow-ups. Due to their proximity to the host stars, these planets are expected to be tidally locked into synchronous rotation \citep[][]{Leconte2015,Barnes2017}.   
GCM are also fundamental benchmarks for  faster lower complexity models, allowing multi-parametric simulations.

Among such simpler models, 1D single-column radiative-convective models including detailed line-by-line radiative transfer (RT) have been used for instance to define the reference classical HZ mentioned above. 
Another class of 1D models are the so-called zonal Energy Balance Models (EBM), which solve a latitudinally-averaged energy balance with a simplified meridional heat diffusion equation \citep[][]{North1981,Spiegel2008}. 
This class of models is still applied to the Earth climate, to be able to explore and isolate the effects of specific processes on the global climate \citep[see e.g.][]{Pierrehumbert2010}. Their flexibility and short computing time can be exploited also for the large parameter space required to simulate exoplanetary conditions, by properly modelling all the terms entering the energy balance equation. By coupling single-column RT atmospheric modelling with an EBM \citep[e.g.][hereafter WK97 and V13, respectively]{WK1997,Vladilo2013,HaqqMisra2022}, and by further elaborating a physically-based description of the meridional transport, \citet[hereafter V15]{vladilo2015} developed a 2D EBM, the Earth-like planet Surface Temperature Model (ESTM), specifically aiming to compute the seasonal and zonal surface temperature of non-tidally locked exoplanets with a large range of non terrestrial (atmospheric and planetary) physical conditions. The range of applicability of this model was thoroughly explored in V15 by comparing with the 3D aquaplanet model by \citet[]{KS2015}. The flexibility and fast computing time of ESTM has been exploited in \citet[]{Murante2020}\footnote{The library of climate models used for this work was extracted from the ESTM-generated ARchive of TErrestrial-type Climate Simulations
(ARTECS) available at https://wwwuser.oats.inaf.it/exobio/climates/. The database is in continuous expansion.} for a statistical study of the multiple equilibrium states affecting climate systems due to non-linear feedbacks (e.g.\ the warm and snowball Earth states, during the latter the Earth would have been tagged as non-habitable). In \citet{Silva2017B} we performed with ESTM a multi-parametric exploration of the habitability for Kepler-452b \citep[][]{Jenkins2015}, currently the only known Earth-twin candidate.    
ESTM, by computing the latitude- and seasonal-dependent surface temperature, allows different operative definitions of habitability to be computed. Given the importance of liquid water for terrestrial life, the liquid-water temperature interval is the commonly adopted definition, and a pressure-dependent, liquid-water habitability index (V13) can be defined. But also biological temperature-
based considerations can provide further HZ definitions and can be all computed for each set of parameter choices \citep[e.g.][]{Silva2017A,VladiloHassanali2018}. These more restrictive definitions, as compared to the liquid water index, may help to increase the probability of selecting surface ambient conditions that maximize the production and detectability of atmospheric biosignatures \citep[a discussion on the necessity and possibly of the non-limiting assumption on searching for terrestrial-like life requirements can be found in e.g.][]{McKay2014,Kasting2014}.  \\

In this paper, we present a new release of the ESTM model, which we call EOS-ESTM. EOS is our new procedure for calculating RT in rocky planetary atmospheres with any pressure, chemical composition, and stellar spectral type \citep[][]{Simonetti2021}. In our previous version we were limited to Earth-like systems. We have introduced and improved on several new parameterizations with respect to the V15 model, in particular for the treatment of the temperature-dependence of the ice coverage over land and ocean, and for the zenith distance-dependence of the surface albedo specifically for any type of surface. We have carefully calibrated EOS-ESTM to reproduce the Earth climate by making use of large recent satellite (CERES-EBAF Ed4.1; \citet{loeb18}) 
and reanalysis (ERA5; \citet{Hersbach2020}) 
datasets, and validated the predictive power of the model through detailed comparison with 1D and 3D models under a large range of physical conditions. We also provide a first exploration of the dependence of the maximum greenhouse distance of the HZ on planetary parameters, as compared to 1D-based values.

The paper is structured as follows. In Section \ref{sec:model}, after a schematic summary of the ESTM model and parameters by V15, we provide a detailed description of each physical input of the model which has been either newly introduced or improved in the new EOS-ESTM release. In Section \ref{sec:reference_earth_model}
we exploit the large amount and good quality of experimental data of the Earth climate system 
to calibrate and validate our Earth's model, the reference for habitable rocky exoplanets.
In Section \ref{sec:model_validation} we present the validation of EOS-ESTM for a large range of non-terrestrial conditions with a comprehensive comparison of the predictive power of our model with several other 1D and 3D models. Our summary and conclusions are finally presented in Section \ref{sec:conclusions}.

%%%%%%%%%%%%%%%%%%%%%%%%%%%%%%%%%%%%%%%%%%%%%%%%

\begin{table*} 
\centering
  \caption{Treatment of the terms in Eq. \eqref{diffusionEq} in classic EBMs, ESTM, and EOS-ESTM}
  \label{tab:EBM_ESTM_comparison}  
  \scalebox{0.8}{%
\begin{tabular}{clcccl}
\hline 
 Term  & Description & Classic EBMs & ESTM & EOS-ESTM & Reference to the most updated prescription\\
\hline  
$C$ & Thermal capacity & $C$ = constant & Ocean, Land, Ice & ESTM + Transient ice 
& This paper \\
$D$ & Meridional transport & $D$ = constant & $D=D(p,g,R_p,RH,\Omega_\text{rot})$ & $D=D(p,g,R_p,RH,\Omega_\text{rot})$ & \citet{vladilo2015} \\
$I$ & Outgoing Longwave Radiation & $I=I(T)$ & CCM3 atmospheric RT & EOS atmospheric RT & \citet{Simonetti2021} \\
$S$ & Insolation & $S=S(t,\phi)$ & $S=S(t,\phi)$ & $S=S(t,\phi)$ & \citet{Vladilo2013} \\
$A$ & Top-of-Atmosphere Albedo & $A=A(T)$ & Surface \& clouds + CCM3 atm. RT & Surface \& clouds + EOS atm. RT 
& This paper; \citet{Simonetti2021} \\
\hline 
\end{tabular} } 
\end{table*}

\section{The climate model}
\label{sec:model}

In accordance with classic EBMs, the planetary surface is divided in a number $N$ of latitude zones and the zonal surface quantities of interest are averaged over one rotation period. In this way, the surface quantities depend on a single spatial coordinate,  the latitude $\phi$.
The thermal state of the surface is described by the temperature $T=T(t,\phi)$. Since the zonal quantities are averaged over one rotation period,
the time $t$ represents the seasonal evolution induced by the orbital eccentricity and  tilt of the rotation axis. 
By assuming that the heating and cooling rates normalized per unit area are  balanced in each  zone, one obtains a set of $N$ zonal energy balance equations
\begin{equation}
C  \frac{\partial T}{\partial t} - 
\frac{\partial}{\partial x}
\left[ D \, (1-x^2) \, \frac{ \partial T}{\partial x} \right]
+ I = S \, (1-A) ~
\label{diffusionEq}
\end{equation}
where we omit the index that runs from 1 to $N$ for simplicity.
The meaning of the terms in this equation can be summarized as follows.

\begin{itemize}

\item
The term $C$ represents the zonal heat storage and is expressed as heat capacity per unit area (J m$^{-2}$ K$^{-1}$).
It is calculated by summing the contributions of lands, $C_l$, oceans, $C_o$, ice over lands, $C_{il}$ and ice over oceans, $C_{io}$. These contributions are weighted according to the zonal coverage of each surface component.

\item
The second term of Eq. \eqref{diffusionEq} describes the meridional energy transport along the coordinate
$x=\sin \phi$. The transport is modelled 
using the formalism of heat diffusion modulated by the parameter $D$ (the diffusion term). 
As a major improvement with respect to classic EBMs, $D$ is expressed as a function of the physical quantities that most affect the meridional transport, such as the planetary radius, rotational angular velocity, surface gravity, and surface atmospheric pressure. A detailed description of the physics behind this formalism can be found in V15.

\item
The term $I$ is the Outgoing Longwave Radiation (OLR), which peaks in the thermal IR  band for typical conditions of habitable planets. At variance with classic EBMs, $I$ is estimated using single-column, radiative-convective calculations. By including the physics of the vertical transport, 
the ESTM becomes a 2D climate model, one dimension sampling the surface as a function of  latitude, as in classic EBMs, the other dimension sampling the atmosphere as a function of height from the surface. 
In practice, we calculate $I$ as a function of $T$ for a given chemical composition and vertical stratification of the atmosphere. Compared to the original ESTM, the calculations of atmospheric radiative transfer that we present here
have been greatly improved (see Section \ref{sec:radiative_transfer}).

\item
On the right hand of the Eq. \eqref{diffusionEq}, the term $S$ represents the insolation,
i.e. the incoming stellar radiation with maximum emission in the visibile/near IR spectral range.
More specifically, the zonal, instantaneous stellar radiation that heats the planet, $S=S (t,\phi)$,
is calculated taking into account the stellar luminosity, the orbital parameters and the inclination of the planet rotation axis. Details on these calculations can be found in V13. 

\item
The term $A$ is the albedo at the top of the atmosphere, i.e. the fraction of incoming photons that are reflected back in space without heating the planet. The calculation of $A$ is extremely more detailed than in classic EBMs and is performed in several steps. First, we calculate the surface albedo, $a_s$, by weighting the albedo contribution of 
lands, $a_l$, oceans, $a_o$, ice on lands, $a_{il}$, and ice on oceans, $a_{io}$, according the respective fractional coverage. 
Then, the total albedo at the bottom of the atmosphere is calculated
by summing the albedo of the clear-sky surface with the albedo of the clouds, weighted according to the fractional coverage of clouds. As an upgrade over the original ESTM, we now calculate the cloud albedo
taking into account the reflection of the underlying surface (Section \ref{s:cloudsAlb}).
Finally, the top-of-atmosphere albedo is calculated as a function of $T$, $a_s$ and stellar zenith distance, Z, for a given chemical composition and vertical stratification of the atmosphere. These calculations are performed with the upgraded recipes of radiative transfer that we present here (Section \ref{sec:radiative_transfer}).
All the albedo prescriptions are calculated as a function of the zonal, instantaneous stellar zenith distance, $Z= Z(t,\phi)$.
In the original ESTM the albedo dependence on $Z$ was considered for oceans, clouds, and atmosphere. 
Here we improve formulas and we introduce this dependence also for lands and ice.

\end{itemize}

In Table \ref{tab:EBM_ESTM_comparison} we summarize how the terms in Eq. \eqref{diffusionEq}
have been upgraded from classic EBMs to the ESTM. 
The main differences between the ESTM and the EOS-ESTM are summarized in Table \ref{tab:model_comparison}. 
In the rest of this section we review the prescriptions that we adopt to model the different components of the climate system, introducing the recipes that have been upgraded in the current EOS-ESTM version. 
Technical details on the solution of Eq. \eqref{diffusionEq} in the course of the climate simulation can be found in V15 (Appendix A). In the present version of the code we adopt 60 latitude zones, a starting temperature $T_0= 300$\,K, and a tighter criterion of convergence for the global mean temperature, $\langle T \rangle$: in practice, after running 20 orbits, the convergence is considered to be achieved when $| \delta \langle T \rangle/\langle T \rangle | < 10^{-5}$ in two consecutive orbits. All these parameters can be changed according to specific needs. For instance, $T_0$ can be varied in studies of climate bistability where two stable solutions (a Snowball state and a warm state) can be found in an appropriate parameter range depending on the initial temperature. As in most EBMs, the original ESTM \citet{Murante2020} and EOS-ESTM produce climate bistability.

\begin{table*} 
\centering
\caption{Main differences between ESTM and EOS-ESTM }
\label{tab:model_comparison}  
\scalebox{0.8}{%
\begin{tabular}{llll}
\hline 
Model prescription   &  ESTM & EOS-ESTM & Reference in this paper \\ 
\hline
Stellar spectrum & Solar & Any spectral type & Section \ref{sec:radiative_transfer} \\
Atmospheric composition & Earth-like  & Variable bulk composition & Section \ref{sec:radiative_transfer} \\ 
Greenhouse gases & Trace amounts of CO$_2$ and CH$_4$ & Significant amounts of any greenhouse gas & Section \ref{sec:radiative_transfer} \\
Surface albedo vs $Z$ & Oceans & Oceans, lands, ice & Sections \ref{s:oceans},\ref{s:lands},\ref{s:ice} \\
Calibration of ice coverage & Based on \citet{WK1997} & Based on Earth's satellite data & Section \ref{s:ice} \\
Albedo \& thermal inertia of transient ice &  Not treated & Function of zonal ice cover & Section \ref{s:trice} \\
Calibration of cloud albedo vs $Z$ &  Based on \citet[]{Cess1976}  & Based on CERES-EBAF satellite data & Section \ref{s:cloudsAlb} \\
Cloud short-wavelength transmission & Not treated & Two-valued function of $T$ & Section \ref{s:cloudsAlb}\\  
Cloud OLR forcing & Constant & Two-valued function of $T$ & Section \ref{s:cloudsOLR} \\  
Cloud coverage over ice & Constant & Decreasing with global ice coverage & Section \ref{s:cloudsCover}\\
\hline
\end{tabular} } 
\end{table*}

\subsection{Oceans}
\label{s:oceans}

\subsubsection{Ocean fraction}

The coverage of oceans on the planetary surface
is parameterized by assigning a fractional area coverage of oceans, $f_o$,  to each latitude zone.
This parameterization is sufficient to test the climate impact of different latitudinal
distributions and oceans, including the extreme cases of ocean  worlds ($f_o=1$ in each zone).
Oceans are characterized by their specific properties of albedo and thermal inertia.

\subsubsection{Ocean albedo}
\label{oceanAlb}

\begin{figure}
    \centering
    \includegraphics[width=75mm]{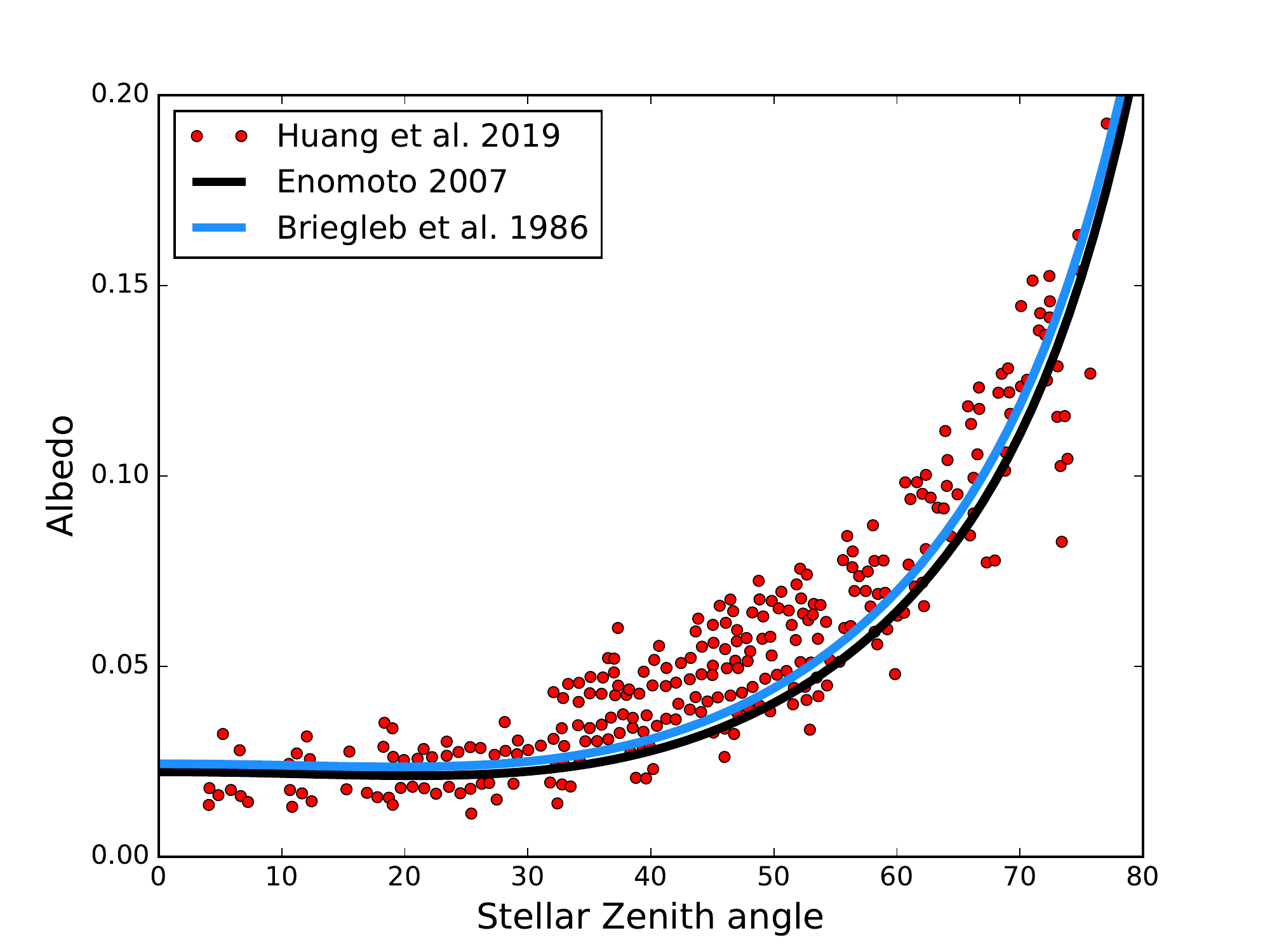}
    \caption{The albedo of oceans as a function of the stellar zenith angle, $Z$. The blue and black lines represent the formulation adopted in \citet{Briegleb1986} and \citet{Enomoto2007}, respectively. Red dots represent the data obtained from  \citet[Fig. 5a]{Huang2018}.}
    \label{fig:albedo_oceani}
    \end{figure}

The surface reflectivity of the oceans is modelled using empirical laws that take into account its dependence on $Z$ and the fact that the water surface is not smooth. 
We compared previous algorithms published in the literature \citep{Briegleb1986,Enomoto2007}
with a recent set of measurements obtained at different values of $Z$ \citep{Huang2018}. The observational data (red dots in Fig. \ref{fig:albedo_oceani}) show a large spread at any value of $Z$ due to the variations of atmospheric transmittance created by scattering and absorption of sunlight in the atmosphere \citep{Payne1972}. To model the ocean albedo we are interested in the data in clear sky conditions, since the transmittance of the atmosphere is accounted for in our radiative transfer calculations (Section \ref{sec:radiative_transfer}). The lower envelope 
of the data in Fig. \ref{fig:albedo_oceani} represent the clear-sky case. The formula proposed by \citet{Enomoto2007} (black line) and \citet{Briegleb1986} (blue line) are also shown in the figure. 
One can see that the expression proposed by \citet{Enomoto2007}, namely 
\begin{equation}
    \small
    a_o=\frac{0.026}{1.1 \mu ^{1.7} + 0.065}+ 0.15 (\mu -0.1) (\mu - 0.5) (\mu - 1.0) ~.
    \label{eq. albedo_oceani}
\end{equation}
(where $\mu = \cos Z$) yields a slightly better match to the lower envelope of the data. 
We therefore adopt this expression as in V15.
We do not propose an expression to match the lowest points of the observational data set
because we do not have information about the measurement errors and we cannot exclude the presence of outliers.

Despite having being calibrated with the Earth's oceans, the empirical law \eqref{eq. albedo_oceani} can be reasonably applied to any exoplanetary ocean, since the zenith dependence basically follows the universal Fresnel formula (WK97), corrected for the roughness of the surface.

\subsubsection{Thermal inertia of oceans}

Due to the high thermal capacity of water, the thermal inertia of oceans, $C_o$, gives a major contribution to the term $C$. 
The full oceanic contribution, which is effective over long time scales (typically decades on Earth), is not treated in the ESTM.
However, the short-term thermal impact of the oceans is accounted for by considering the contribution of the mixed layer, i.e. the surface layer of  water that exchanges heat with the overlying atmosphere \citep[hereafter P10]{Pierrehumbert2010}.
In the original version of ESTM, we adopted a mixed-layer contribution $C_\text{ml}=C_\text{ml50}$
(WK97,P10), corresponding to thermal inertia of a 50-m, wind-mixed ocean layer (see Table \ref{tab:model_parameters_C}). 
In Section \ref{sec:reference_earth_model} we present a new tuning of this parameter based on the short-term monthly variations of Earth's surface temperatures. For exoplanets with shallow oceans, $C_\text{ml}$ can be changed to simulate the impact of water layers of different depths.

\subsection{Lands}
\label{s:lands}

\subsubsection{Land fraction}

The surface fraction of continents is described by assigning a fractional area of land $f_l=1-f_o$, to each latitude zone.
This parametrization is sufficient to test the climate impact of extreme distributions of continents, such as polar or equatorial continents, or desert worlds ($f_o=0$ in each zone).
Continents are characterized by their specific properties of albedo and thermal inertia. 

\subsubsection{Land albedo}
\label{landAlb}

To model the albedo of land we adopt a formulation proposed by \citet{Briegleb1992}, namely
\begin{equation}
        a_x(\mu) = a_x(0.5) \frac{1+d}{1+2 \, d \, \mu}
        \label{eq:Briegleb92}
\end{equation}
where $a_x(0.5)$ is the albedo of a surface $x$ when $\mu$=0.5 ($Z = 60^{\circ}$), and the parameter $d$ regulates the dependence on stellar zenith distance  
($d=0.1$ for a ``weak" dependence; $d=0.4$ for a ``strong" dependence). 
These parameters can be varied according to the type of surface (desert, basalt, vegetation, etc.)
in order to model planets with specific characteristics \citep[Table 3 therein]{Coakley2003}. 
The adopted values should be representative of clear-sky conditions, since ESTM takes into account the effects of the atmospheric albedo separately (Section \ref{sec:radiative_transfer}).

\begin{table*}
  \centering
  \caption{Adopted terms for thermal inertia}
  \label{tab:model_parameters_C}
  \scalebox{0.8}{%
  \begin{tabular}{llll}
    \hline
    Parameter & Description & Adopted value & Comments\\
    \hline
    $C_\text{ml50}$ & \small{Thermal inertia of the water mixed layer} & $210 \times  10^6$ \small{J m$^{-2}$ K$^{-1}$} &  \small{Equivalent to a 50-m water layer \citep{Pierrehumbert2010} }\\ 
    $C_{\text{atm},\circ}$ & \small{Thermal inertia of the  atmosphere} & $10.1 \times 10^6$ \small{J m$^{-2}$ K$^{-1}$} & \small{Equivalent to a 2.4-m water layer  \citep{Pierrehumbert2010}} \\
    $C_\text{solid}$ & \small{Thermal inertia of the solid surface} & $1 \times 10^6$ \small{J m$^{-2}$ K$^{-1}$}  & \small{Equivalent to a 0.5-m rock layer \citep{Vladilo2013}} \\ 
    \hline
  \end{tabular}}
\end{table*}

\subsubsection{Thermal inertia}

The solid surface has a negligible thermal capacity compared to that of the oceans and even compared to that of a relatively thin, Earth-like atmosphere. The value of land thermal inertia that we adopt (Table \ref{tab:model_parameters_C}) is
representative of a layer of rock with thickness of 0.5 m \citep{Vladilo2013}.  
Even if small, this value becomes important in planets without oceans and with
extremely thin atmospheres.

\begin{figure*}
    \centering
    \includegraphics[width=75mm]{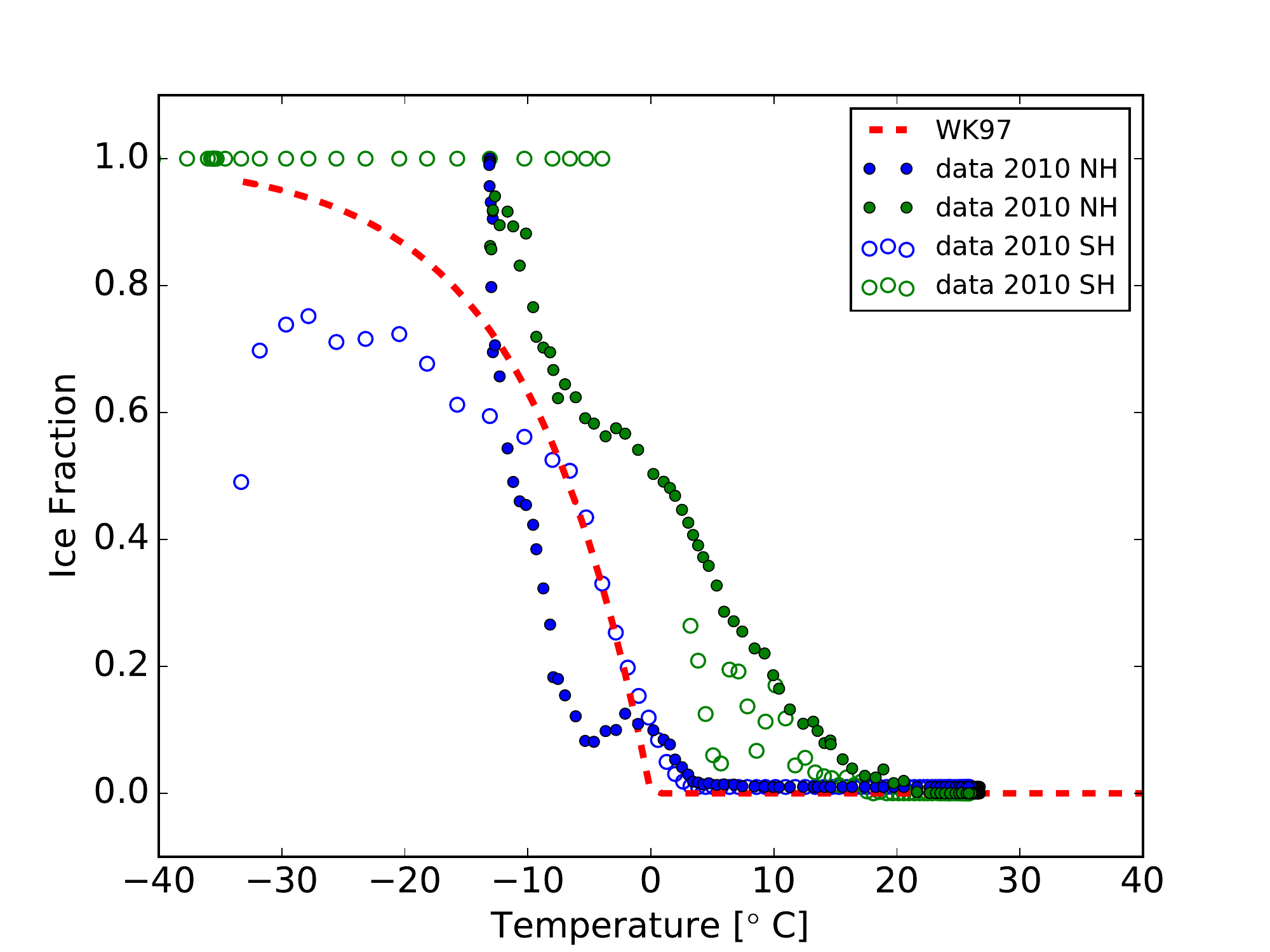}
     \includegraphics[width=75mm]{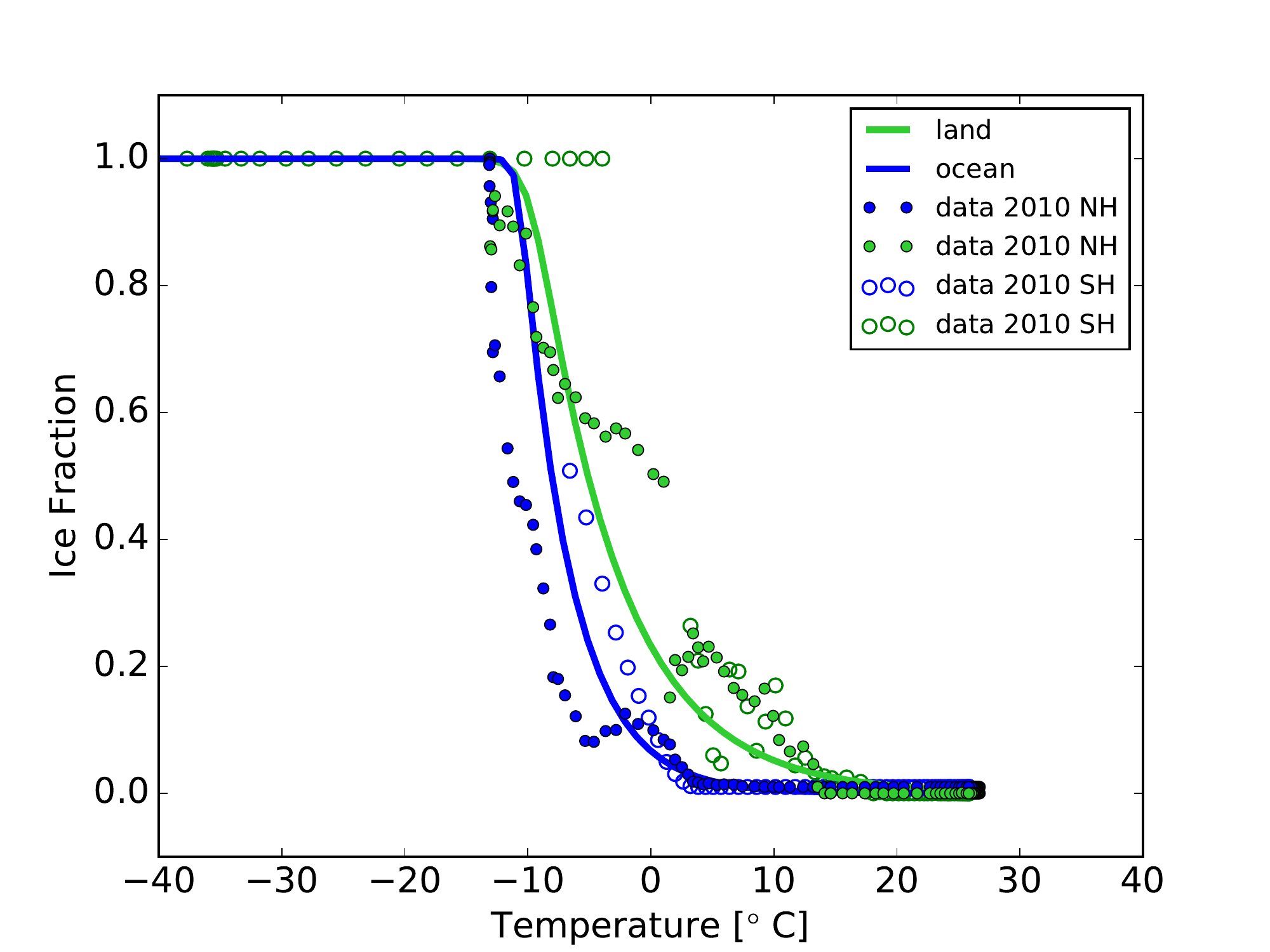}
    \caption{Mean-annual values of Earth's ice cover over lands and oceans plotted as a function of the mean annual zonal surface temperature.
    Left panel:  Observational data for lands (green symbols) and oceans (blue symbols) obtained with ERA5 reanalysis and NASA's Terra and Aqua Satellites averaged over the 2010; filled and empty circles represent data collected from the Northern and Southern hemisphere, respectively; red dotted line: prescription for the ice cover adopted in previous versions of ESTM.
    Right panel: same as in the left panel, but excluding data with 
    altitude above the freezing level  and the ocean data at the edge of Antarctica; 
    solid lines: prescriptions adopted in this work for lands (green line) and oceans (blue line);
    land data influenced by local orographic conditions specific of the Earth have been ignored; see Section \ref{s:icefraction}}. 
    \label{fig:ice_cover} 
\end{figure*}

\subsection{Ice}
\label{s:ice}

\subsubsection{Ice fraction} 
\label{s:icefraction}

ESTM calculates the fractional coverage of ice over lands, $f_{il}$, and oceans, $f_{io}$, making use of temperature-dependent algorithms. 
These prescriptions are critical because the ice coverage plays a key role in the albedo-temperature feedback and affects the lower temperature limit of liquid-water habitability. 
We paid special attention to re-calibrate the algorithms by searching for (i) a new set of experimental data and (ii) an appropriate functional dependence on $T$.

\begin{enumerate}

\item
The distribution of ice on the Earth surface was derived from measurements obtained from  NASA's Terra and Aqua satellites\footnote{See https://modis.gsfc.nasa.gov/} in the period 2005-2015. 
To find a trend with temperature, we associated the mean annual temperature of each latitude zone to the corresponding fraction of ice. The temperature data were obtained from the ERA5 dataset \footnote{See https://climate.copernicus.eu/climate-reanalysis}  \citep{Hersbach2020}  in the same period. This exercise was done separately for ice on lands and on oceans. To minimize the impact of orographic/oceanographic conditions specific of the Earth, we considered only land data in areas 
unaffected by local mountains and altitude below the freezing level 
 and we excluded ocean data 
at the edge of Antarctica\footnote{The combined effect of katabatic winds and ocean currents form mesoscale areas of open water near the Antarctica coastline, known as polynyas \citep{Stringer1991}. These areas fringe the edge of the continent  owing to the opening waterways (known as flaw leads) produced by the interconnection between themselves \citep[]{Meredith2016}.
}.
The results are shown in Fig. \ref{fig:ice_cover}, where the data  (filled and empty circles) show that the dependence on $T$ is quite different for lands and oceans (green and blue colors, respectively). 
The empirical trends show two features: (1) a very sharp rise of ice coverage below the water freezing point; (2) the existence of a small fraction of ice coverage at $T$ slightly above the freezing point.  The exponential model adopted in WK97 and in previous versions of the ESTM (dashed line in the left panel) is not able to reproduce these two features, indicating the need of a new type of functional dependence.

\item
After trying different types of functions,
we found that the empirical trends of Fig. \ref{fig:ice_cover} can be well approximated using a 
generalized logistic function \citep{Richard59}. 
Based on the physical boundary conditions of our problem,
 we chose a function that 
vanishes at very high $T$ and tends to 1 at very low $T$:
\begin{equation}
    f_{ix}(\overline{T})
    ={\frac{1}{\left[ 1+\xi_x \, e^{\theta_x \, (\overline{T}-T_{\circ,x})}\right]^{1/\xi_x}}}
\label{eq_icecover}
\end{equation}
where $\overline{T}$ is the zonal temperature averaged over a time $\tau_\text{ice}$ representative of the time scale of ice growth/melting; 
the index $x$ refers to the type of surface
underlying the ice cover ($x=o$ for oceans and $x=l$ for lands);
$T_\circ$ is the temperature turning point for the liquid-solid transition of water; 
$\theta$ is the growth rate; and $\xi$ is the shape parameter.

The zonal temperature $\overline{T}$ is averaged over the interval $\tau_\text{ice}$ that precedes the current time of the climate simulation. 
For consistency with the data shown in Fig. \ref{fig:ice_cover}, which have been averaged over one year, we adopt $\tau_\text{ice}=12$\, months. 
The parameters $T_\circ$, $\theta$ and $\xi$ were tuned using: (i) the data versus temperature shown in the right panel of Fig. \ref{fig:ice_cover}, and (ii) the data versus latitude of the Earth's reference model (bottom right panel in Fig. \ref{fig:earthannualzonal}). The adopted parameter values
are listed in Table \ref{tab:ice_cover_parameters}. The
resulting logistic functions (solid curves in the right panel of Fig. \ref{fig:ice_cover}) are able to reproduce  
the observed sharp rise below the turning point $T_\circ$ and the existence of a small ice fraction slightly above
$T_\circ$.

\end{enumerate}

\subsubsection{Albedo of frozen surfaces}
\label{iceAlb}

The albedo of frozen surfaces (ice and snow) shows a remarkable scatter in Earth measurements, with temporal variations that take place on different time scales. In ESTM we adopt representative values based on average conditions.
As an upgrade with respect to the previous version, we model the albedo of ice with Eq. \eqref{eq:Briegleb92}.
For stable ice on lands and oceans we adopt 
$a_{il}(0.5) = 0.70$ $a_{io}(0.5) = 0.55$, respectively.
In both cases we set $d=0.1$, i.e. a ``weak'' dependence on $\mu$ \citep{Briegleb1992}. 
These values provide a good match to Earth's zonal albedo (Section \ref{sec:reference_earth_model}), but can be changed to model exoplanets with specific properties of frozen surfaces. 

\subsubsection{Thermal inertia}

The contribution to thermal inertia of ice is important only if the planet lacks oceans and has an extremely thin atmosphere. For icy surfaces we adopt the same representative value adopted for any solid surface
(Table \ref{tab:model_parameters_C}). 
For icy surfaces over oceans, following WK97, we add a small contribution (10.5 $\times$ $10^6$ J m$^{-2}$ K$^{-1}$) representative of the thermal inertia of the underlying water.

\begin{table}
  \centering
  \caption{Parameters adopted for the ice coverage function \eqref{eq_icecover}.}
  \label{tab:ice_cover_parameters}
  \begin{tabular}{lll}
    \hline
    Parameter & Description & Value\\
    \hline
    \textbf{Land} & &  \\
    $T_{\circ \,l}$ & \small{Temperature turning point} & 265.15 K \\  
    $\theta_l$ & \small{Growth rate} &  1.2 \\
    $\xi_l$ & \small{Shape parameter} &  8.0 \\
    \textbf{Ocean} & &  \\
    $T_{\circ \,o}$ & \small{Temperature turning point} &  263.15 K \\
    $\theta_o$ & \small{Growth rate} &  3.0 \\
    $\xi_o$ & \small{Shape parameter} &  12.0 \\
    \hline
  \end{tabular}
\end{table}

\subsection{Transient ice} 
\label{s:trice}
 
The albedo and thermal capacity of transient ice
(ice that is forming or melting) are different from those of stable ice.
A possible way to take into account this effect is to introduce specific ice parameters in a temperature range around the water freezing point (WK97). However, this approach requires the introduction of several parameters not easy to quantify (the albedo and thermal capacity of unstable ice and the temperature range where the transition takes place). To avoid this additional parametrization we adopt
a prescription that provides a gradual change of the albedo and thermal capacity from the case in which the ice
is totally absent, to the case in which the ice is stable.  

\subsubsection{Albedo of transient ice} 
\label{triceAlb}

The albedo of stable ice over lands, $a_{il,s}$, and of stable ice over oceans, $a_{io,s}$,
is higher than the albedo of the underlying surface. When the temperature increases and the ice becomes more and more patchy, the albedo of unstable ice gets closer and closer to the albedo
of the underlying surface. To simulate this transition we assume that the fractional
coverage of ice \eqref{eq_icecover} is a reasonable estimator of the patchiness of the ice and
we adopt the expressions  
\begin{equation}
    a_{il}(\overline{T}) = a_l + (a_{il,s}-a_l) \, f_{il}(\overline{T})
    \label{albedo_lands_partial_ice}
\end{equation}
for ice over lands and
\begin{equation}
    a_{io}(\overline{T}) = a_o + (a_{io,s}-a_o) \, f_{io}(\overline{T}) 
    \label{albedo_oceans_partial_ice}
\end{equation}
for ice over oceans. 
In this way the albedo attains the high values typical of stable ice only when the ice coverage is complete. 
When the ice coverage is absent, the albedo equals that of lands ($a_l$) or oceans ($a_o$) without ice. 
The parameters in the above equations are calculated for $\mu=0.5$
(see Table \ref{tab:model_parameters_albedo})
and the albedo dependence on $\mu$ is modelled as explained in Section \ref{iceAlb}.

\subsubsection{Thermal inertia of transient ice}  

For the thermal inertia of transient ice we follow the same approach used for the albedo of transient ice. 
For the oceans, which provide the main contribution to the thermal inertia, we adopt the relation
\begin{equation}
    C_{io}(\overline{T}) = C_o + (C_{io,s}-C_o) \, f_{io}(\overline{T}) ~,
    \label{C_oceans_partial_ice}
\end{equation}
where $C_{io,s}$ is the thermal inertia of stable ice over ocean.
For land, the contribution is very small, and there is no need
to adopt a similar relation since in our parametrization $C_{il,s}=C_{l}$.

\subsection{Clouds}
\label{s:clouds}

The complexity of the physics of cloud formation and the lack of fluidodynamics and 3D capabilities of ESTM prevent us to model the spatial distribution and physical properties of clouds in a self-consistent way.
Clouds are treated as bottom-of-the atmosphere features with parametrized values of coverage, albedo, and OLR forcing. 
Given their critical role in the climate energy budget, we introduced upgraded algorithms for all these features.  

\subsubsection{Cloud fraction}
\label{s:cloudsCover}

The cloud fraction is estimated  with the expression
\begin{equation}
 f_c= f_o \, \left[ (1-f_{io}) \, f_{co} + f_{io} \, f_{ci} \right] +   
      f_l \, \left[ (1-f_{il}) \, f_{cl} + f_{il} \, f_{ci} \right]       
     \label{eq.cloud_cover}
\end{equation}
where $f_{co}$, $f_{cl}$, and $f_{ci}$ are representative values of the cloud coverage over oceans, lands, and ice, respectively (see Table \ref{tab:astroplan}).
These values are based on Earth data but, in principle, can be tuned for other planets.
In the original version of ESTM a constant cloud coverage over ice is adopted.
This approach is reasonable when ice is not dominant, as in the case of the present-day Earth.
However, when the planet undergoes a transition to a snowball state, the  global fraction of clouds is expected to decrease.
To capture this effect, we introduce a parameter  $f_{c,sb}$, representative of the cloud coverage in a snowball planet, and we adjust $f_{ci}$
in the course of the climate simulation 
introducing a dependence on the globally-averaged ice fraction, $\langle f_\text{ice} \rangle $.
In practice we adopt the expression
\begin{equation}
       f_{ci} = \left( f_{ci,\earth}-f_{c,sb} \right) \left( \frac{ 1 - \langle f_\text{ice} \rangle }{1 - \langle f_\text{ice} \rangle_{\earth} } \right) + f_{c,sb} ~~, 
       \label{eq_cloud_cover_ice}
\end{equation}
where $f_{ci,\earth}$ is the cloud coverage over ice calibrated with Earth's data.
With this prescription, $f_{ci}=f_{c,sb}$ when the planet enters a hard snowball state ($\langle f_\text{ice} \rangle=1$). The parameter $f_{c,sb}$ can in principle be tuned from results of GCM simulations of snowball planets \citep{Abbot14}. The Earth model is not affected by the choice of $f_{c,sb}$ because
$f_{ci}=f_{ci,\earth}$ when 
$\langle f_\text{ice} \rangle=\langle f_\text{ice} \rangle_{\earth}$.

\subsubsection{Cloud albedo} 
\label{s:cloudsAlb}

To upgrade the prescriptions for the albedo of the clouds we: (i) used an updated set of Earth satellite data; (ii) adopted a new functional form for the dependence of the cloud albedo on $\mu$; and
(iii) introduced
a dependence of the effective cloud albedo, $a'_c$, on the albedo of the underlying surface, $a_s$.

\begin{enumerate}

\item
To upgrade the experimental data, 
we use the recent set of top-of-atmosphere (TOA) albedo data obtained from the CERES-EBAF satellite \citep{loeb18}.  
Following Eq. (8) by \citet{Cess1976},
we estimate the zonal TOA albedo of the clouds with the expression
\begin{equation}
    A_\textrm{c,obs}=\frac{A_\textrm{obs}-A_\textrm{obs,clear}\, (1-f_c)}{f_c} ~,
    \label{eq. asc}
\end{equation}
where $A_\text{obs}$ are the TOA albedo data obtained in a given zone, $A_\text{obs,clear}$, the TOA albedo data obtained in the same zone in clear-sky conditions, and $f_{\textrm{c}}$ is the fractional cloud coverage in the latitude zone of interest. 
To extract the dependence on zenith distance, the derived values of $A_\textrm{c,obs}$
are associated to the mean annual value of $\mu=\cos Z$ of the zone of interest, $\overline{\mu}$.
In Fig. \ref{fig:CERES_Cess} we compare the cloud albedo versus $\overline{\mu}$ that was obtained by \citet{Cess1976} from the  dataset by \citet{EH1976}, with the results that we obtain inserting  the CERES-EBAF data for the period 2005-2015 in Eq. \eqref{eq. asc}.   One can see that the use of the updated and extensive dataset provided by CERES leads to significant differences. In particular, the cloud albedo becomes weaker in the equatorial regions. We use this updated data set to improve the description of the cloud albedo with respect to previous parametrizations (WK97, V15)
%\citep{WK1997,vladilo2015}
, which were based on \citet{Cess1976}.

\begin{figure}
    \centering
    \includegraphics[width=75mm]{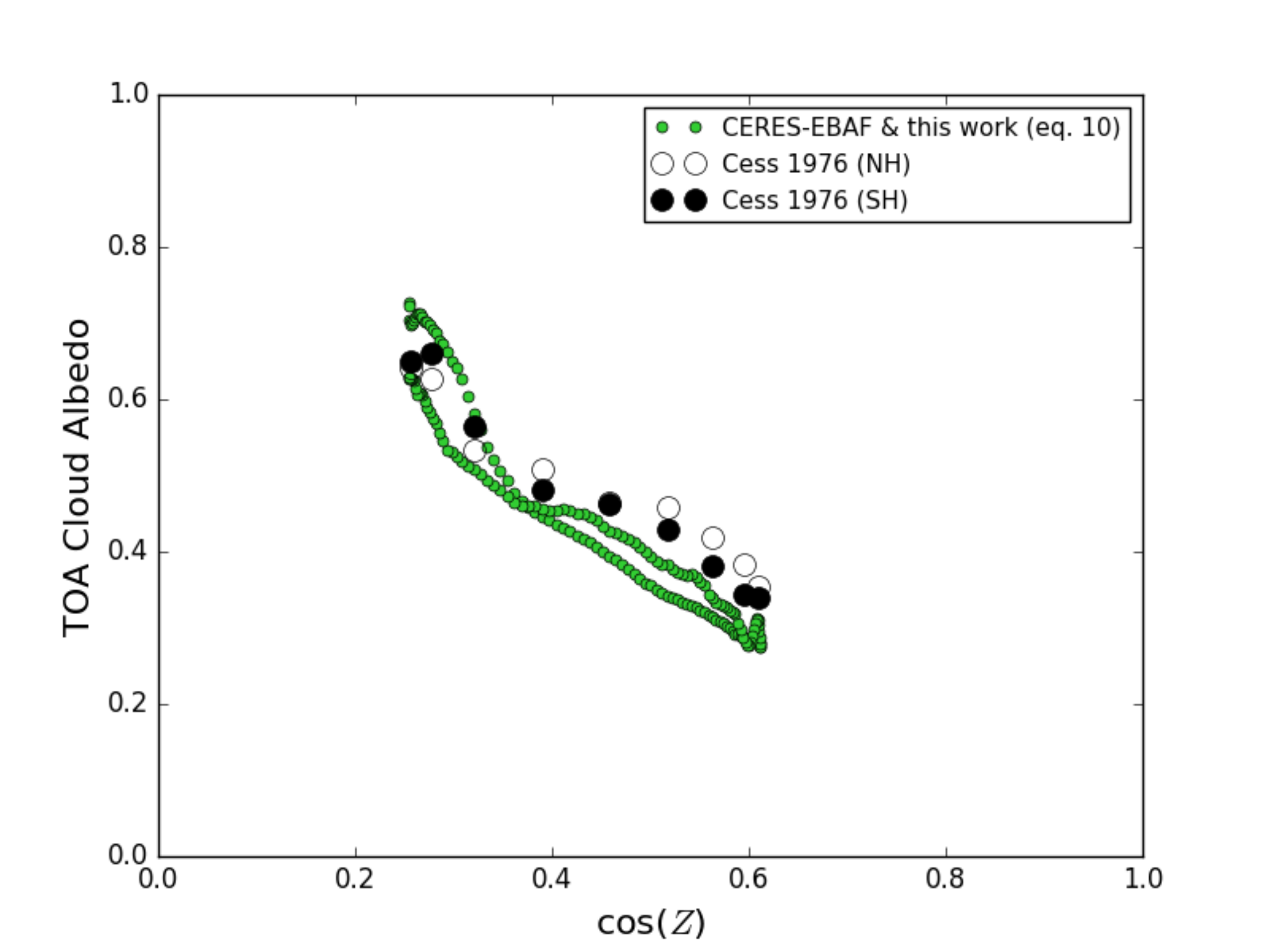}
    \caption{TOA cloud albedo profile obtained using the data collected from CERES-EBAF (green dots) in the period 2005-2015 compared with \citet{Cess1976} data for the NH (empty circles) and SH (black circles). The albedo is plotted versus the mean-annual Solar zenith angle of each latitude zone.}
    \label{fig:CERES_Cess}
\end{figure}

\item

In previous work, the dependence of the cloud albedo on the zenith distance was modeled using the linear form $a_\text{c}= \alpha + \beta \, Z$ 
(WK97,V13).
To prevent the existence of negative values of albedo at low $Z$, 
V15 introduced a third parameter
(the minimum value of cloud albedo at low zenith distances, $a_\text{c,min}$), and used the expression 
$a_c= \max\{a_\text{c,min},(\alpha+\beta Z)\}$. 
Here, for consistency with the description of the albedo of land, ice and ocean, we adopt a dependence on $\mu$, rather than on $Z$, also for the cloud albedo. In practice, we adopt
\begin{equation}
    a_c(\mu) = a_c(0.5) + m_c \, ( \mu - {\frac{1}{2}} )
    \label{eq. vladilo 2021}
\end{equation}
where $m_c$ is the slope and $a_c(0.5)$ is the cloud albedo at $\mu$=0.5. As can be seen in Fig. \ref{fig:cloud_albedo_parameterization}, this new function (solid line) 
yields positive values at low $Z$ without the need of an extra parameter, and  
yields a smoother dependence on $Z$ than the previous prescription (dashed line).

\begin{figure}
    \centering
    \includegraphics[width=75mm]{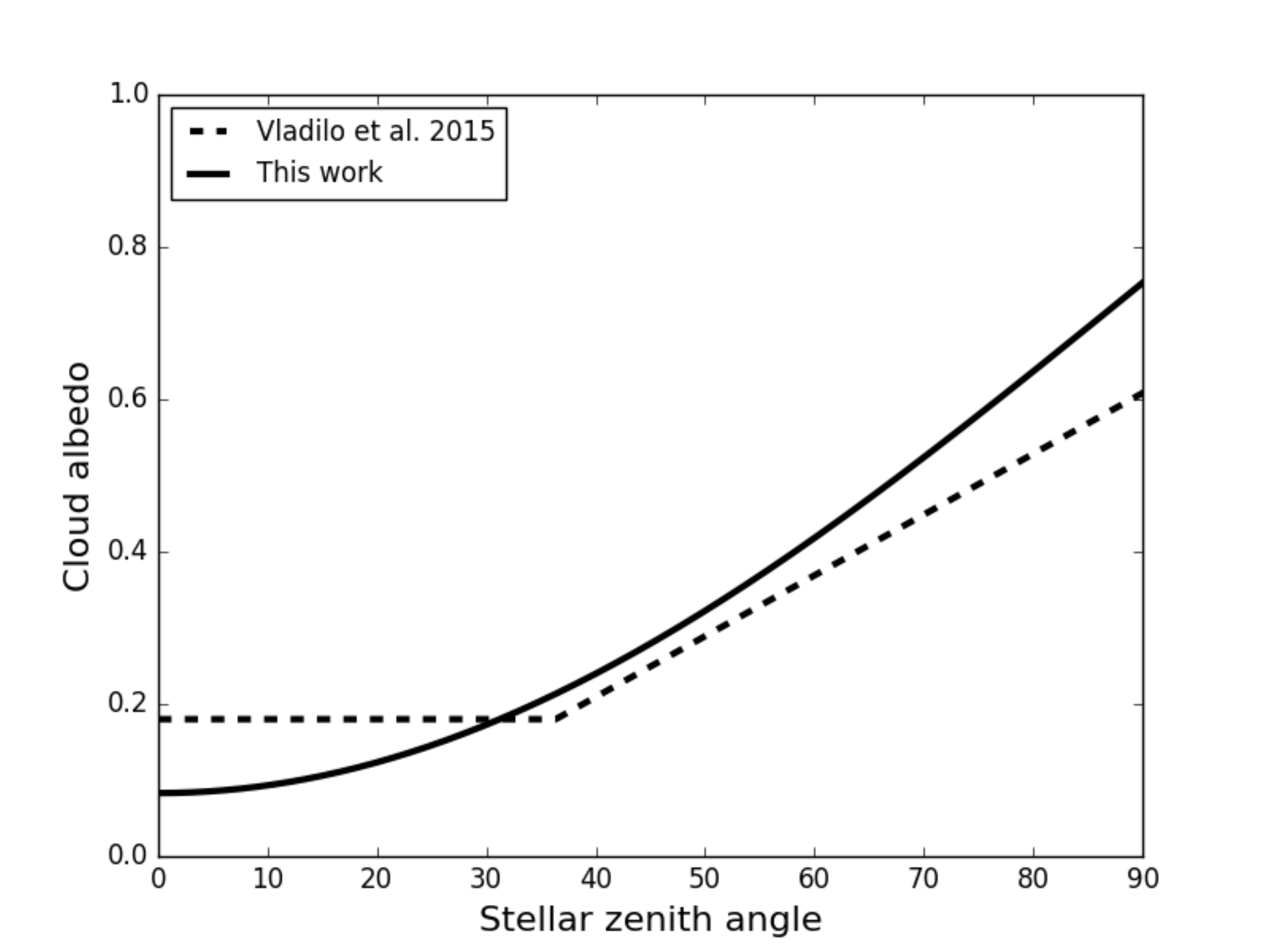}
    \caption{Comparison of the cloud albedo models adopted in V15 (dashed line) and in the present work (solid line). See Section \ref{s:cloudsAlb}.}
    \label{fig:cloud_albedo_parameterization}
\end{figure}

\item   
Polar clouds have a relatively high transmittance of short-wavelength photons, and part of the photons are reflected 
by the underlying surface, making their path up to the outer space through the clouds and the atmosphere. 
This effect becomes particularly important when the underlying icy surface is very reflective. 
The impact of this effect can be appreciated in Fig. \ref{fig:CERES_Cess}, where one can see that at low $\mu$, in correspondence with the ice polar caps, the slope of the Earth's cloud albedo versus $\mu$ becomes steeper.
To incorporate this effect in our model we follow \citet{ThomposonBarron1981} and, based on their Eq. (A14), we calculate the effective cloud albedo over reflective surfaces,
\begin{equation}
    \small
    a'_c =a_c+\frac{(1-a_c)
    \, (1-a^{*}_c)}{a^{*}_c}
    \times [(1-t^2\, a^*_\text{s}\, a^{*}_c)^{-1}-1]
    \label{eq. t6b81}
\end{equation}
where $a_c$ is the cloud albedo over a non-reflective surface, $a^{*}_c$ the cloud albedo for diffuse radiation, $a^*_\text{s}$ is the surface albedo for diffuse radiation; the values of diffuse albedo are estimated from the corresponding direct albedo calculated at $\mu$=0.5; 
the parameter $t$ is an estimator of the transmittance, 
i.e. the radiation fraction not absorbed between the cloud top and the surface. 
We model $t$ with the function
\begin{equation}
     t=0.90-0.05 \, \tanh { \left( \frac{T- 263.15 \,\text{K}}{10 \,\text{K}}  \right) }
     \label{cloud_transmittance}
\end{equation}
which provides a smooth transition between the typical transmittance of 
thin polar clouds ($t \simeq 0.95$; \citet{ThomposonBarron1981})
and the lower transmittance in regions without surface ice.
In these regions the surface albedo is low, 
the term $t^2\, a^*_\text{s}\, a^{*}_c$ is small, and
the exact choice of $t$ has a modest impact on the calculation of $a'_c$ with Eq. \eqref{eq. t6b81}.

\end{enumerate}

\subsubsection{Cloud OLR forcing}
\label{s:cloudsOLR}

In the original version of the ESTM the cloud OLR forcing had a constant value.
This approach is unsatisfactory since the OLR measurements of terrestrial clouds show an extremely large seasonal and latitudinal scatter \citep[Fig. 10 therein]{Hartmann1992}.  
As an attempt to introduce a variable value by capturing a dependence on $T$,
we tried to scale the cloud OLR according to the water vapour content
of the atmospheric column, which is a function of $T$ at constant relative humidity. 
However, this attempt did not provide a good match to the experimental data. 
This is not too surprising, given the complexity of the physics of cloud formation and cloud 
radiative transfer, which depends on a variety of thermodynamical and microscopic factors
not treated in our model. 
Despite the negative result of this attempt, we decided to take into account the properties of clouds over icy regions as an upgrade of our model. Terrestrial clouds over icy regions show a very small OLR forcing, typically one order of magnitude smaller than the average value \citep[Fig. 10 therein]{Hartmann1992}. 
To capture this effect, we calculate the cloud OLR forcing (also called cloud radiative effect, {\it CRE}) with the following expression
\begin{equation}
 CRE = CRE_\circ 
 \left[ 0.60+0.40 \, \tanh { \left( \frac{T- 263.15 \, \mathrm{K}}{10 \,\text{K}}  \right) } \right]
          \label{cloudOLRfactor}
\end{equation}
where $CRE_\circ$ is a representative value that can be calibrated with terrestrial clouds (see Section \ref{cloud}) and the term in square brackets varies smoothly from 1, at high $T$, to 0.2, at very low $T$. 
The above expression provides a smooth transition with decreasing $T$ that
mirrors the increase of cloud transmittance adopted in Eq. \eqref{cloud_transmittance}.
This simplified formalism is justified by the fact that it 
improves the match between predicted and observed zonal OLR in the Northern polar regions (see Section \ref{sec:reference_earth_model}).

\subsection{Atmosphere and top of atmosphere}

\subsubsection{Clear-sky radiative transfer} 
\label{sec:radiative_transfer}

The vertical radiative transport of energy throughout the atmosphere has a strong impact on  the $I$ and $A$ terms in Eq.~\eqref{diffusionEq} and therefore  plays a central role in the climate simulations.
In the previous version of ESTM, the terms $I$ and $A$ were estimated using the radiative transfer model developed as part of the Community Climate Model 3 \citep[CCM3,][]{kiehl98}, which is based on the HITRAN 1992 spectroscopic repository data \citep[]{rothman92}. CCM3 is a band model tailored for an  Earth-like atmosphere illuminated by  solar-type radiation. As such, the concentrations of greenhouse gases can only be varied in trace abundances, the list of greenhouse gases cannot be expanded, and it is not possible to model stellar spectra different from the solar one.  To overcome these limitations and to use an updated repository of spectroscopic data,  the new version of ESTM uses the EOS radiative transfer procedure \citep[]{Simonetti2021} to calculate the terms $I$ and $A$.

EOS is a line-by-line procedure based on the publicly available opacity calculator HELIOS-K \citep[]{grimm15,grimm21} and the radiative transfer code HELIOS \citep[]{malik17,malik19}. Line absorption from N$_2$, O$_2$, H$_2$O, CO$_2$ and CH$_4$ are calculated using data from the HITRAN 2016 \citep[]{gordon17} repository. The continuum of H$_2$O is included via the standalone version of the CNTNM routine of the LBLRTM code \citep[]{clough05}, which runs the MT\_CKD v3.4 opacity model \citep[]{mlawer12}. For CO$_2$-dominated atmospheres, the Collision-Induced Absorption (CIA) and the sub-Lorentzian absortion lines shape of CO$_2$ are also taken into account and calculated from, respectively, HITRAN data and the recipes in \citet[]{Perrin89}. The EOS model has the advantage of not being tied to a specific type of atmosphere via e.g.~gas opacity parameterizations, thus allowing a far greater flexibility in choosing the composition of the atmosphere. Radiative transfer calculations are performed on a 60-layer atmospheric column logarithmically spaced in pressure, from the $\sim$ 1 bar surface to the 1 $\mu$bar TOA level (10 layers per order of magnitude). The OLR is evaluated as a function of $T$ every 20 K below 280 K, every 10 K between 280 K and 310 K and every 5 K up to 360 K. The TOA albedo is evaluated every 20 K in the entire $T$ range up to 360 K, for $Z \in (0^\circ,30^\circ,45^\circ,60^\circ,70^\circ,75^\circ,80^\circ,83^\circ,85^\circ,87^\circ,88^\circ,89^\circ)$ and for $a_x \in (0.0,0.15,0.30,0.60,0.90)$. We adopted a non-equally spaced grid in order to sample more precisely the regions in which the OLR and the TOA albedo change their slopes. Multilinear interpolation on both OLR and TOA albedo tables is then carried out by ESTM. Thanks to the fact that HELIOS and HELIOS-K run on GPU processors (also known as graphic cards or accelerators) it is possible to calculate the line-by-line radiative transfer in a reasonable amount time even on desktop machines. As reported in \citet{grimm21}, GPU-based codes can be more than an order of magnitude faster than CPU-based ones.
Starting from the HITRAN line parameters files, EOS requires $\sim$70 hours to calculate the OLR and TOA albedo tables for a specific atmosphere\footnote{Estimates obtained using a workstation equipped with an nVidia RTX 2080 graphic card, a 7200 rpm HP/Seagate hard drive and an Intel Xeon Silver 4108 CPU. The reading-writing times from the storage memory (hard drive or solid state memory) and the CPU efficiency play important roles in determining the final amount of time required by the procedure, which can be consistently different on other machines.}.
Thanks to the modularity of the EOS procedure, part of the results obtained in the early steps can be reused, reducing the total time for the calculations of new cases.

\subsubsection{Top-of-atmosphere albedo}

The top-of-atmosphere (TOA) albedo is tabulated as a function of $T$ for a given set
of atmospheric parameters, using radiative transfer calculations in the short wavelength (visible, near IR)
spectral range. Since the TOA albedo also depends on the surface albedo, $a_s$,
and the stellar zenith distance, $Z$, the calculations
of the TOA albedo must sample the parameter space ($T$, $a_s$, $Z$).
The wavelength dependence of the short wavelength scattering implies that the albedo is sensitive
to the spectral distribution of the host star. 
The new radiative transfer procedure that we adopt allows us to tabulate $A$ for planetary atmospheres illuminated by stars of different spectral type.

\subsubsection{OLR}

For a given set of atmospheric parameters (surface gravity, surface pressure, chemical composition, relative humidity, vertical structure), the OLR in the thermal IR band is computed as a function of surface temperature $T$ by using a reverse calculation of radiative transfer where the temperature of the lowest atmospheric layer is forced to be equal to $T$. The resulting OLR tables, calculated in clear-sky conditions, are given in input to the climate simulation. 
The OLR forcing of clouds calculated with Eq. \eqref{cloudOLRfactor} is then subtracted from the clear-sky OLR.

\subsubsection{Thermal inertia of the atmosphere}

The atmospheric thermal inertia is much smaller than the oceanic one and can be neglected in
planets with oceans and thin, Earth-like atmospheres. However, in general, the atmospheric thermal inertia must be taken into account, since habitable planets may lack oceans and/or may have thick atmospheres. For this reason, in our formulation we scale the thermal inertia of the Earth's atmosphere (Table \ref{tab:model_parameters_C}) according to the thermal capacity and columnar mass of the planetary atmosphere (V15). The thermal capacity is calculated for the specific atmospheric composition; assuming hydrostatic equilibrium, the atmospheric columnar mass is calculated as $p/g$, where $p$ is the surface atmospheric pressure and $g$ the surface gravitational acceleration. The atmospheric contribution to thermal inertia is summed to the ocean, land, and ice contributions described above.

\begin{table*}
  \centering
  \caption{Astronomical and planetary Earth data}
  \label{tab:astroplan}
  \scalebox{0.9}{%
  \begin{tabular}{llll}
    \hline
    Parameter & Description & Adopted value & Reference/Comments\\
    \hline
    $S_\circ$ &  \small{Mean annual insolation} & 1361.0 W m$^{-2}$ & \small{CERES-EBAF (2005-2015)} \\
    $e$   & \small{Orbital eccentricity} & 0.01671022 & \\
    $\epsilon$ & \small{Axis obliquity} & 23.43929  & \\
    $g$ & \small{Surface gravity acceleration} & 9.81 m s$^{-2}$  & \\
    \hline
  \end{tabular}}
  \begin{tablenotes}
      \small
      \item {a. Adopted Earth's values can be changed 
      to model exoplanets with different types of stellar, orbital and planetary properties.}
  \end{tablenotes}
\end{table*}

\begin{table*}
  {\centering
  \caption{Earth satellite data and results of the Earth reference model}
  \label{tab:eraceresdata}
  \begin{tabular}{lllll}
    \hline
    Quantity & Description & Earth value & Model & Units\\
    \hline
    \texttt{$\langle T \rangle$} & Global surface temperature & 287.5$^a$ & 288.7 & K \\
    \texttt{$\langle T \rangle_\text{NH}$} & Mean surface temperature of Northern hemisphere & 288.4$^a$ & 288.5 & K \\
    \texttt{$\Delta T_\text{PE}$} &  North Pole-Equator temperature difference& 38.9$^a$ & 41.1 & K \\
    \texttt{$\langle h \rangle_\text{NH}$} & Fraction of habitable surface (Northern hemisphere) & 0.866$^b$ & 0.855 & ... \\
    \texttt{$\langle A \rangle$} & Global top-of-atmosphere albedo & 0.314$^c$ &  0.315 & ... \\
    \texttt{$\langle A \rangle_\text{NH}$} & Mean top-of-atmosphere albedo of Northern hemisphere & 0.310$^c$ & 0.314 & ... \\
    \texttt{$\langle OLR \rangle$} & Global outgoing longwave radiation & 240.2$^c$ &  241.4 & W m$^{-2}$ \\
    \texttt{$\langle OLR \rangle_\text{NH}$} & Mean outgoing longwave radiation of Northern hemisphere & 240.8$^c$ & 241.6 & W m$^{-2}$ \\
    \texttt{$\langle f \rangle_{c}$} & Global cloud fraction & 0.674$^c$ & 0.666 & .. \\
    \texttt{$\langle f \rangle_{c,\text{NH}}$} & Mean cloud fraction of Northern hemisphere & 0.644$^c$ & 0.646 & .. \\
    \texttt{$\Phi_\text{max}$} & Peak of atmospheric transport at mid latitudes  & 5.0$^d$ & 5.0 & PW \\
    \hline
  \end{tabular}}
  \begin{tablenotes}
      {\small
      \item {a. Average ERA5 temperatures in the period 2005-2015.}
      \item {b. Average fraction of planet surface with temperature satisfying the liquid water criterion.}
      \item {c. Average CERES-EBAF data in the period 2005-2015.}
      \item {d. \citet[]{Trenberth2001}}}
  \end{tablenotes}
\end{table*}

\begin{table*}
  \centering
  \caption{Parameters of surface and cloud albedo}
  \label{tab:model_parameters_albedo}
  \scalebox{0.9}{%
  \begin{tabular}{llrl}
    \hline
    Parameter & Description & Adopted value & Comments\\
    \hline
    $a_{l}$ & \small{Albedo of lands (at $\mu$=0.5)} & 0.20      &   \small{Tuned to match zonal albedo profile  (Fig.\,\ref{fig:earthannualzonal})} \\
    $a_{il,s}$ & \small{Albedo of stable frozen surfaces (at $\mu$=0.5)} & 0.70 &  \small{Tuned to match zonal albedo profile}  (Fig.\,\ref{fig:earthannualzonal}) \\  
    $a_{io,s}$ & \small{Albedo of stable ice on ocean (at $\mu$=0.5)} &  0.55    &  \small{Tuned to match zonal albedo profile  (Fig.\,\ref{fig:earthannualzonal})} \\
    $a_{c}$ & \small{Albedo of clouds (at $\mu$=0.5)} &  0.44 &  \small{Tuned to match CERES-EBAF data (Fig. \ref{fig:cloud albedoESTM})} \\
    $m_{c}$  &  \small{Slope of cloud albedo equation} & $- 0.67$   &  \small{Tuned to match CERES-EBAF data (Fig. \ref{fig:cloud albedoESTM})} \\
    $f_{cw}$ &  \small{Cloud coverage on water} & 0.72 &  \small{\citet{king2013}   }\\
    $f_{cl}$ &  \small{Cloud coverage on land} & 0.55 &   \small{\citet{king2013} } \\ 
    $f_{ci}$ &  \small{Cloud coverage on ice}   & 0.56 & \small{Tuned to match the cloud coverage of Earth's North Hemisphere} \\
    \hline
  \end{tabular}}
\end{table*}

\section{The reference Earth model}
\label{sec:reference_earth_model}

In this Section we present the calibration and the validation tests of the model applied to Earth, which represents the reference for modelling habitable exoplanets of terrestrial type.
The large amount and good quality of experimental data of the Earth climate system provide the best way for adjusting many important model parameters and testing the new recipes that we have introduced in the previous section.

\subsection{Astronomical and planetary quantities}

The values of astronomical and planetary quantities that we adopt for the reference Earth model are listed in Table \ref{tab:astroplan}. Global values of planetary temperature, OLR and albedo taken from  satellite observations (CERES-EBAF and ERA5) are summarized in Table \ref{tab:eraceresdata}.  Unless differently specified, all data were averaged for the period 2005-2015. For consistency, the insolation $S$ and the volumetric mixing ratio of CO$_2$ and CH$_4$ were estimated for the same time period. 
For the solar constant we adopt $S_0$ = 1361 W m$^{-2}$ in accordance with the mean annual insolation value measured from CERES-EBAF-EBAF (1361.16 W m$^{-2}$).

\subsubsection{Surface albedo}

The adopted values of albedo are listed in Table \ref{tab:model_parameters_albedo}.
For the albedo of lands
we adopt a value representative of the Earth continents, namely $a_l(0.5) = 0.20$,
which is an intermediate value between bare and vegetation-covered soil. 
As far as the dependence with $\mu$ is concerned, we adopt a ``weak" dependence ($d=0.1$), which is representative of most types of Earth's continental surfaces \citep{Briegleb1992,Coakley2003}.

\subsubsection{Clouds}
\label{cloud}

The parameters adopted for the clouds fraction on land, ocean and ice are shown in Table \ref{tab:model_parameters_albedo}.
With respect to V15, the coverage over land and ocean, $f_{cl}$, $f_{co}$, were updated following the experimental data by \citet{king2013}, while the adopted value of coverage over ice, $f_{ci}$, was estimated from the CERES-EBAF satellite data.

The value of the cloud radiative forcing for longwave radiation, which acts as a parameter in the model, was tuned in order to obtain a better match in the OLR profiles in Figs. \ref{fig:olr_vs_temp} and \ref{fig:earthannualzonal}. 
By adopting $CRE_\circ$ = 26.1 W\,m$^{-2}$ in Eq. \eqref{cloudOLRfactor}, we obtain an average value $\langle CRE_\text{olr} \rangle = 25.5$ W\,m$^{-2}$ for the Earth model. This is in excellent agreement with the mean value of the Earth, 25.8 W\,m$^{-2}$, obtained from CERES-EBAF Ed4.1, also considering the still large uncertainty on this quantity found in the literature.

\begin{table*}
  \centering
  \caption{Atmospheric radiative transfer parameters for the Earth's reference model}
  \label{tab:HELIOS}
  \scalebox{0.9}{%
  \begin{tabular}{llll}
    \hline
    Quantity & Description & Adopted value & Reference/Comments\\
    \hline
    $p_\text{dry}$ & Surface pressure dry air & $10^5$ Pa  &  \\
    $r$  & Relative humidity & 60\% & \citet{vladilo2015}\\
    $c_{CO2}$ & Atmospheric concentration of CO$_2$ & 350 ppm  & See section \ref{earthatmodata} \\
    $c_{CH4}$ & Atmospheric concentration of CH$_4$ & 1.7 ppm  & See section \ref{earthatmodata} \\
    $T_\text{tp}$ & Temperature of tropopause & 200 K & \citet{Seidel2001}, \citet{Kuell2005}\\
     $CRE_\mathrm{\circ}$ & TOA longwave forcing of clouds & 26.1 W m$^{-2}$  & tuned to match the OLR profiles in Figs. \ref{fig:olr_vs_temp} and \ref{fig:earthannualzonal}  \\
    \hline
  \end{tabular}}
\end{table*}

\subsubsection{Atmospheric quantities}
\label{earthatmodata}

Table \ref{tab:HELIOS} shows the atmospheric quantities adopted in the EOS radiative transfer calculations of the Earth model. 
Following \citet{Seidel2001} and \citet{Kuell2005} we adopt a temperature for the tropopause $T_\text{tp}=$ 200 K.
We adopt a relative humidity of $60$\%, in agreement with the global relative humidity measured on Earth, a surface pressure of dry air of $p_\text{dry}$=1.00 $\times$ 10$^5$ Pa and a volumetric concentration for CH$_4$ of 1.7 ppmv.
The CO$_2$ gas concentration of the reference period 2005-2015 ($c=390$ ppm, derived from the NOAA database \footnote{\url{https://gml.noaa.gov/ccgg/trends/}}) has been corrected to compensate the net TOA radiative imbalance of $\Delta F \simeq$ 0.6 Wm$^{-2}$\citep{Wild2013} observed in the current transient climate, since we will be performing constant-forcing simulations. We used the simplified analytical expression linking CO$_2$ concentration changes to the resulting radiative forcing change \citep[]{Myhre1998}:
%in Table \ref{tab:DFexpressions}
\begin{equation}
    \Delta F= \alpha \, \text{ln}(\frac{c}{c_0})  
\end{equation}
with $\alpha$ = 5.35 W m$^{-2}$, leading to a corrected volumetric mixing ratio for CO$_2$ of $c_0=350$ ppm.

\begin{table*}
  \centering
  \caption{Parameters for the meridional transport}
  \label{tab:meridionaltransport}
  \scalebox{0.80}{%
  \begin{tabular}{llll}
    \hline
    Quantity & Description & Adopted value & Reference/Comments\\
    \hline
    $D_\circ$ & Coefficient of latitudinal transport   & 0.66 W m$^{-2}$ K$^{-1}$  & \small{Tuned$^a$ to match the zonal  temperature profile (Fig.\,\ref{fig:earthannualzonal})} \\
    ${R}$ & Modulation of latitudinal transport & 1.4 &  \small{Tuned to match the zonal temperature  profile (Fig.\,\ref{fig:earthannualzonal})} \\
    $\Lambda_\circ$ &  Ratio of moist over dry eddie transport &  0.7 & \small{V15; Fig.\,2 in \citet{KS2015}} \\
    \hline
  \end{tabular}}
  \begin{tablenotes}
      \small
      \item {a. $D_\circ$ is also tuned to match the Earth's peak of atmospheric transport at mid latitudes, $\Phi_\mathrm{max}$ (Table \ref{tab:eraceresdata})}
  \end{tablenotes}
\end{table*}

%\subsection{Fine tuning of parameters}

\subsubsection{Meridional transport}

The parameters adopted for the meridional transport are shown in Table \ref{tab:meridionaltransport}.
The parameter $D_\circ$ was tuned to match the Earth's peak of atmospheric transport at mid latitudes, $\Phi_\mathrm{max}$ (5.0 PW, \citet{Trenberth2001}) and the temperature-latitude profile in Fig. \ref{fig:earthannualzonal}.
We refer to V15 for a full description of the parameters listed in the table.

\subsubsection{Depth of the mixed ocean layer}

\begin{figure}
    \centering
    \includegraphics[width=75mm]{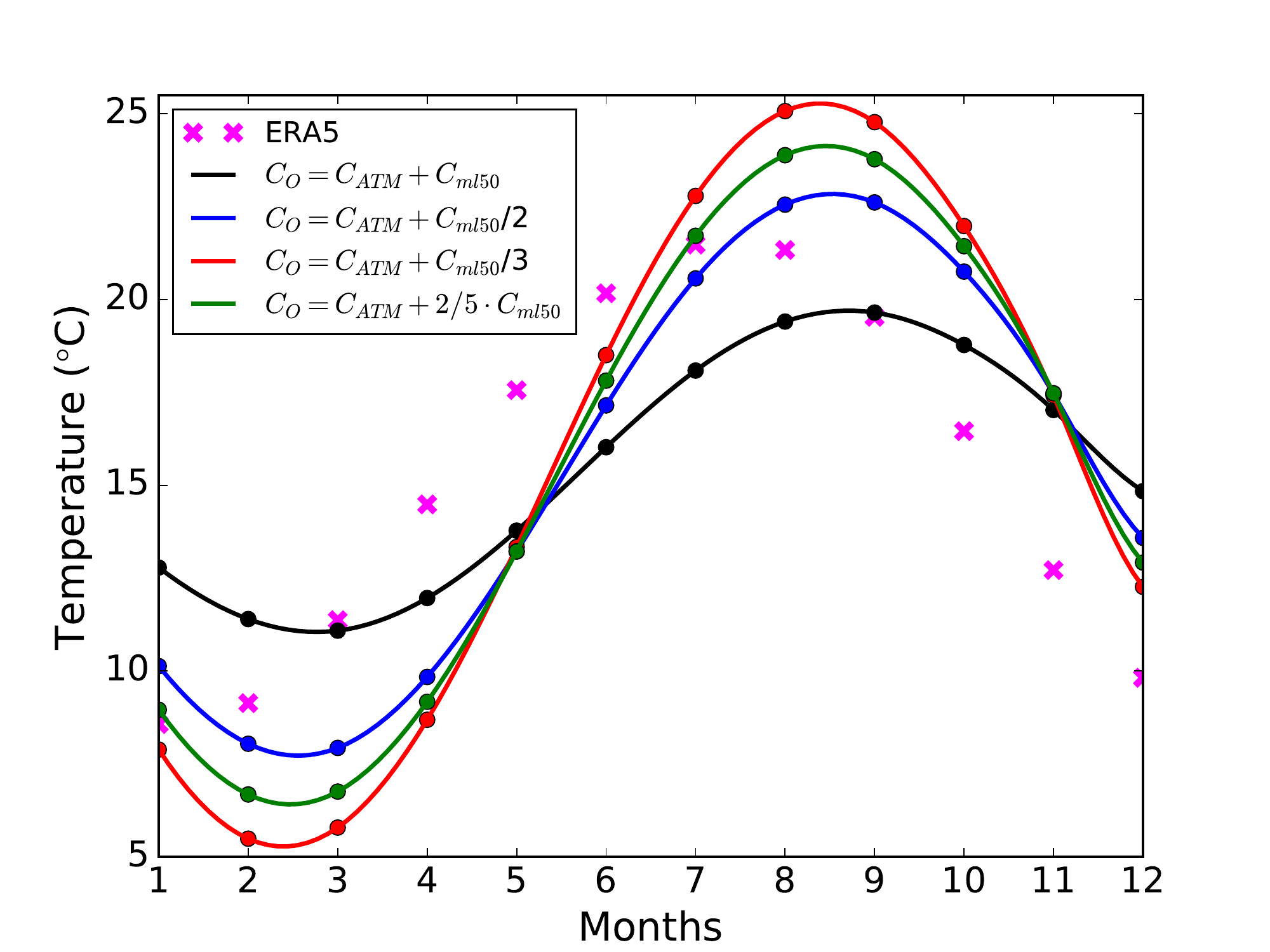}
    \caption{Seasonal evolution of mean surface temperature as a function of the effective thermal capacity. The black, blue, green and red lines represent the thermal capacities of oceans $C_o$ with $C_{ml50}$= 210, 105, 70 and 84 $\times$ 10$^{6}$ J m$^{-2}$ K$^{-1}$, respectively. Magenta Crosses: averaged ERA5 temperatures in the period 2005-2015 for the North Hemisphere (NH).}
    \label{fig:confronto_diversi_CO}
    \end{figure}

To tune the mixed ocean layer parameter, $C_\text{ml}$, 
we investigated the monthly excursions of the Earth global surface temperature.
In Fig. \ref{fig:confronto_diversi_CO} we compare the annual evolution  of  this quantity observed in the northern hemisphere (crosses) with the predictions of our model obtained for different choices of $C_\text{ml}$ (solid lines).
One can see that by increasing $C_\text{ml}$, the maximum annual excursion of monthly surface temperatures, $\Delta T_\text{max}$, becomes smaller. A small time lag between the predicted and observed peak is also present and increases with decreasing $C_\text{ml}$.  
The value of thermal capacity that better reproduces the observed trend is found at $C_\text{ml} \simeq C_{ml50}/2$. We therefore adopt this value, which provides a small time lag and a  difference of only 0.8 K between the observed and predicted value of $\Delta T_\text{max}$.

\subsubsection{Cloud albedo parameters} 
 
\begin{figure}
    \centering
   \includegraphics[width=75mm]{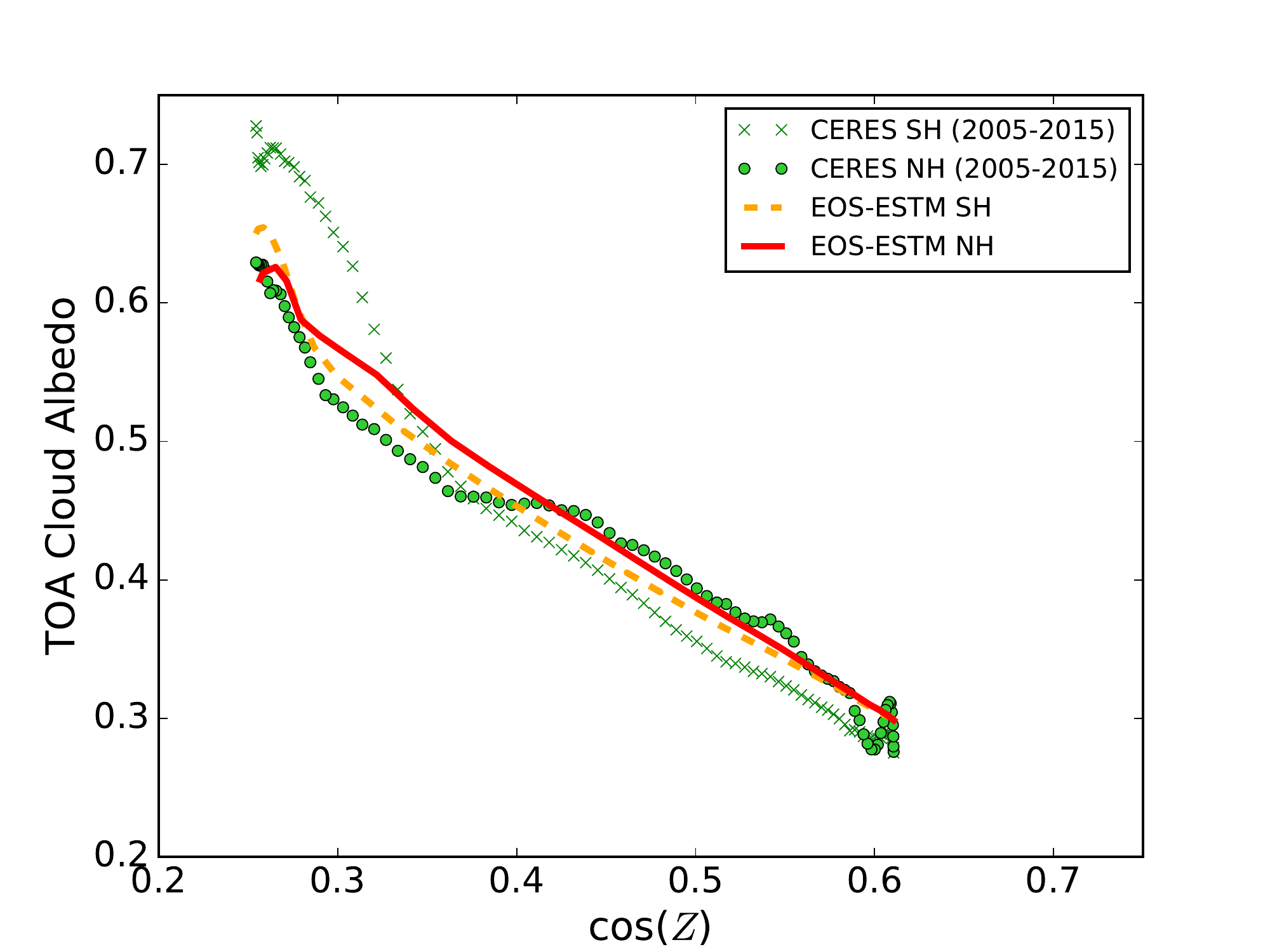}
    \caption{Cloud albedo versus $\mu=\cos Z$ for the present-day Earth (Northern and Southern hemispheres). Dark green circles and crosses: observational data obtained from Eq. \eqref{eq. asc} for the Northern and Southern hemisphere, respectively. Red-solid and orange-dashed lines: ESTM predictions obtained by converting to TOA values the cloud albedo calculated from Eq. \eqref{eq. t6b81} for the Northern and Southern hemisphere, respectively. }
    \label{fig:cloud albedoESTM}
\end{figure}

To tune the cloud albedo parameters $a_c(0.5)$ and $m_c$ in Eq. \eqref{eq. vladilo 2021}
we used the TOA cloud albedo estimated from satellite data with Eq. \eqref{eq. asc}.
The corresponding model predictions were then calculated by applying the EOS radiative transfer calculations
to the effective cloud albedo at the bottom of the atmosphere  estimated with Eq. \eqref{eq. t6b81}.
In this process we fine-tuned the cloud transmittance defined in Eq.\eqref{cloud_transmittance}.  
In Fig. \ref{fig:cloud albedoESTM} we compare the observational data set (green circles) with the final result of this modelization (orange curve).
One can see that the new calibration of cloud albedo provides a good match to the measured trend 
of TOA cloud albedo versus $\mu$,
including the sharp rise observed at low $\mu$, i.e. over the Earth polar caps.
From this figure it is clear that the reflectivity of the underlying surface becomes fundamental at the poles. 
The previous version of the ESTM is unable to reproduce these features.  

\begin{figure*}
    \centering
    \includegraphics[width=75mm]{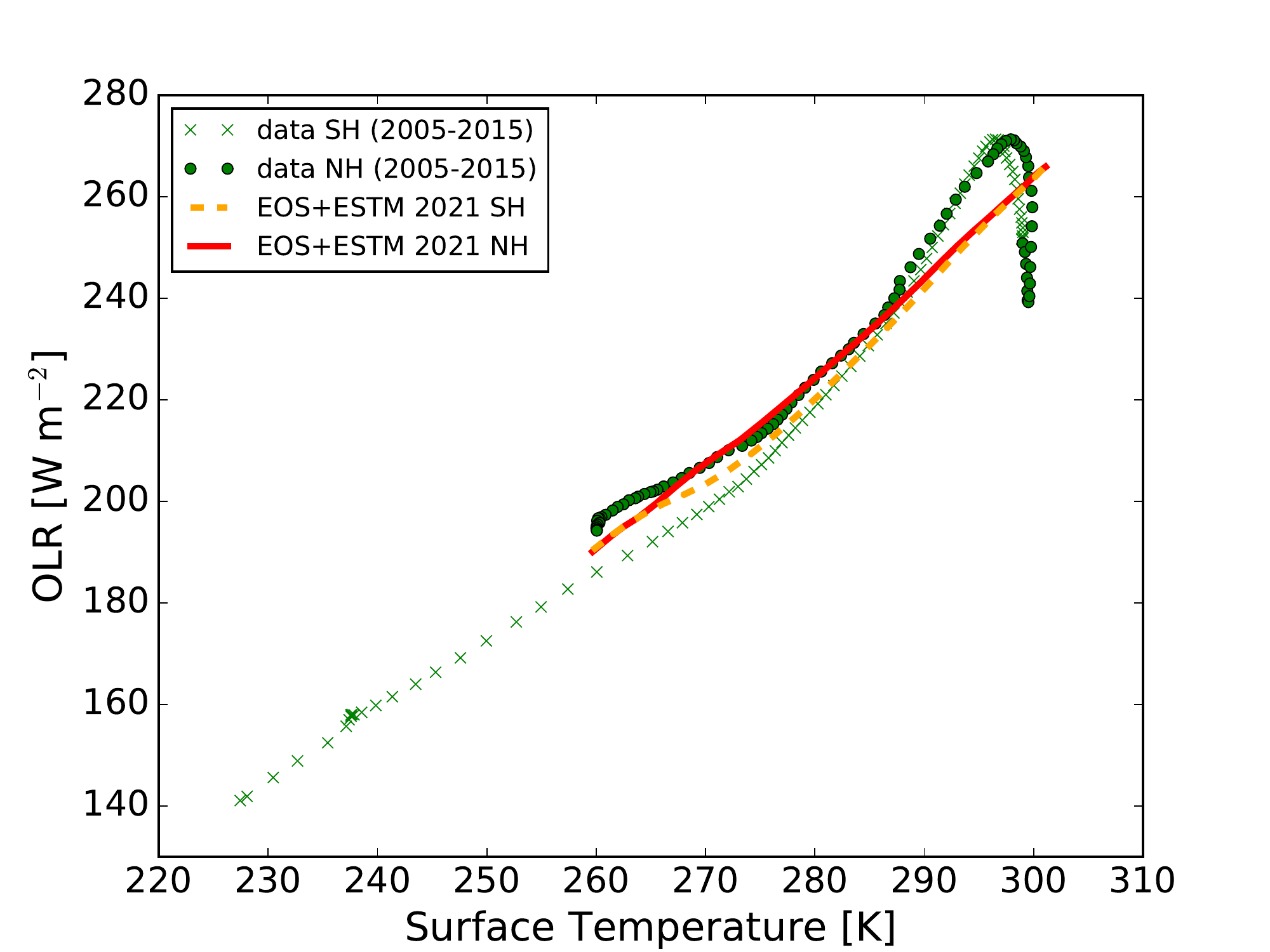}
     \includegraphics[width=75mm]{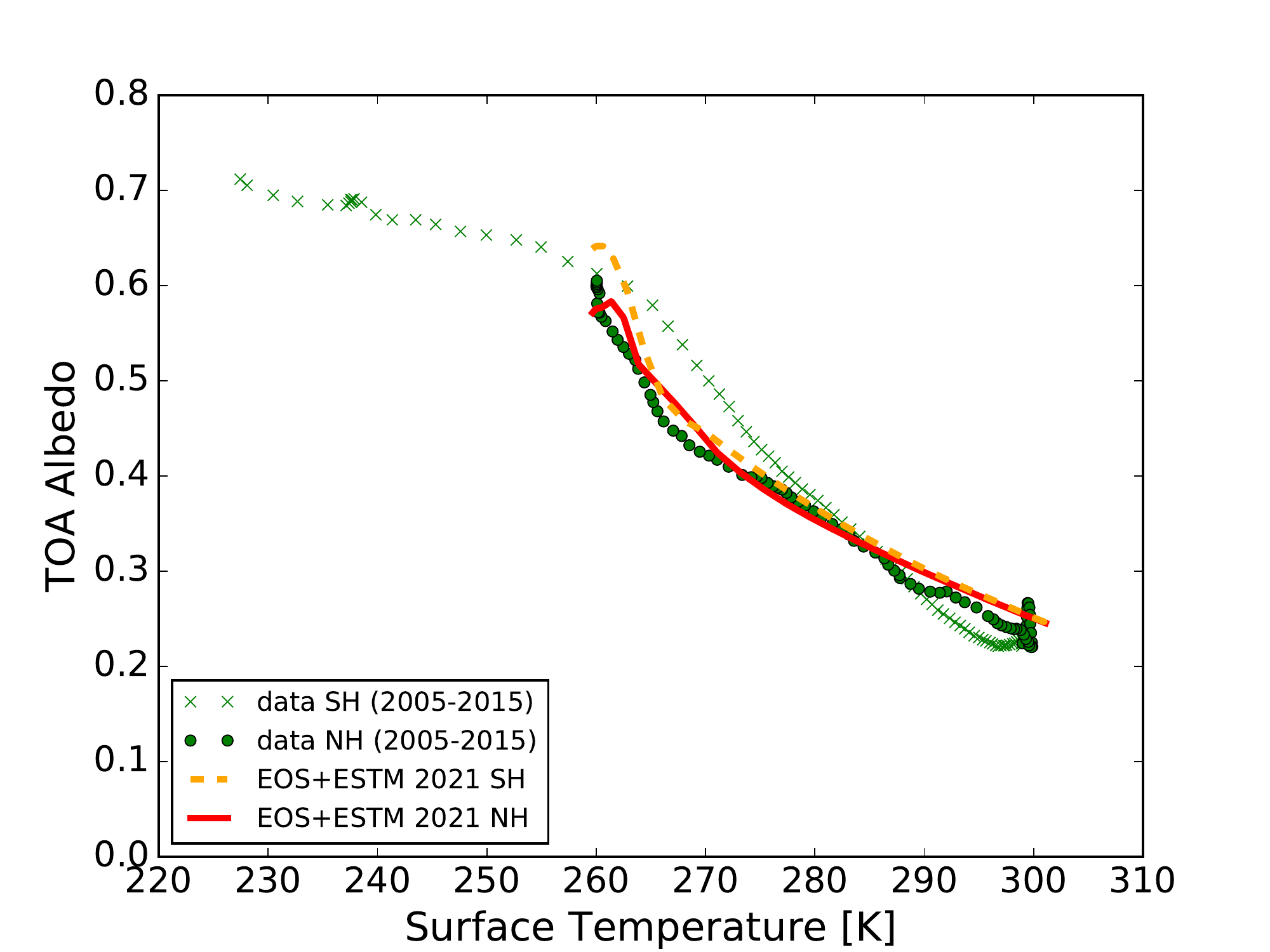}
    \caption{Mean annual values of OLR (left panel) and TOA albedo (right panel) plotted versus surface temperature for the present-day Earth. Data at temperatures below 260 K are related to the South Polar cap. Green dots and crosses: observational data obtained with the ERA5 reanalysis and CERES-EBAF satellite data averaged over the period 2005-2015, for the Northern and Southern hemisphere, respectively. Red-solid and orange-dashed lines: predictions of the Earth's reference model obtained with the radiative transfer calculations specified in Section \ref{earthatmodata}, for the Northern and Southern hemisphere, respectively.
    The OLR peak at $\simeq$ 296 K in the left panel is due to emission at the edges of tropical regions, which are connected with the presence of large deserts and of low clouds with warm tops; the decrease around the equator is associated with the presence of deep convective clouds with cold tops.
    }
    \label{fig:olr_vs_temp}
\end{figure*}

\begin{figure*}
    \centering
    \includegraphics[width=75mm]{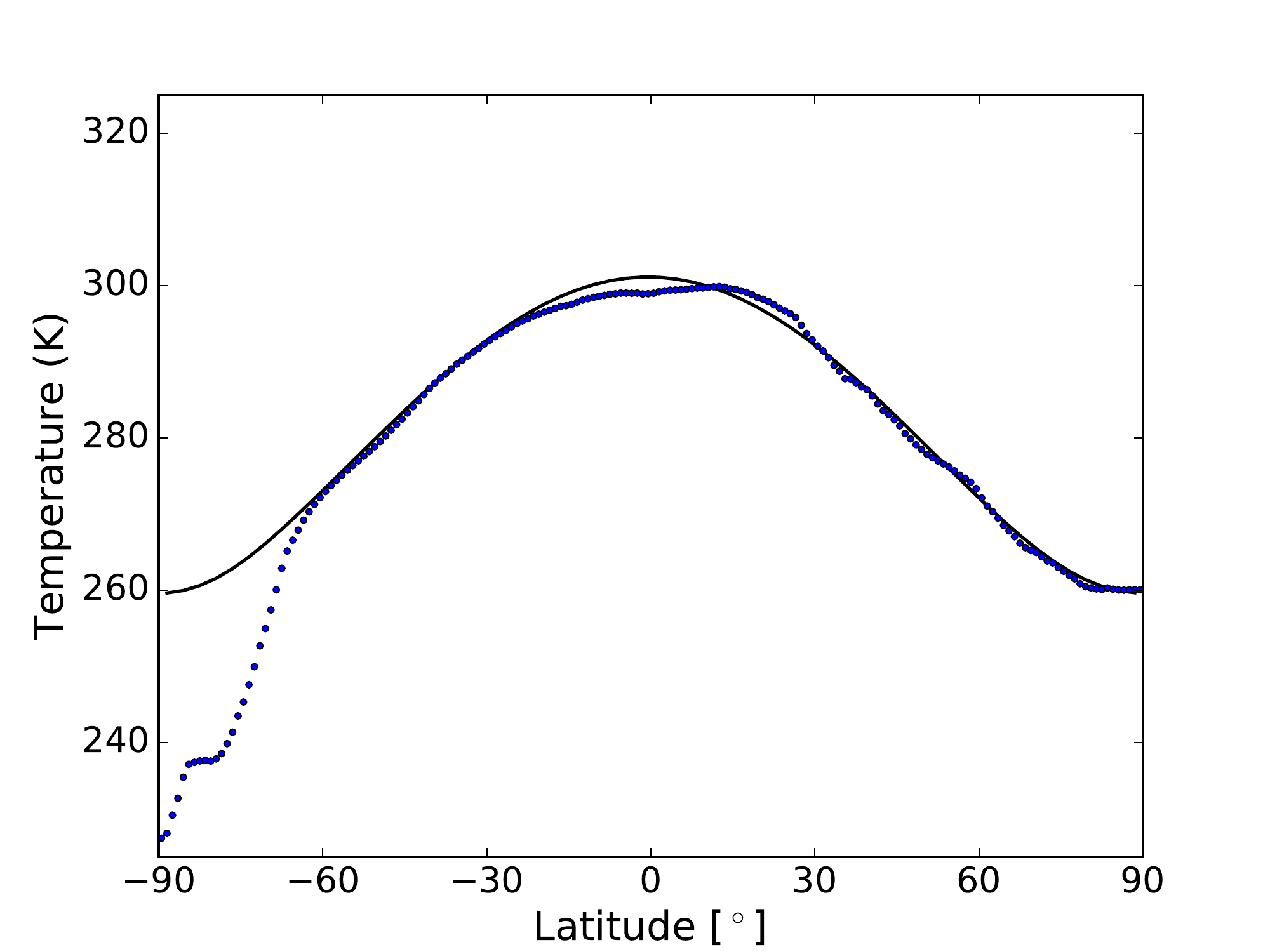}\quad\includegraphics[width=75mm]{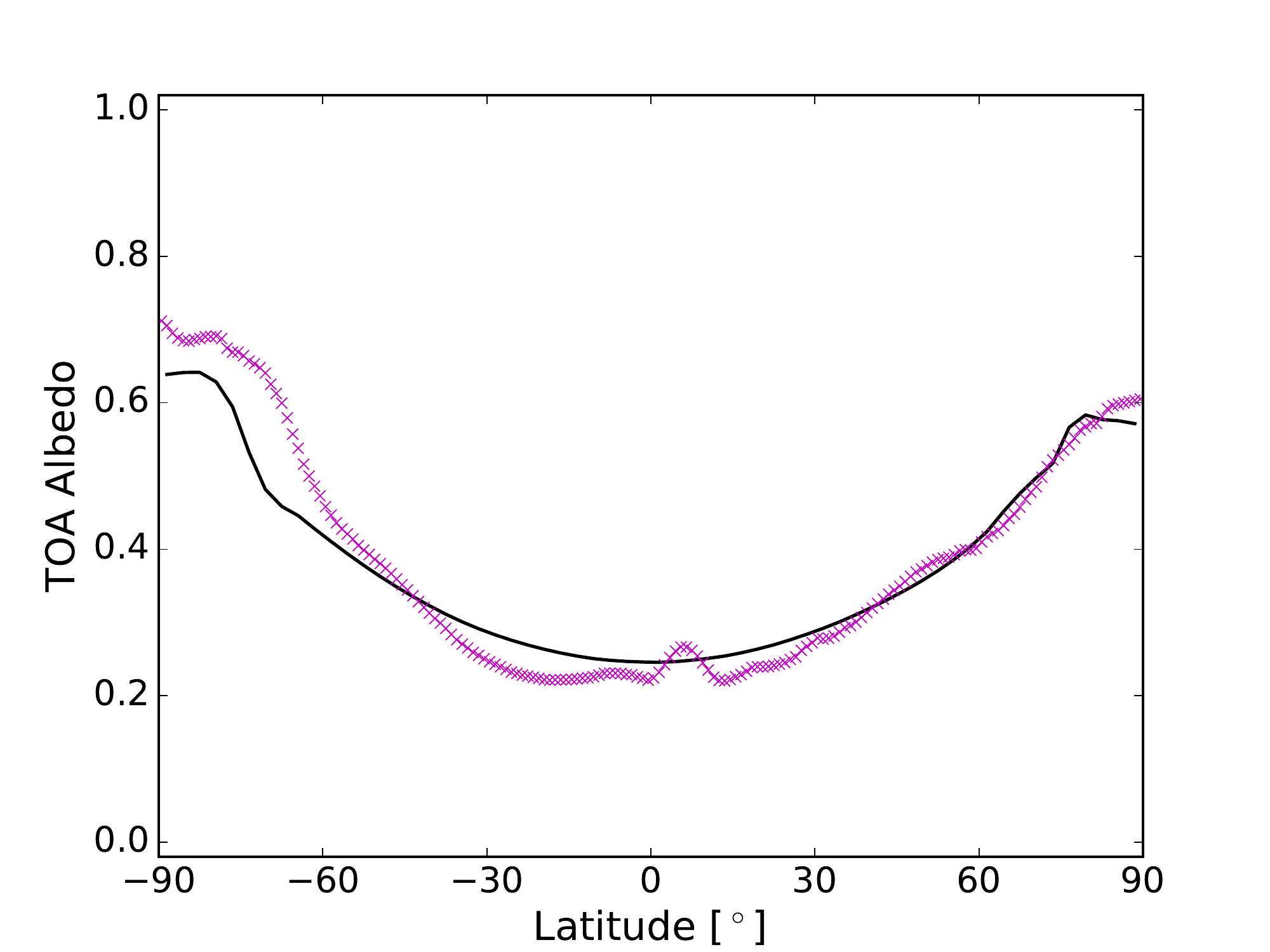}\quad\includegraphics[width=75mm]{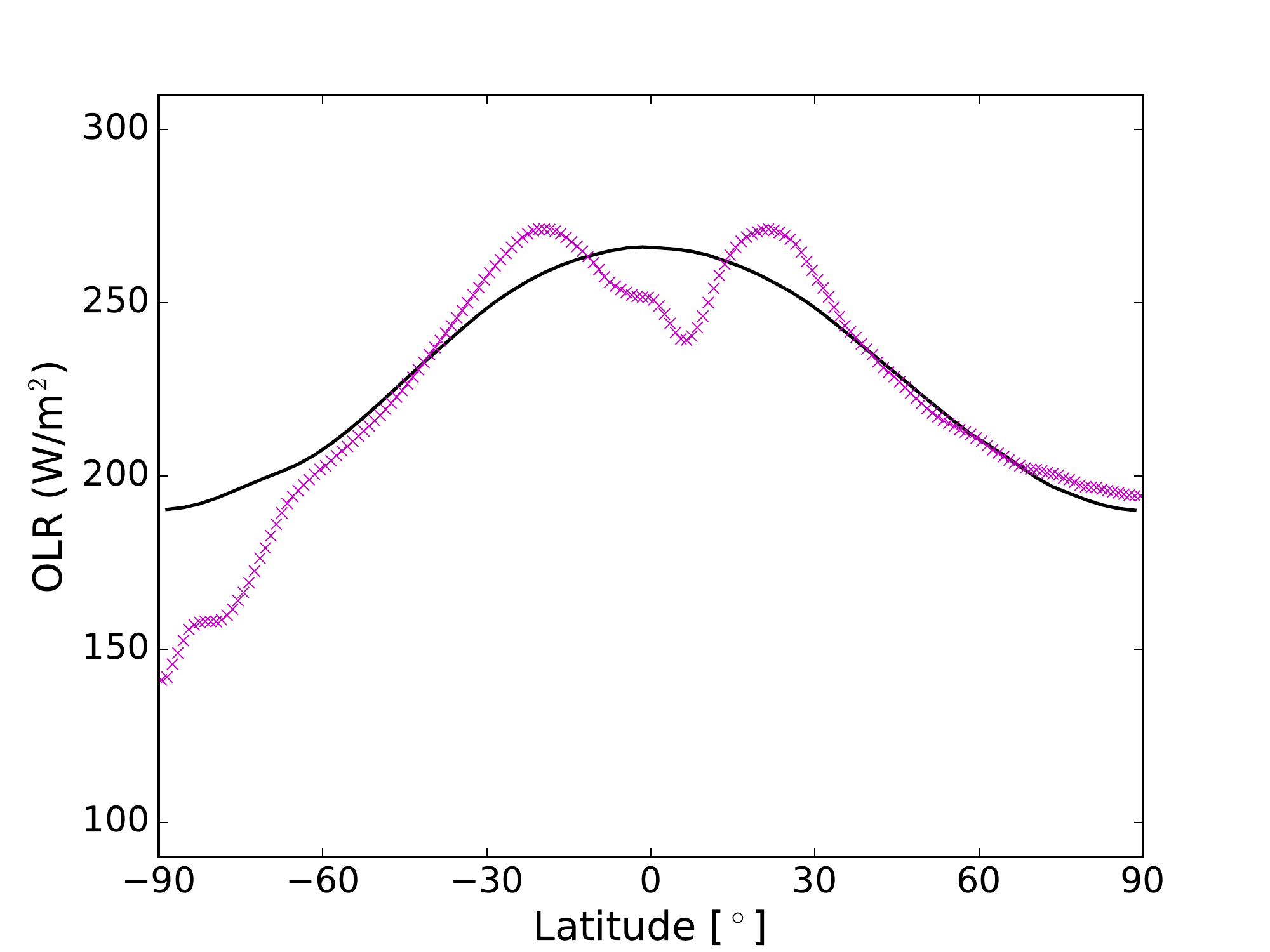}\quad\includegraphics[width=75mm]{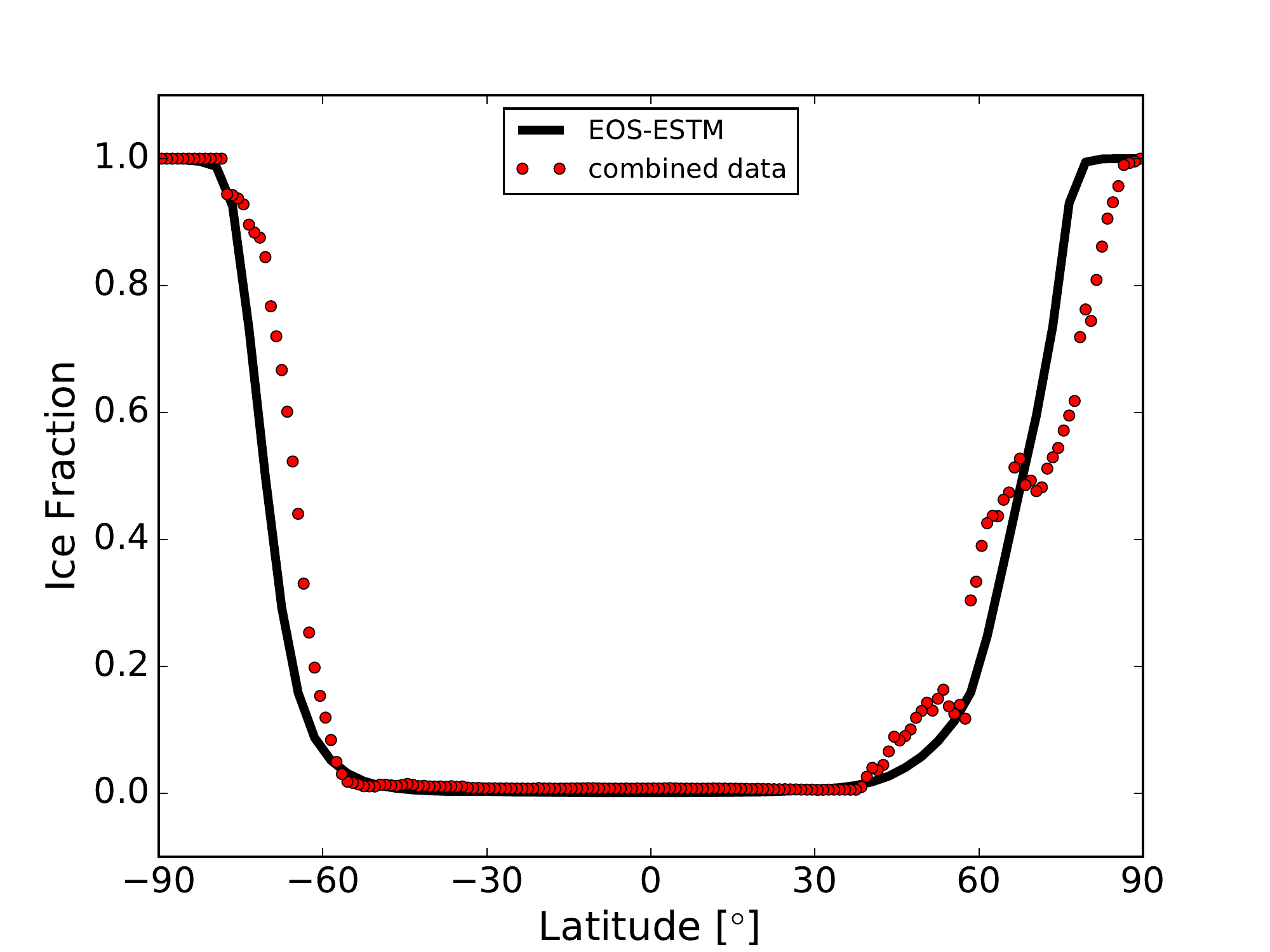}
    \caption{Mean annual latitude profile of surface temperature, albedo, outgoing longwave radiation (OLR), and fractional ice coverage predicted by the reference Earth model (solid lines). Top left panel: the temperature profile is compared with ERA5 temperatures averaged in the period 2005-2015 (blue dots). Top right panel: the albedo profile is compared with CERES-EBAF data averaged in the period 2005-2015 (pink crosses). Bottom left panel: the OLR profile is compared with CERES-EBAF data averaged in the period 2005-2015 (pink crosses). Bottom right panel: the model profile is compared with the mean ice coverage, obtained by weighting the land and ocean data (averaged in the period 2005-2015) in each zone according to the zonal coverage of lands and oceans. 
    }
    \label{fig:earthannualzonal}
\end{figure*}

\subsection{Diagnostic tests of the Earth model}

To test the predictions of Earth model we first investigated
 the temperature dependence of two key energy balance quantities, namely the OLR and TOA albedo.
We then compare global and zonal planetary data with the model predictions.

\subsubsection{OLR and TOA albedo}

In Fig. \ref{fig:olr_vs_temp}  we plot the OLR (left panel) and TOA albedo (right panel) versus surface $T$
obtained from the Earth's reference model (black lines) and Earth's satellite data (green circles).
To obtain these plots, we plotted the mean annual OLR and TOA albedo of each latitude zone 
versus the corresponding zonal value of mean annual surface temperature. 
With this procedure we obtain two independent sets of data versus $T$, one for the  Northern and the other for the Southern hemisphere.
One can see that the EOS/ESTM calculations match well the Earth data,
despite the simplified nature of the model. The agreement is better in the Northern hemisphere,
which is less affected by the peculiar orography of Antarctica. A better match would require a 3D climate model with orography and a physical description of the atmospheric and oceanic fluidodynamics and clouds. 

\subsubsection{Global and zonal data}

In Table \ref{tab:eraceresdata} we display the globally averaged values of planetary quantities 
predicted by the Earth's reference model (second to last column).
The comparison with the corresponding experimental data (previous column) shows an excellent agreement,
with relative differences  below 2\% for the albedo and much smaller for the temperature and OLR.
The match of model and observed data is particularly good in the Northern Hemisphere.
The experimental data of the Southern hemisphere are significantly influenced 
by the high altitude of Anctartica, which is not accounted in our model without orography.

In Fig. \ref{fig:earthannualzonal} we compare the mean annual zonal values of surface temperature, TOA albedo, OLR, and ice cover obtained 
from the reference Earth model (solid lines)
and Earth's satellite and reanalysis data (symbols). 
The agreement of the surface temperature curve (top left panel) 
is excellent, with an area-weighted rms deviation of 1.0 K for the Northern hemisphere. 
The main difference arise above the South Polar cap, where the observed temperature 
is $\sim 20$\,K lower than predicted due to the lack of orography of the model.
This difference is consistent with the $\sim 2$\,km thickness of the ice sheet and a dry lapse rate
of $\simeq 10$\,K/km.

The albedo curve (top right panel) shows an excellent agreement in the Northern Hemisphere at mid-high latitudes, where the gradual change of the albedo of transient ice provided by Eqs. \eqref{albedo_lands_partial_ice} and \eqref{albedo_oceans_partial_ice} yields a better match to the data than in the original ESTM. 
In the equatorial regions the agreement is reasonable, considering the existence of albedo factors that can only be treated in 3D models, such as the atmospheric circulation, which affects the clouds distribution. In the Antarctic region, the model underestimates the albedo
due to the lack of orography. 

The OLR profile (bottom left panel) shows a general agreement, with strong deviations in Antarctica and in the tropical regions. The OLR excess predicted by the model over Antarctic regions is due to the temperature excess that we have already discussed.  
The bumps of OLR emission measured at the edges of tropical regions 
are connected with the presence of large deserts and of low clouds with warm tops, while the reduced equatorial OLR is connected with the presence of deep convective clouds with cold tops \citep[e.g.][]{Hartmann2016}.
Neither of these two features can be captured by our model which performs a sort of average that provides the correct global value of OLR (Table \ref{tab:eraceresdata}). 
The good match between model and observations in the North polar region is an improvement with respect to the original ESTM, and is due to the fact that the scaling factor \eqref{cloudOLRfactor}
rises the planetary OLR emitted from frozen regions.

In the bottom-right panel we show a diagnostic test on ice coverage
that was not performed by V15. 
The mean annual zonal coverage of ice predicted
is in general agreement with the area-weighted lands and oceans data (red dots). 
This implies that the new algorithm that we have introduced, based on Eq. \eqref{eq_icecover},
is able to capture the main characteristics of ice coverage, using the dependence
on a single parameter, namely the surface temperature. A treatment of the physics
of ice formation and melting is, at this moment, beyond the scope of our model.

\section{Testing non-terrestrial conditions}
\label{sec:model_validation}

By changing the input parameters that describe the stellar, orbital and planetary properties,
the ESTM can be in principle applied to simulate a broad spectrum of exoplanetary climates.
In this context, validation tests are required to assess the limits of validity
of the model in non-terrestrial conditions. 
At present time, however, the climate systems of exoplanets are poorly constrained by observations and of no use for validating the model. Given this situation, the best way to test EOS-ESTM is to perform a comparison with the predictions obtained by other models that have been developed to investigate non-terrestrial planetary climates. 
Of particular interest is the comparison with 3D and 1D climate models, given the fact that ESTM is a 2D model, in the sense that we have clarified in Section \ref{sec:model}. 
Below we provide the results of some preliminary comparison tests, starting from a simple simulation
that we have performed using a 3D model of intermediate complexity.
We then describe several tests that we performed using predictions published in the literature.
So far, the space of stellar/planetary parameters that affects exoplanetary climates has been covered only partially in previous work. Therefore
a comparison of different model predictions is only possible for a limited number of cases. 
Here we focus our attention on the models and published results summarized in Table \ref{tab:fig2Godolt2016}.

\subsection{Earth-like aquaplanet}
\label{subsec:plasim}

\begin{figure*}
    \centering
    \includegraphics[width=75mm]{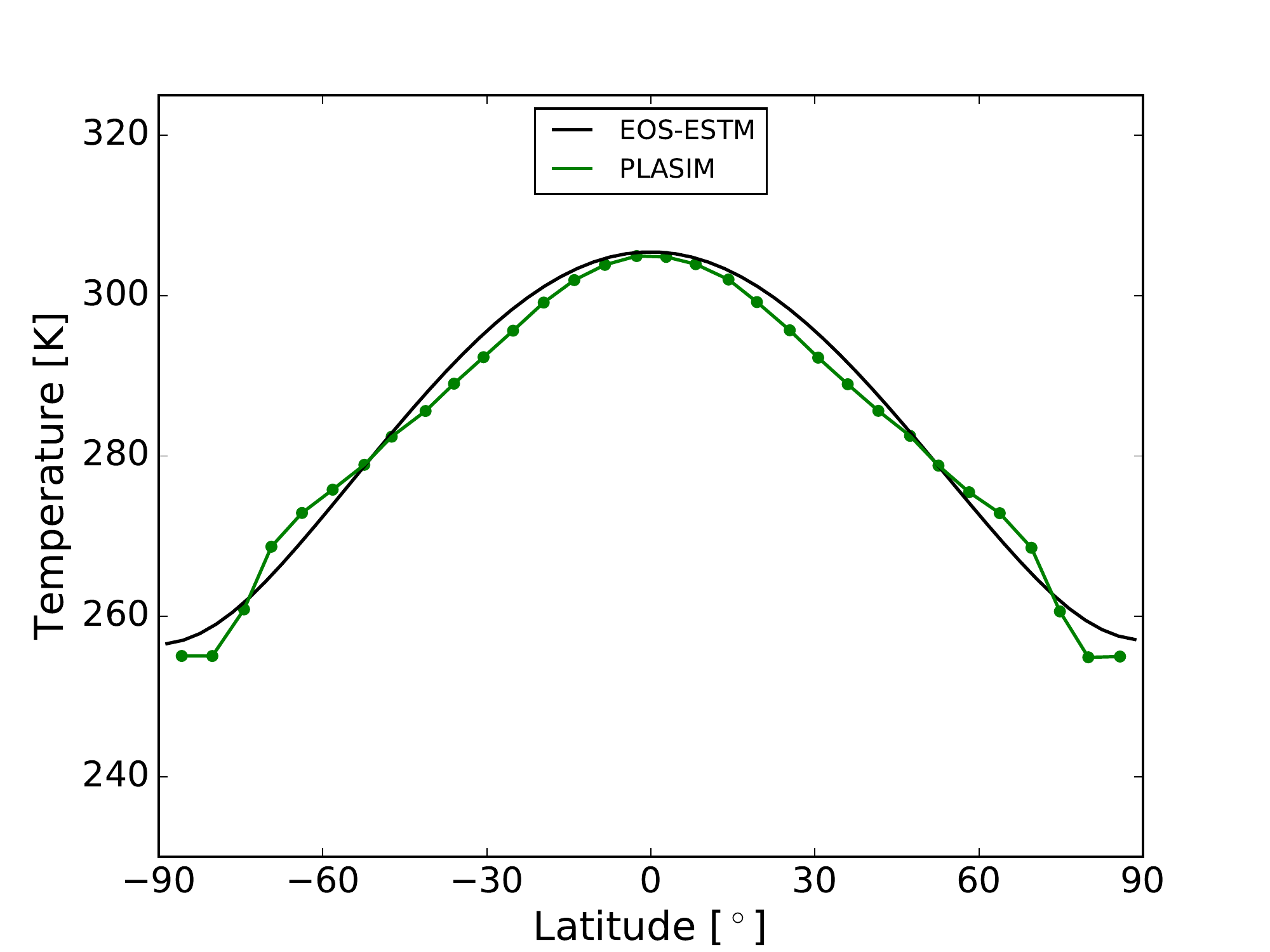}
    \quad\includegraphics[width=75mm]{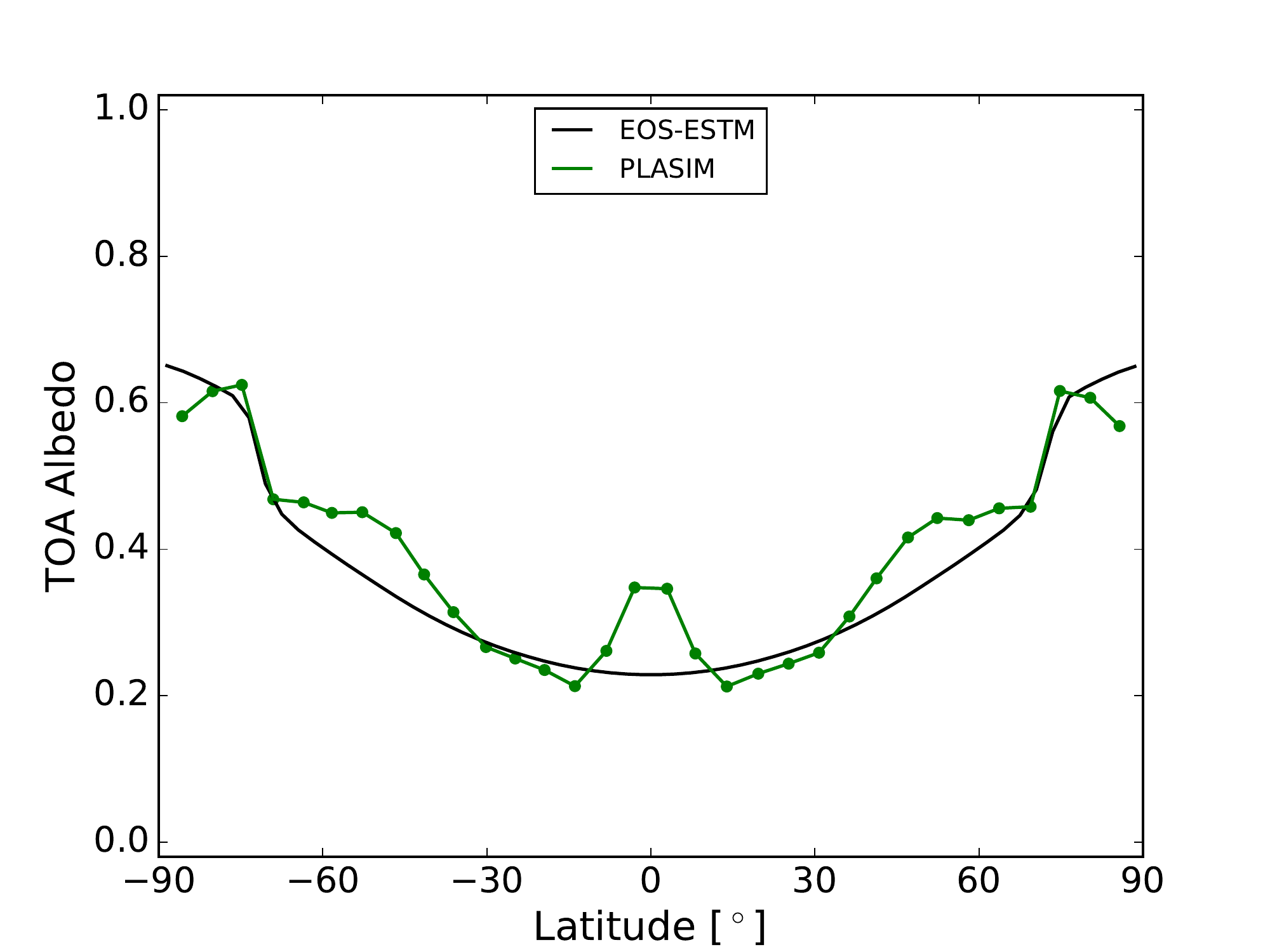}
    \caption{Comparison of model predictions obtained for an Earth-like aquaplanet using the EOS-ESTM (black curves) and  the 3D climate of intermediate complexity PlaSim (green curves). Left panel: mean annual surface temperature versus latidude. Right panel: mean annual top-of-atmosphere albedo versus latitude.
      See Section \ref{subsec:plasim}.    } 
    \label{fig:plasim}
\end{figure*}

As a preliminary comparison test with a 3D climate model, we used the global climate model of intermediate complexity, PlaSim \citep{Fraedrich2005,Angeloni2020}. Specifically, we tested the case of an aquaplanet with rotational spin aligned with the orbital spin ($\epsilon=0$), the remaining parameters being equal to those of the Earth. For the sake of comparison with EOS-ESTM, the PlaSim simulation was run without oceanic transport. The resulting mean annual latitude profiles of surface temperature and top-of-atmosphere albedo are shown in Fig. \ref{fig:plasim}. Despite the 3D nature of PlaSim and the  different  prescriptions of surface features, clouds, and ice between the two models, 
one can see that the results are in general agreement. 
This is true, in particular, for the temperature profile (left panel). 
The differences found in the albedo profiles (right panel) are due to the cloud distribution, which follows the atmospheric circulation pattern that can be modelled in PlaSim, but not in EOS-ESTM. 

Setting PlaSim parameters to non terrestrial conditions is not trivial.
This is true, in general, for all 3D models, particularly for the most complex ones.
For this reason, extending the comparison tests with PlaSim
to cover a broader space of parameters will be the subject of a separate work. 

\begin{table*}
\centering
\caption{Summary of the test cases represented in Fig. \ref{fig:Godolt+2016}, \ref{fig:WT2015} and \ref{fig:Shields2014}. The second and third columns give the radiative transfer (RT) model and the atmospheric properties adopted. The last column reports the references.
\citet{Kunze2014} adopted the RAD4ALL model \citep{Nissen2007} for the shortwave transport and the RRTM model \citep[]{Mlawer1997} for the longwave transport.}
\label{tab:fig2Godolt2016}
\begin{tabular}{llllll}
\hline
Model Name & RT model & Atmosphere  & Reference\\
\hline
ESTM & EOS  & 1.013 bar & This work \\
    & & N$_2$, CO$_2$ (360 ppm), CH$_4$ (1.8 ppm) and H$_2$O & \\
1DGodolt2016 & K84\footnote{\citet{Kasting1984}}  & 1 bar &  \citet{Godolt2016} \\
    & & N$_2$, O$_2$, CO$_2$ (355 ppm), CH$_4$ (1.64 ppm), O$_3$ and H$_2$O & \\
3DLeconte2013 & LMDG  & 1 bar &  \citet{Leconte2013b} \\
    & & N$_2$, CO$_2$ (376 ppm), and H$_2$O & \\
3DWolf\&Toon2015 & CAM 4  & 0.983 bar  &  \citet{WolfToon2015} \\
    & & N$_2$, CO$_2$ (367 ppm), and H$_2$O & \\
3DWolf\&Toon2014 & CAM 3  & 0.983 bar  &  \citet{WT2014a} \\
    & & N$_2$, CO$_2$ (367 ppm), and H$_2$O & \\
EBMShields+2013 & SMART & present-day Earth  &  \citet{Shields2013} \\
    & & CO$_2$, O$_2$ and H$_2$O & \\
\hline
\end{tabular}
\end{table*}

\subsection{Variations of stellar insolation}

To test the model response to variations of insolation, $S$, we run a set of simulations 
aimed at reproducing similar climate experiments 
performed with 1D \citep{Godolt2016} and 3D models \citep{Leconte2013b,WT2014a,Shields2014,WolfToon2015}.
In all cases, an Earth-like atmosphere was considered, with properties 
described in Table \ref{tab:fig2Godolt2016}.

\begin{figure*}
    \centering
    \includegraphics[width=80mm]{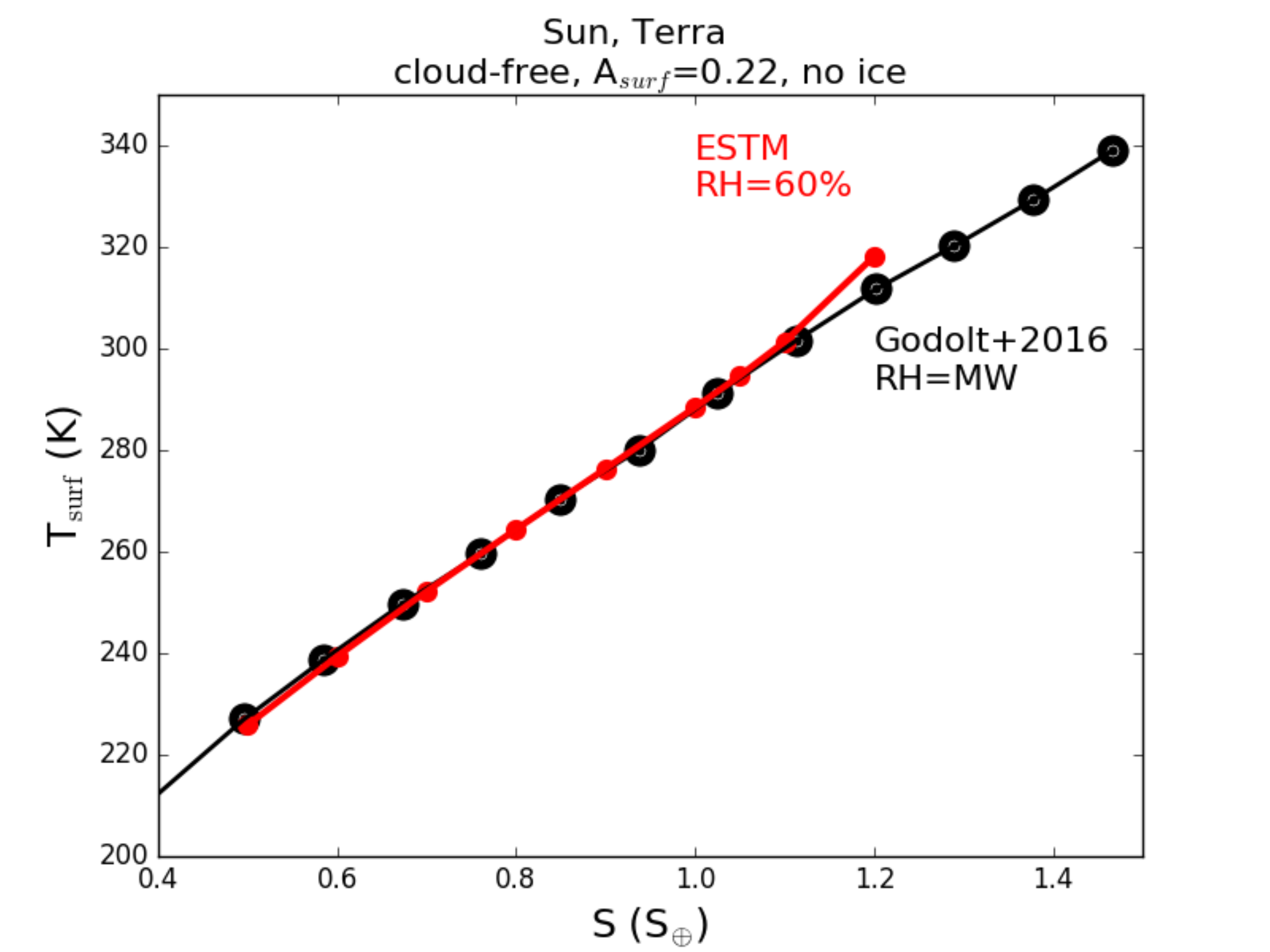}
    \includegraphics[width=80mm]{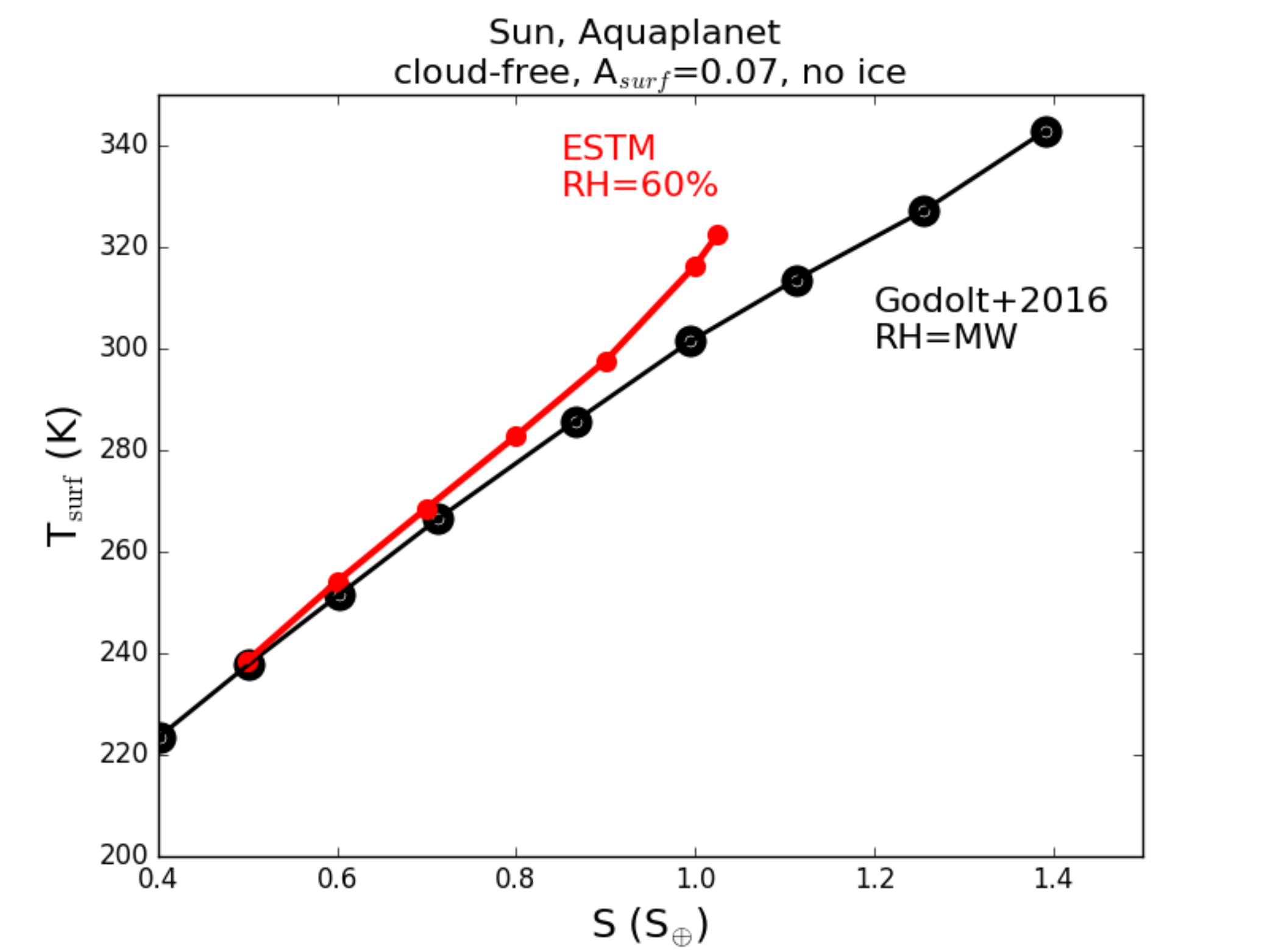}
    \caption{Surface temperatures an Earth-like planet (left panel) and of an aquaplanet (right panel) orbiting a Sun-like star at various stellar insolations, $S$. In both cases, a cloud-free configuration is considered with no ice formation and a fixed value of surface albedo: 0.22 (left panel) and 0.07 (right panel). Black solid line: results obtained  by \citet[]{Godolt2016} with a 1D model with relative humidity specified by \citet{MW1967}. Red solid line: results obtained in this work, where we adopt RH=60\%.}
    \label{fig:Godolt+2016}
\end{figure*}

The comparison with the 1D model is shown in Fig. \ref{fig:Godolt+2016}, where we plot the mean surface temperature as a function of $S$ for two cases considered by \citet{Godolt2016}. 
The first case (left panel) is a cloud-free Earth-like planet with a fixed albedo  $A_\texttt{surf}$ = 0.22, a value 
that, according to \citet{Godolt2016}, reproduces the mean surface temperature of Earth in a cloud-free model. 
The second case (right panel) is an Earth-like aquaplanet with $A_\texttt{surf}$ = 0.07, as adopted by \citet{Godolt2016}, which is representative of the ocean albedo. In both cases ice was not considered. 
One can see that, despite the existence of some differences in the parametrizations (Table \ref{tab:fig2Godolt2016}),
the EOS-ESTM results (red solid lines) are in good agreement with those provided by \citet{Godolt2016} (black solid lines). Departures between the models arise for high values of surface temperatures. This effect is emphasised in the aquaplanet scenario, where the deviations start to be present above $\sim$ 280 K (right panel). 
Since the differences become important at high temperature, when the atmospheres contain more water vapor, this effect may be induced by a different treatment of the relative humidity (RH).  
Indeed, in our model we adopt a constant value, RH=60\%, representative of the mean global value of the Earth,
whereas \citet[]{Godolt2016} used 
a parametrization proposed by \citet{MW1967}, representative of the RH vertical profile of the Earth.

\begin{figure*}
    \centering
    \includegraphics[width=80mm]{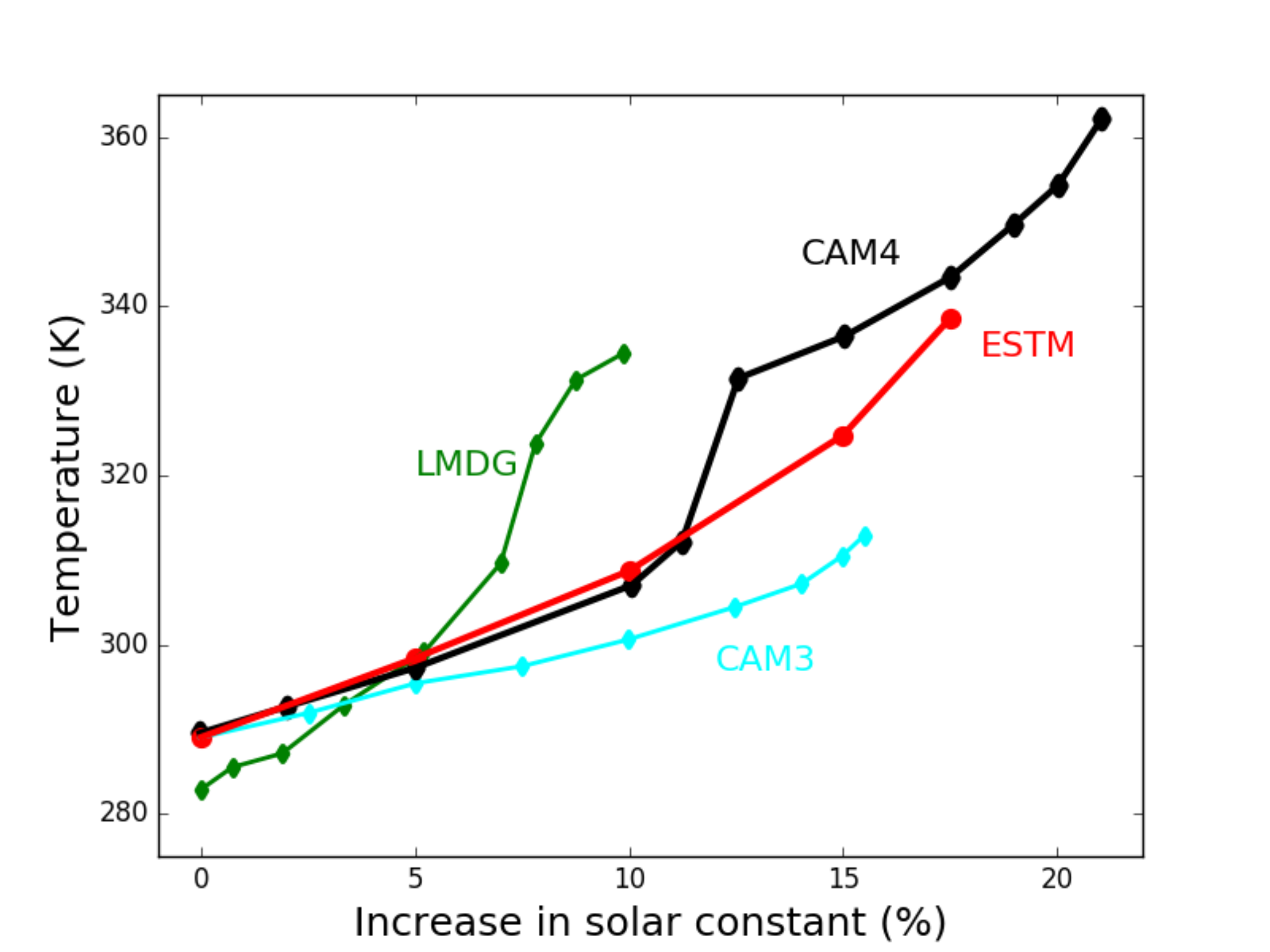}
    \includegraphics[width=80mm]{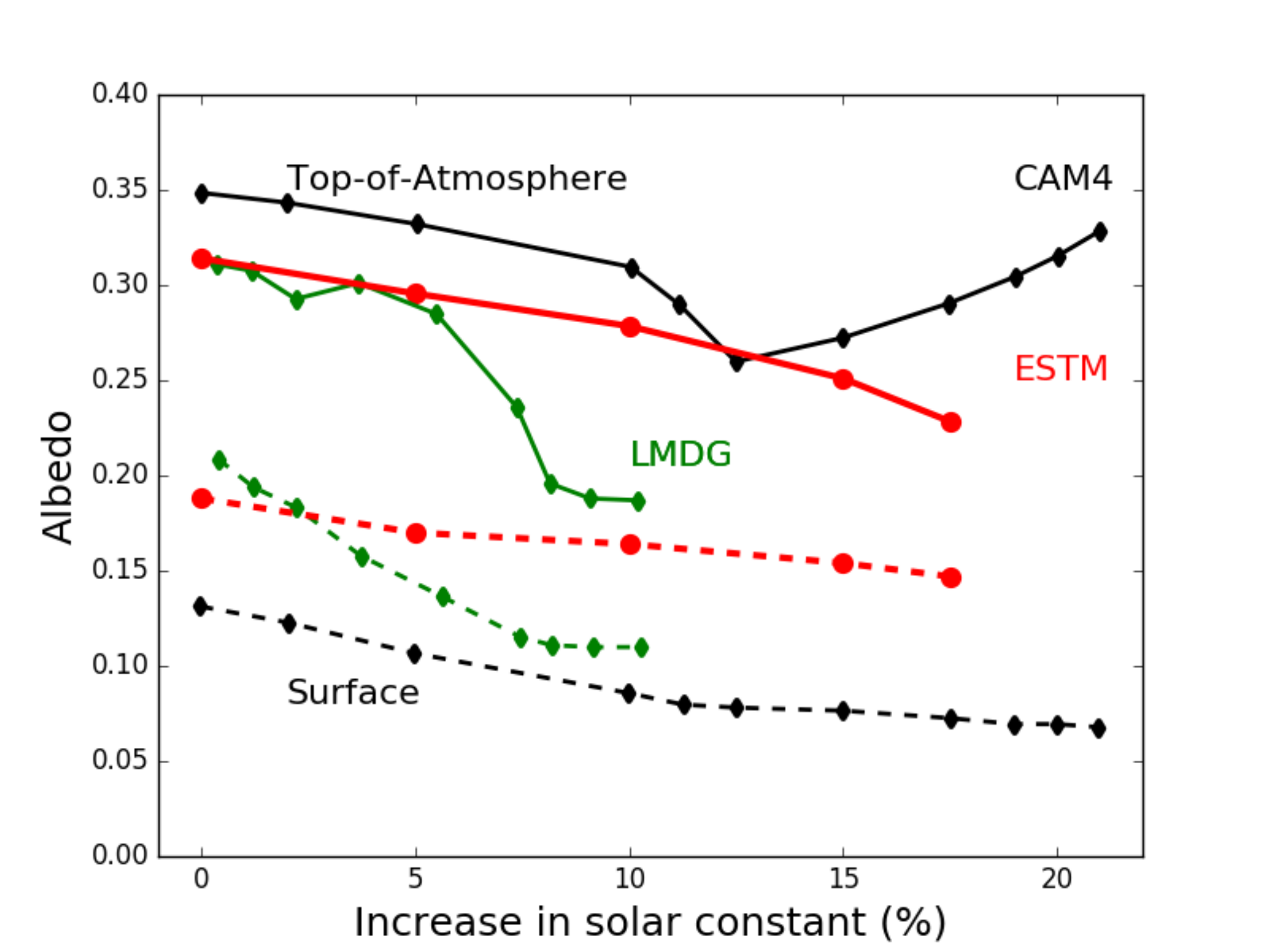}
    \caption{Comparison of global and annual  mean surface temperature (left panel) and  TOA albedo (right panel)  obtained from different climate Earth's models by increasing the solar constant. Red, solid line: EOS-ESTM (this work). Black, solid line: 3D model CAM4 \citep{WolfToon2015}. Green solid line: 3D model by \citet[]{Leconte2013b}. Dashed lines in the right panel represent the surface albedo of these three models, indicated with the same color coding. 
 }
    \label{fig:WT2015}
\end{figure*}

In Fig. \ref{fig:WT2015} we compare the mean annual global surface temperature (left panel) and TOA albedo (right panel) obtained  with the 3D climate models. 
When the increase of insolation is modest, the results obtained with our model (red line) are in general agreement with those provided by the 3D models.
However, discrepancies with \citet{Leconte2013b} (green solid lines) and \citet{WolfToon2015} (black solid lines) appear at higher insolation, when the surface temperature rises above $\sim$ 290 K. 
Important differences are likely to arise from the different RT model adopted (Table \ref{tab:fig2Godolt2016}). 
As shown in \citet[fig. 10a therein]{Simonetti2021} and in \citet[fig. 3a therein]{Yang2016}, differences in the impact of the water vapor absorption predicted by different RT models start to become important for temperatures higher than 290 K. In particular, due to the onset of the runaway greenhouse instability, the slope of the OLR vs $T_\texttt{surf}$ in the CAM4 and LMDG models flattens more than in EOS. Therefore, under equal insolation conditions, EOS-ESTM features lower surface temperatures. 
On the other hand, the CAM3 model (solid cyan line) exhibits a lower surface temperature in response to higher insolations. We believe that this behaviour can be associated to the temperature dependence of the OLR and TOA albedo. In fact, for temperatures above $\sim$300 K, the CAM3 model features the largest value of OLR \citep[Fig. 10a therein]{Simonetti2021} and, at the same time, a slightly higher TOA albedo compared to other models \citep[Fig. 10b therein]{Simonetti2021}. 

Besides the different RT recipes, deviations in the predictions are expected because our model does not incorporate a 3D physical treatment of the cloud and water vapor feedbacks, even though it does reproduce the essential features of the ice-albedo feedback and, to some extent, the rise of water vapor with temperature.  
We suggest that the sharp transitions of surface temperature (left panel) and TOA albedo (right panel) found by \citet[]{Leconte2013b}  and \citet[]{WolfToon2015} 
can be associated to variations in the cloud fraction in response to the increase of insolation. This interpretation is consistent with the fact that such transitions are not found for the surface albedo (dashed lines, right panel), for which the cloud/atmospheric effects are not relevant.

\subsection{Variation of stellar spectra}

The ice-albedo feedback is a well known mechanism that affects the planetary climate with a de-stabilizing effect that, in the most extreme cases,  may lead to  an ice-covered planetary state, called “snowball” \citep[]{Kirschvink1964}. Owing to the wavelength dependence of the albedo,
the impact of this effect will depend on the spectral energy distribution (SED) of the central star.
G-type stars, like our Sun, emit a far greater fraction of their radiation in the visible light spectrum, whereas smaller and cooler M-dwarfs exhibit their peak output in the $\sim$0.8 to 1.2 $\mu$m range \citep[Fig. 1a therein]{Shields2013}.
The fact that these stars emit a significant fraction of their radiation above 1 $\mu$m, combined to the reduction of the albedos of snow and ice at the same wavelengths \citep[Fig. 1b therein]{Shields2013}, implies that the albedos of frozen surfaces are lower on planets orbiting M-type stars than on Earth.
Calculations of broadband albedo\footnote{The ratio of the surface upward radiation ﬂux to the downward radiation ﬂux within a certain wavelength range \citep[]{Kokhanovsky2021}.}, performed taking into account the stellar SEDs and the wavelength-dependence albedo of snow and ice, 
indicate that the ice-albedo feedback is weaker around M-type stars  \citep[]{Joshi2012}.
The atmospheric contribution to the albedo in these stars was studied by \citet[]{vonParis2013}:
the presence of trace amounts of H$_2$O and CH$_4$ in the atmosphere, as well as high CO$_2$ pressures, damps the ice-albedo feedback in planets around M-type stars.

\begin{figure*}
    \centering
    \includegraphics[width=80mm]{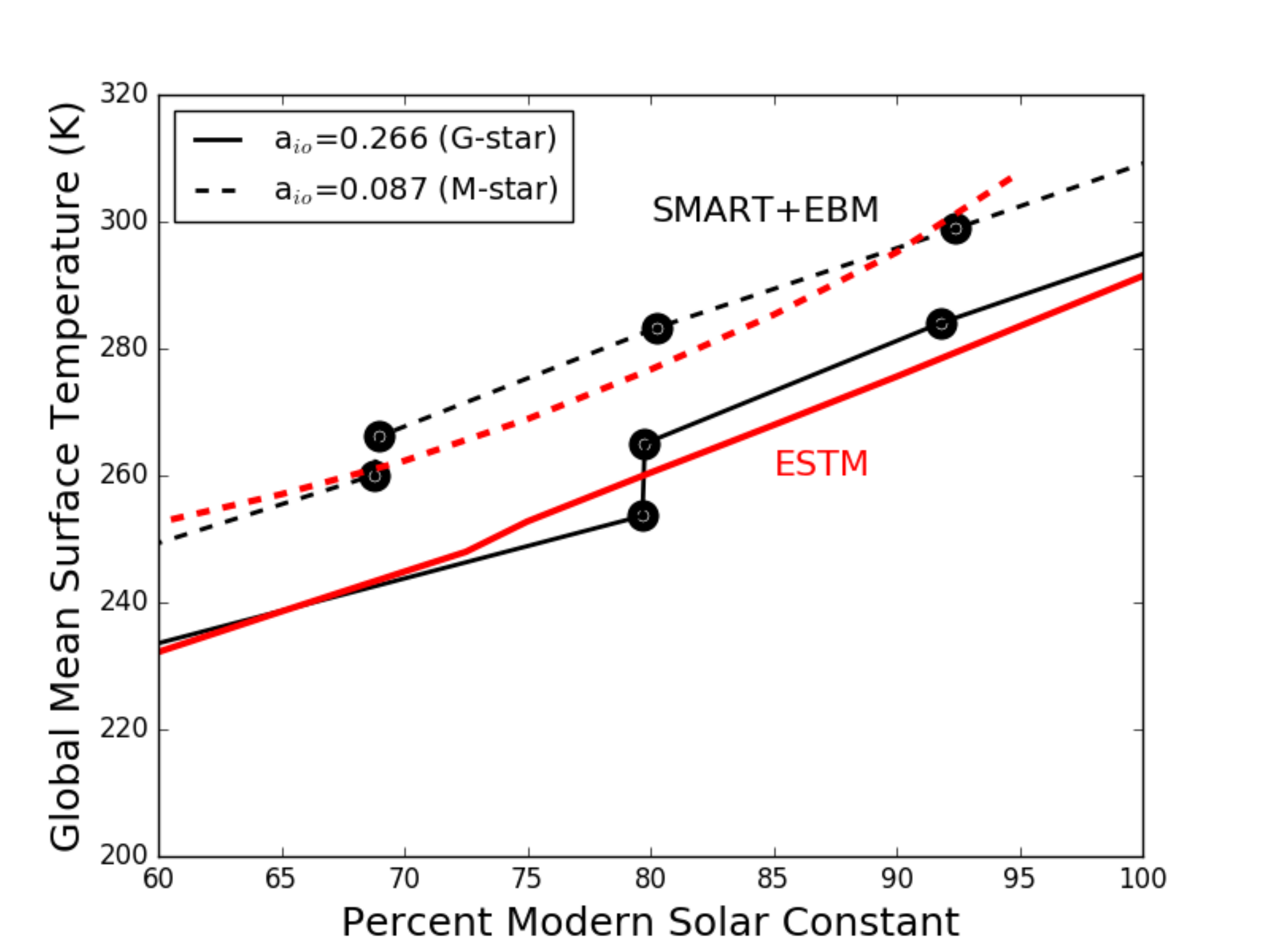}
    \caption{Mean global surface temperature versus stellar flux for an aquaplanet orbiting  an M-type (dashed lines) and a G-type (solid lines) star. Due to the different stellar SED, the albedo of ice over the ocean, $a_{io}$,
    is varied as indicated in the legenda.
    Black lines: predictions obtained by \citet[]{Shields2013} with SMART in combination with an EBM adapted from \citet[]{North1979}. Red lines: results obtained in this work  adopting a warm start for consistency with \citet[]{Shields2013}. }
    \label{fig:Shields2014}
\end{figure*}

To test the EOS-ESTM predictions at different stellar SEDs we performed a comparison with the work by \citet[]{Shields2013}. These authors used a 1D radiative transfer model (SMART) to calculate the broadband planetary albedo, given the spectrum of the central star and that of the surface albedo. 
Then, they included the resulting broadband albedo into a 1D EBM to calculate the mean global surface temperature as a function of insolation
for an aquaplanet orbiting a G-type star (the Sun) and an M-type star (AD Leo).
Following their prescriptions, we considered an aquaplanet with an axis obliquity of 23$^{\circ}$, zero orbital eccentricity and a present-day Earth atmospheric composition.
For consistency with their work, we considered a spectral distribution representative of AD Leo, which is a M3.5-type star, with a M$_{\star}$ = 0.42 M$_{\odot}$ \citep{Reiners2009}.
Since at decreasing insolation the aquaplanet undergoes a transition towards a “snowball” state, 
we dedicate special attention to select the value of the albedo of ice over ocean, $a_{io}$. 
Among the different values of “blue ice“\footnote{According to \citet[]{Shields2013} blue marine ice results from freezing of liquid marine water and not from glacier ice.} calculated by \citet[Table 2]{Shields2013} 
we adopt the value for the case with no gases and clouds and no Rayleigh scattering. 
This is the case more appropriate for the surface albedo in our model, since the EOS-ESTM calculates the contribution of the atmosphere and clouds in its own way. 
The results are shown in Figure \ref{fig:Shields2014}, where the planet orbiting the M-type dwarf appears less susceptible to “snowball” states, since the ice is particularly absorptive in the NIR, as well as the atmosphere.
In spite of small differences associated to the transition to a complete “snowball” state, expressed as a sudden decrease of global surface temperature, the results obtained with EOS-ESTM (red lines) and by \citet[]{Shields2013} (black line) show an overall agreement.  
The main difference between the trends found in the two models may arises from the different parameterization of ice: the smooth transition in our model is probably due the gradual temperature dependence of the ice coverage
described in Section \ref{s:ice}.

\subsection{Variations of planet radius and rotation rate}

A critical difference between ESTM and GCMs is the treatment of the meridional transport. As explained in V15 (see Sec. 2.1), we model the $D$ term in Eq. \eqref{diffusionEq} as a scaling relation between planetary quantities
that are involved in the physics of the meridional transport. To test the reliability of this parameterization, we run a set of simulations varying one parameter at a time and compared our results with similar tests performed with GCMs by other authors.
Specifically, we varied the planetary radius and rotation rate and performed a comparison with results published by \citet[]{KS2015} and  \citet[]{KA19}. 

The mean annual equator-to-pole temperature difference, $\Delta T_\text{EP}$, is a good indicator of the efficiency of the meridional transport and is expected to be higher in planets with fast rotational velocities or large radii. 
The Coriolis forces resulting from planetary rotation tend to inhibit the transport from the tropics to the poles, leading to a higher gradient $\Delta T_\text{EP}$.
Quantifying these effects with 3D models is important because changes of rotational angular velocity may affect the location of the inner edge of the habitable zone  \citep{Yang2014,YangKomacekAbbot2019}. 
Also variations of planetary radius affect the meridional gradient: as the radius increases, so does the physical distance between equator and poles, leading to a less efficient meridional heat distribution, i.e. a larger $\Delta T_\text{EP}$.  
In Fig. \ref{fig:KS2015} we show how $\Delta T_\text{EP}$ is predicted to change as a function of planetary radius (left panels) and rotation period (right panels) for different models that we describe below. 
Following \citet[]{KA19}, we normalize $\Delta T_\text{EP}$ to the values predicted by each model
for Earth's values of rotation rate and radius. 
The results obtained by \citet[]{KS2015} and \citet[]{KA19} are rather different, despite both being based on 3D models. We refer to the latter paper for a discussion on these differences, which may be due to different physico-chemical assumptions and to the fact that the model of \citet[]{KA19} had not been tuned to match the Earth. 
Here we compare our results with those obtained in these two papers.

\subsubsection{Comparison with \citet[]{KS2015}}
\label{Subsec:Comparison with KS2015}
\begin{figure*}
    \centering
    \includegraphics[width=80mm]{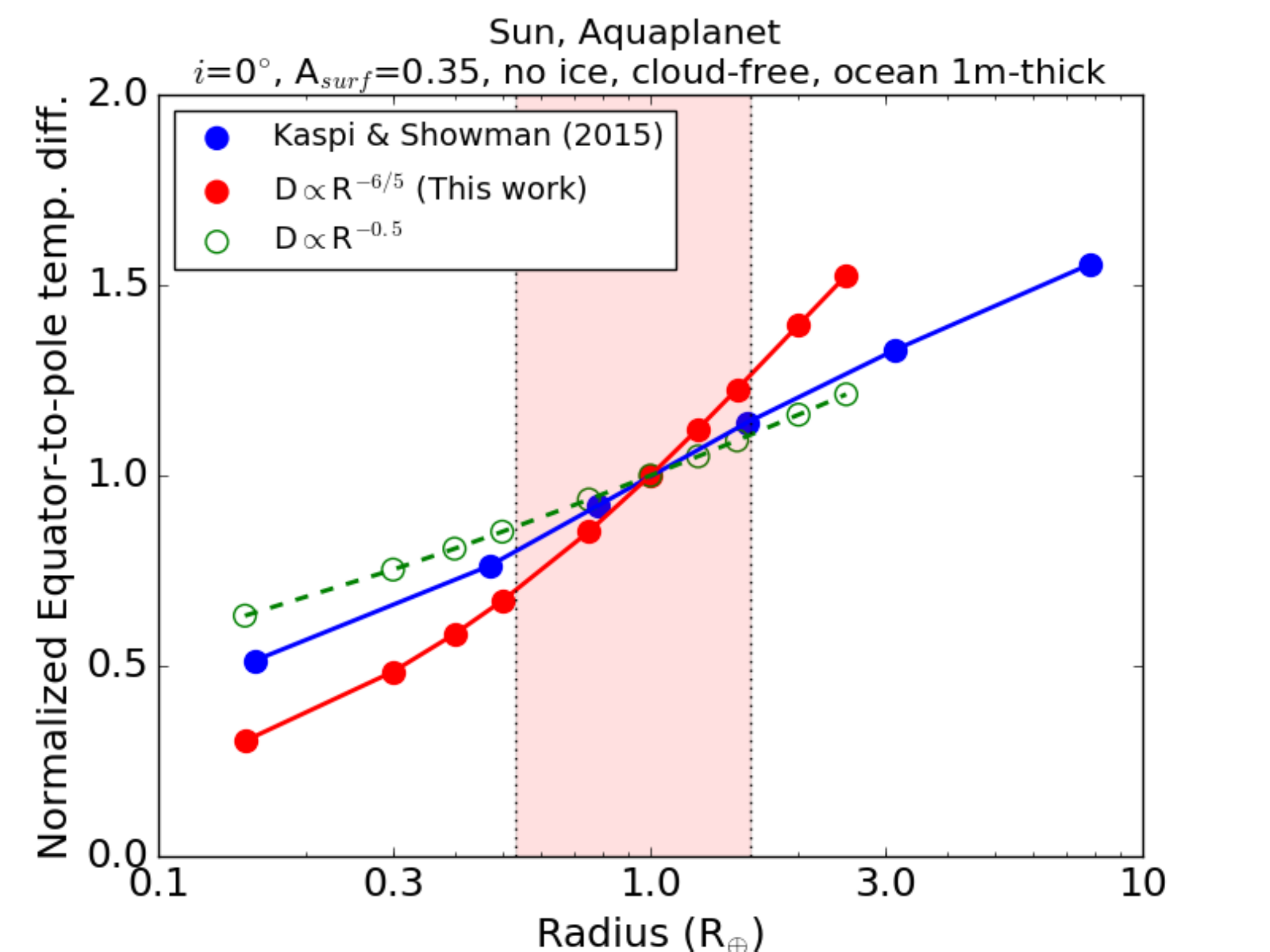}
    \includegraphics[width=80mm]{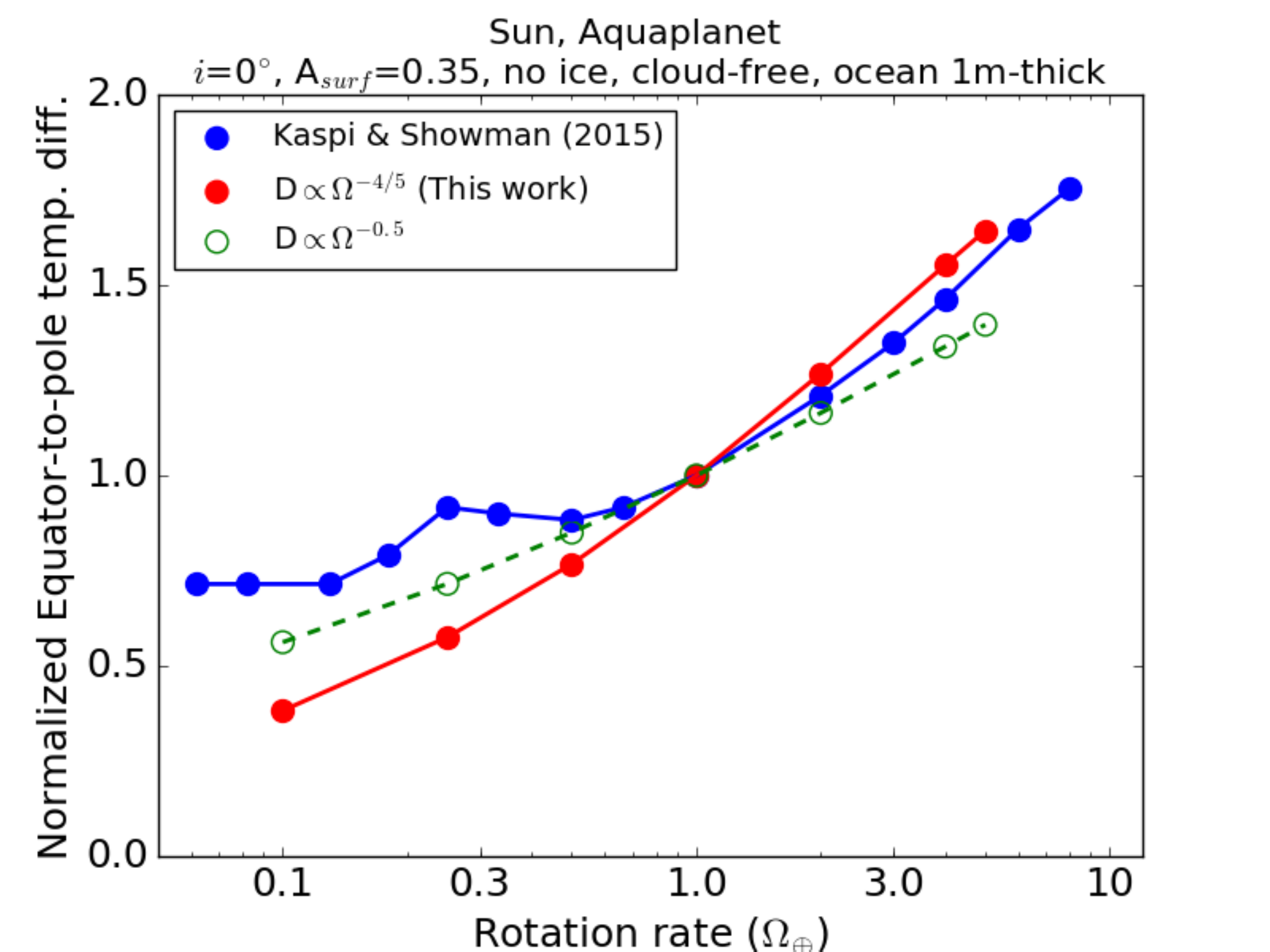}
    \includegraphics[width=80mm]{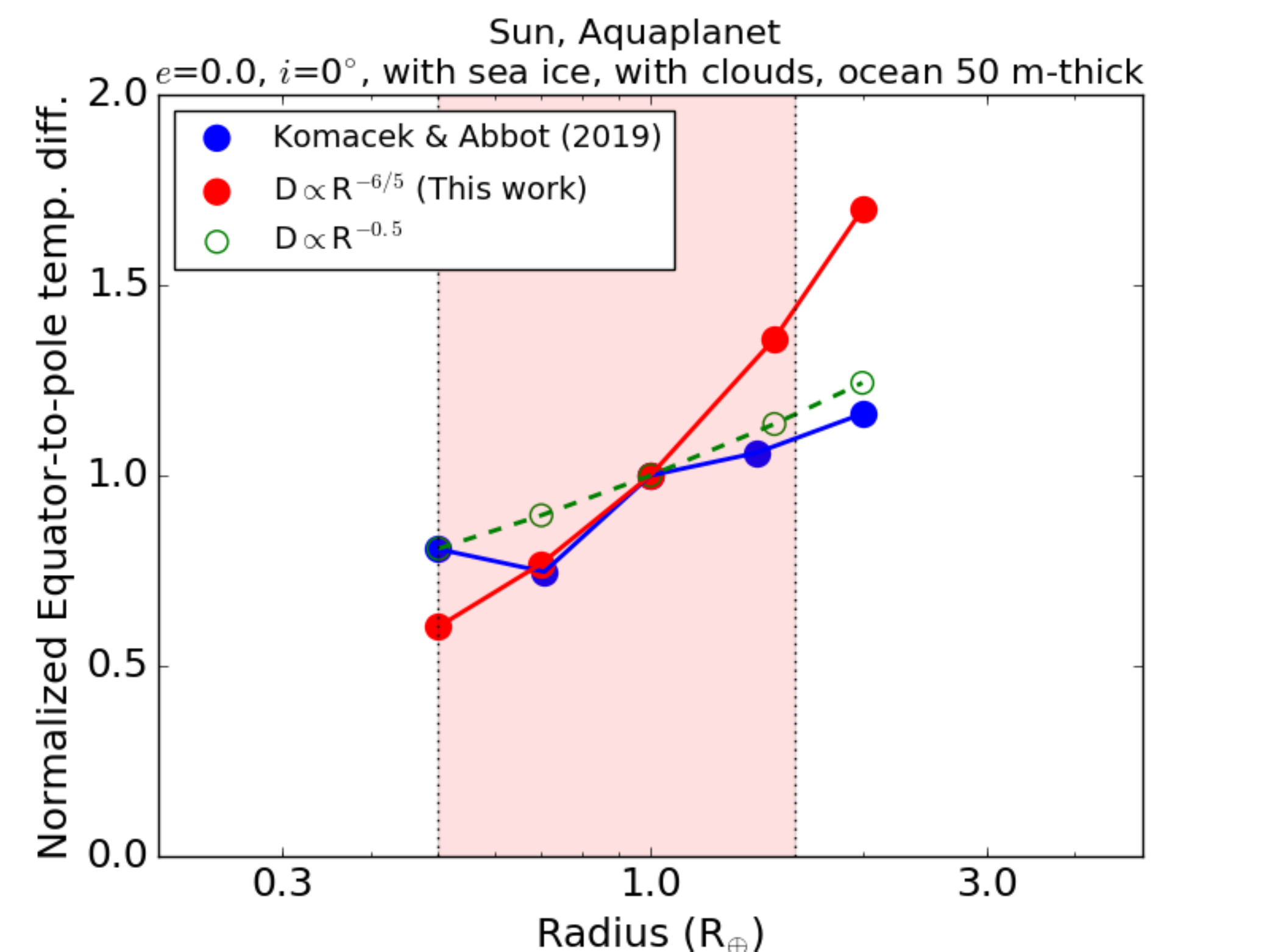}
    \includegraphics[width=80mm]{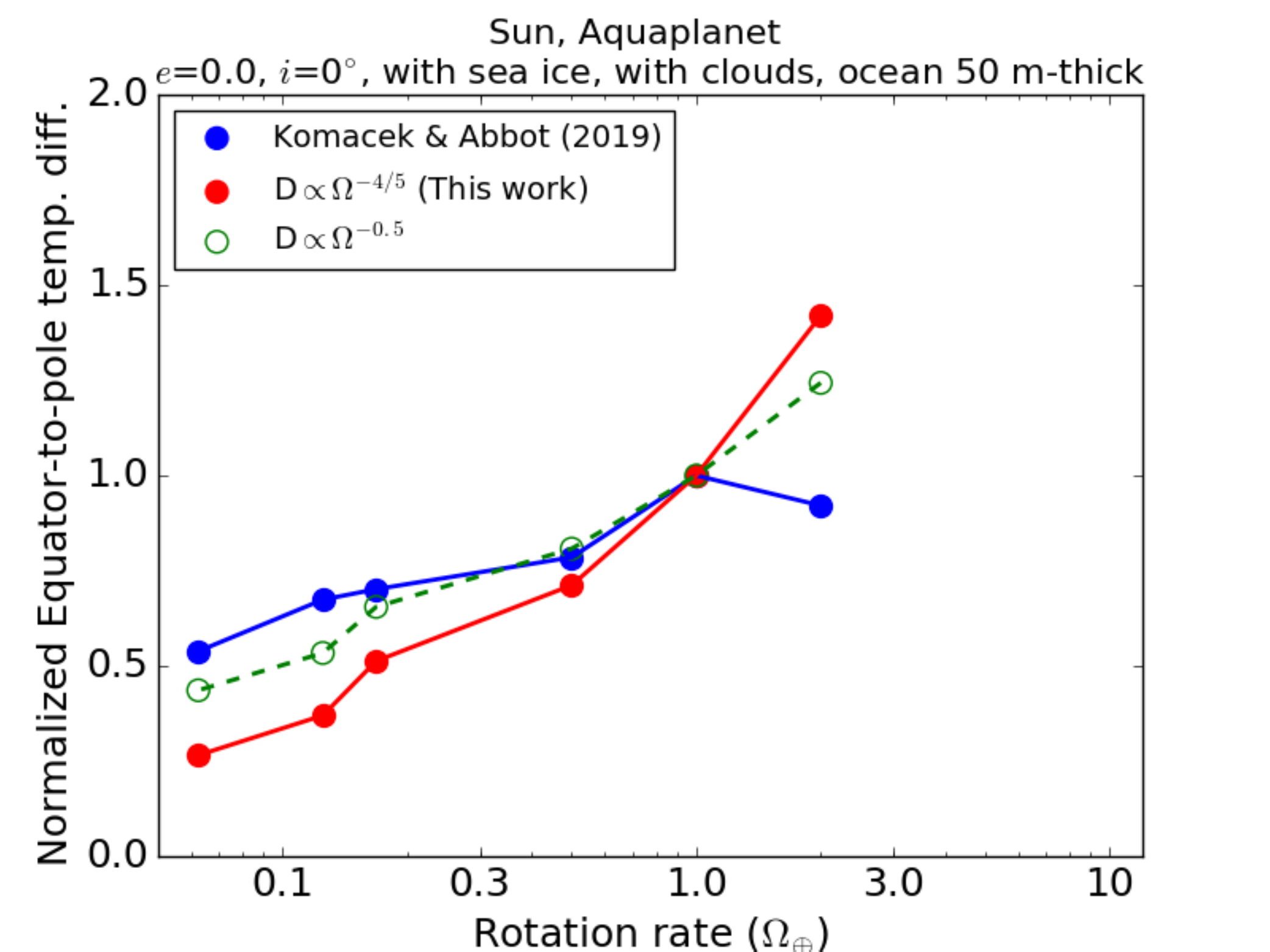}
    \caption{Normalized equator-to-pole temperature difference as a function of planet radius (left panels) and planet rotation rate (right panels) for an Earth-like aquaplanet.
    \textit{Top panels}: comparison between the results obtained in this work (red lines) and those obtained by \citet[]{KS2015} (blue lines) for a cloud-free aquaplanet slab ocean with a depth of 1- meter, an axis obliquity $i$ = 0 $^{\circ}$, a fixed albedo (A=0.35) and no sea ice. 
    \textit{Bottom panels}: comparison between the results obtained in this work (red lines) and by \citet[]{KA19} (blue lines) for an aquaplanet slab ocean with a depth of 50- meter where the effects of clouds, nongray radiative transfer, and sea ice are included; axis obliquity $i=0$ and the eccentricity $e=0$.
    \textit{Shaded red area}: range of radius (0.5<R/R$_{\oplus}$<1.6) at which an exoplanet is more likely to be composed of rock and metal \citep[]{Meadows2018}. }
    \label{fig:KS2015}
\end{figure*}

To investigate the atmospheric dynamics over a wide range of planetary parameter space \citet[]{KS2015} adopted a 3D GCM with a scheme similar to that of \citet[]{Frierson2006} both for the radiative transfer and the surface boundary-layer: a standard two-stream gray radiation and an uniform 1-meter water-covered slab, with an albedo of $A = 0.35$, respectively. 
They modelled an idealized aquaplanet at perpetual equinox with an Earth-like reference atmosphere. The effects of clouds, sea-ices and continents were not accounted for. 
For the sake of comparison we adopted the same set of conditions in EOS-ESTM. 
In the top panels of Fig. \ref{fig:KS2015} one can see that, in spite of differences at the low- and high-radius and rotation rate regimes, the EOS-ESTM predictions (red symbols and lines) reproduce the trends obtained by \citet[]{KS2015} with the 3D aquaplanet (blue symbols and lines). 
The two sets of results are consistent
as long as the planets have radii and rotation rates sufficiently close to those of the Earth.
For habitability studies we are interested in the range of radii expected for rocky planets,
shown as shaded red areas in the left panels of Fig. \ref{fig:KS2015}.
In this range, the predictions of the two models are comparable.

\subsubsection{Comparison with \citet[]{KA19}}
More recently, \citet[]{KA19} investigated how the atmospheric circulation and climate of planets orbiting Sun-like stars vary when planetary parameters are changed. They used the state-of-the-art GCM ExoCAM (a modified version of the Community Atmosphere Model version 4), to simulate an idealized aquaplanet with a 50-m water slab without oceanic transport and an atmosphere of N$_2$ and H$_2$O.  
At variance with  \citet[]{KS2015}, they included the effects of clouds, non-gray radiative transfer, and sea ice. 
To test the EOS-ESTM predictions making use of their results, we adopted, for consistency, a uniform 50-meter thick water-covered slab, the same atmosphere of N$_2$ and H$_2$O,and the same ice albedo; we set to zero both the axis obliquity and orbital eccentricity. In the bottom panels of Figure \ref{fig:KS2015} we compare the normalized $\Delta T_\text{EP}$ obtained with our model (red symbols and lines) with those obtained with the GCM (blue symbols and lines). 

In the bottom-left panel one can see that the trend of increasing $\Delta T_\text{EP}$ that we find (red line) is consistent, but steeper than that found with the GCM (blue line), the departures becoming significant above the radius limit of rocky planets.
The smoother trend found with the GCM suggests that the large scale, 3D circulation, not present in our model, may enhance the heat distribution.
Our trend is somewhat steeper than the one that we found in the previous test (top-left-panel),
indicating how the inclusion of clouds and ice impacts our results.

The bottom-right panel of Figure \ref{fig:KS2015} shows a consistent trend for angular velocity lower than $\Omega_{\oplus}$, with a discrepancy at high angular velocity ($\Omega$ = 2 $\Omega_{\oplus}$). 
This discrepancy is surprising, because our
algorithm for meridional transport is expected to be more realistic for fast-rotating planets (see V15),
and this indication is supported by the comparison with \citet[]{KS2015} shown in the top-right panel. 
Clearly, the two 3D models that we are using for comparison show remarkable differences between them and should be taken with some caution. At low rotation speed, where we know that our assumptions
are more critical (see V15), our model seems to underestimate $\Delta T_\text{EP}$, as in the comparison with \citet[]{KS2015} shown in the top-right panel. These results suggest that the 3D circulation may be able to redistribute the heat efficiently, with a weak dependence on the planet rotation rate.

\subsubsection{Future improvements of the model}

In our parametrization of the meridional transport, the term $D$ scales  as $\Omega^{-4/5}$ and   $R^{-6/5}$ (see V15, Section 2.1).
Taking advantage of the flexibility of our model, we varied the exponents of these power laws, searching for a better agreement with the trends obtained by the 3D models shown in Fig. \ref{fig:KS2015}.
The dashed green lines plotted in all panels of that figure show that a better match 
with the 3D results is achieved when adopting a more moderate dependence
for both the angular velocity, $\propto \Omega^{-0.5}$ and radius, $\propto R^{-0.5}$. 
This exercise shows that, in principle, one could recalibrate the exponents of the scaling relations that we adopt for $D$, making use of specifically designed tests performed with GCMs. 
To this end it would be important to use realistic 3D models for cross validation.
Realistic models should include the main components of the climate system and should be calibrated to match the Earth data.
Setting state-of-the art GCM models to simulate non-terrestrial conditions is not a straightforward task.
However, this is the way to proceed for expanding the range of application of flexible models such as EOS-ESTM and exploring the parameter space that allows habitable climates to exist.

\subsection{The outer edge of the habitable zone}
\label{outeredge}

In classic studies of the habitable zone (HZ) the locations of the inner and outer edge are calculated making use of single-column, cloud-free atmospheric climate models \citep[]{Kasting1993,Kopparapu2013a}.
In recent years several studies have proven the critical role of planetary properties on the position and extension of the circumstellar HZ \citep[]{Yang2014,Rushby2019,YangKomacekAbbot2019,Zhao2021}.
Here we take advantage of the flexibility of EOS-ESTM to investigate how the location of the outer edge is affected by variations of planetary parameters.
Establishing the exact location of the outer edge would require a study of clouds effects \citep[e.g.][]{ForgetPierrehumbert1997,Selsis2007,Kitzmann2017} and the possible presence of other greenhouse gases, such as CH$_4$\citep[see][]{Ramirez2018}.
However, for the climate experiments that we present here, we simply adopt the ``maximum greenhouse'' limit, defined as the maximum distance at which a cloud-free planet with an atmosphere dominated by CO$_2$ can maintain a surface temperature of 273\,K \citep[]{Kasting1993}.
Beyond this limit, the greenhouse effect due to a further rise of CO$_2$ is offset by the rise of atmospheric albedo due to the Rayleigh scattering of CO$_2$ molecules. 
To explore the impact of planetary parameters on the location of the outer edge, we considered a cloud-free, CO$_2$-dominated atmosphere with a dry surface pressure of 7.3 bar\footnote{Kopparapu and collaborators also had an additional bar of N$_2$ in all their models, thus they evaluated the CO$_2$ \textit{partial} pressure. Taken alone, it would produce a surface pressure of 7.3 bar.}, which is the value identified by \citet{Kopparapu2013a} as the maximum greenhouse limit. 
To build the pressure-temperature profile of the atmosphere we followed the recipes in the appendix of \citet{Kasting1991}, with a H$_2$O-saturated lower troposphere, a CO$_2$-saturated upper troposphere and a 154 K isothermal stratosphere.
We varied the insolation of a planet with Earth-like parameters and solar-type central star, searching for the limit at which the planet undergoes a transition to a snowball state. 

For simplicity, we considered only the solutions obtained with warm initial conditions, i.e. starting with $T_0$=300 K, which provide a conservative outer limit to the habitable zone.

\begin{figure*}
    \centering
    \includegraphics[width=80mm]{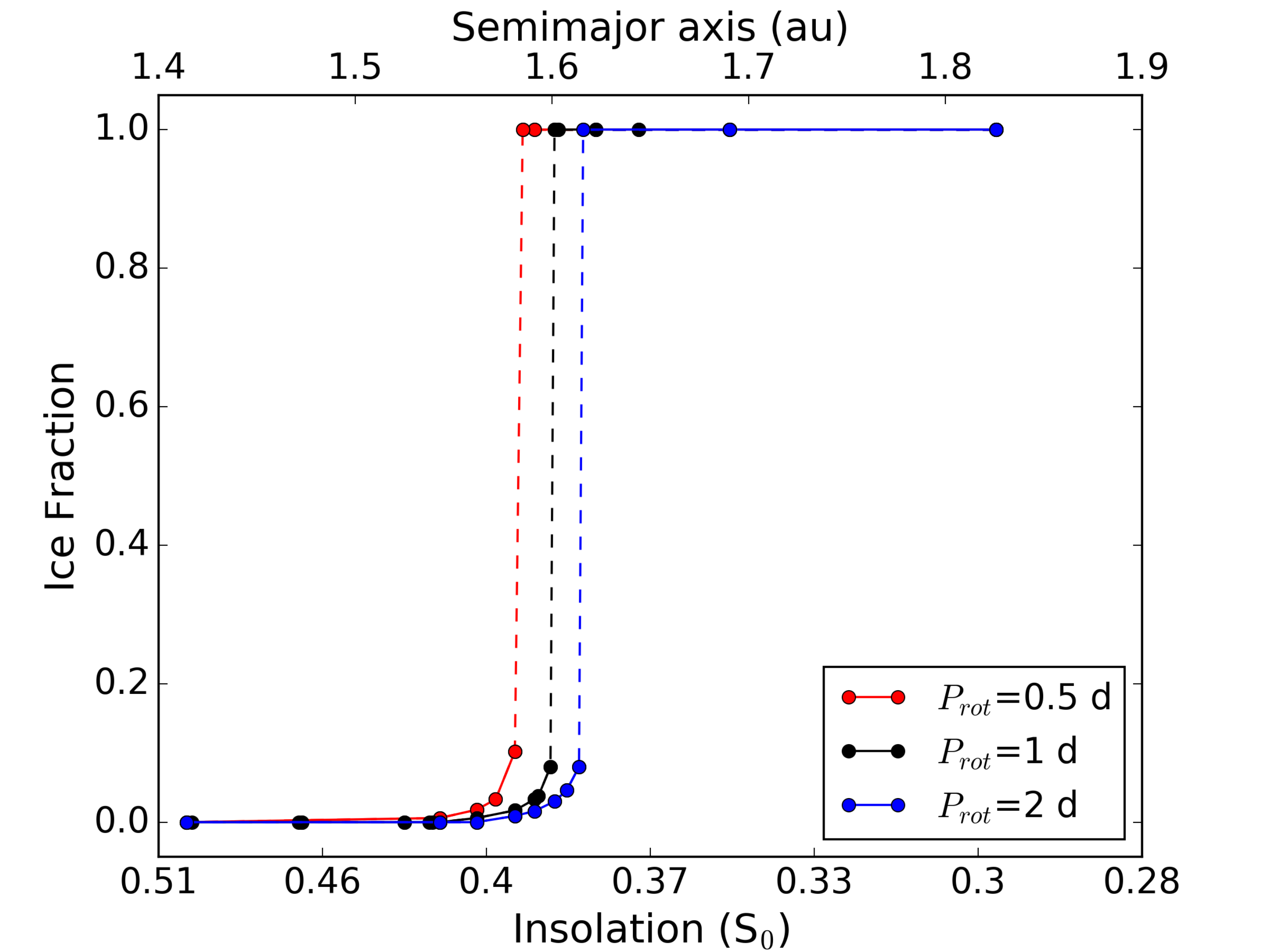}
    \includegraphics[width=80mm]{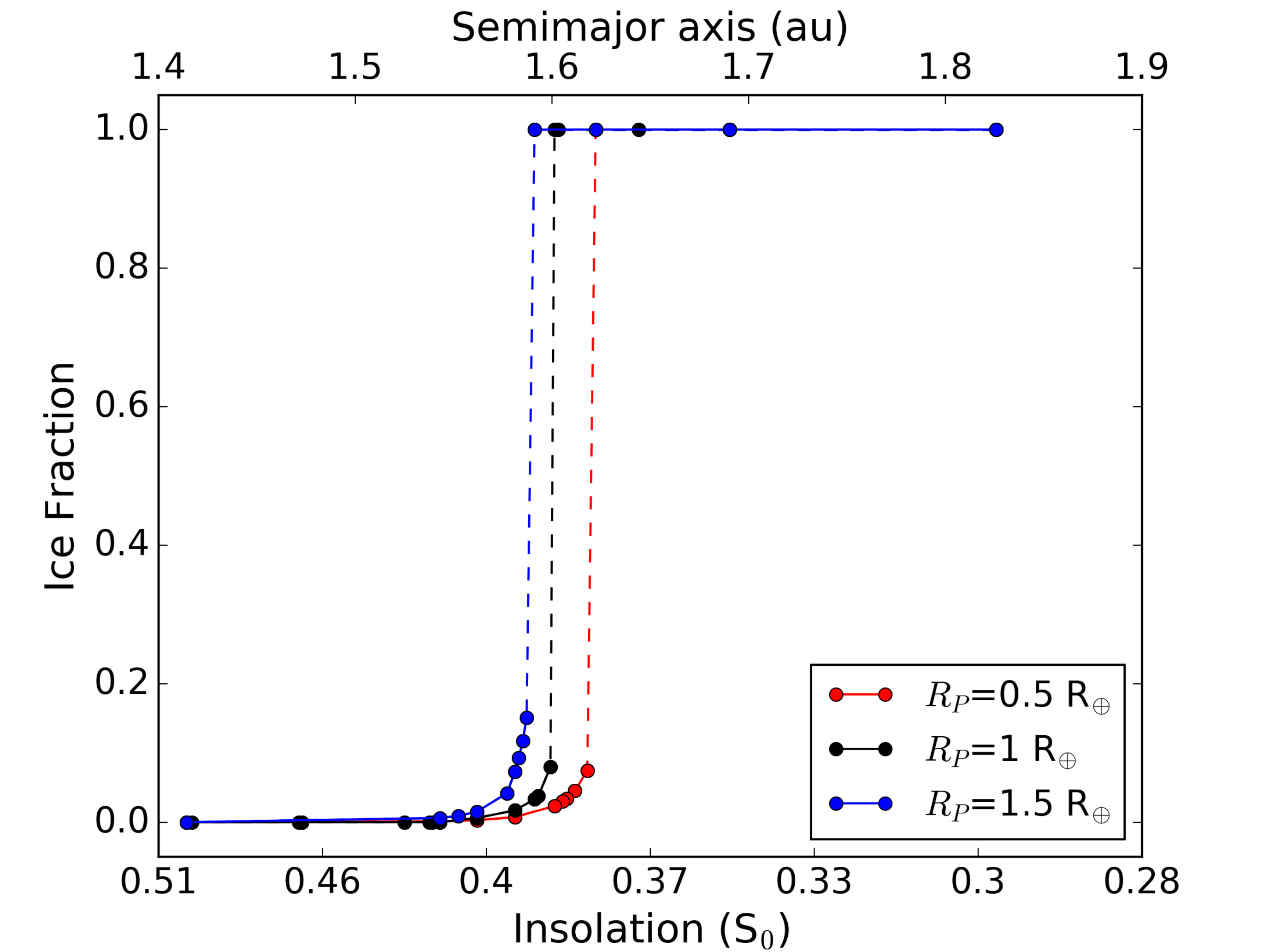}
    \includegraphics[width=80mm]{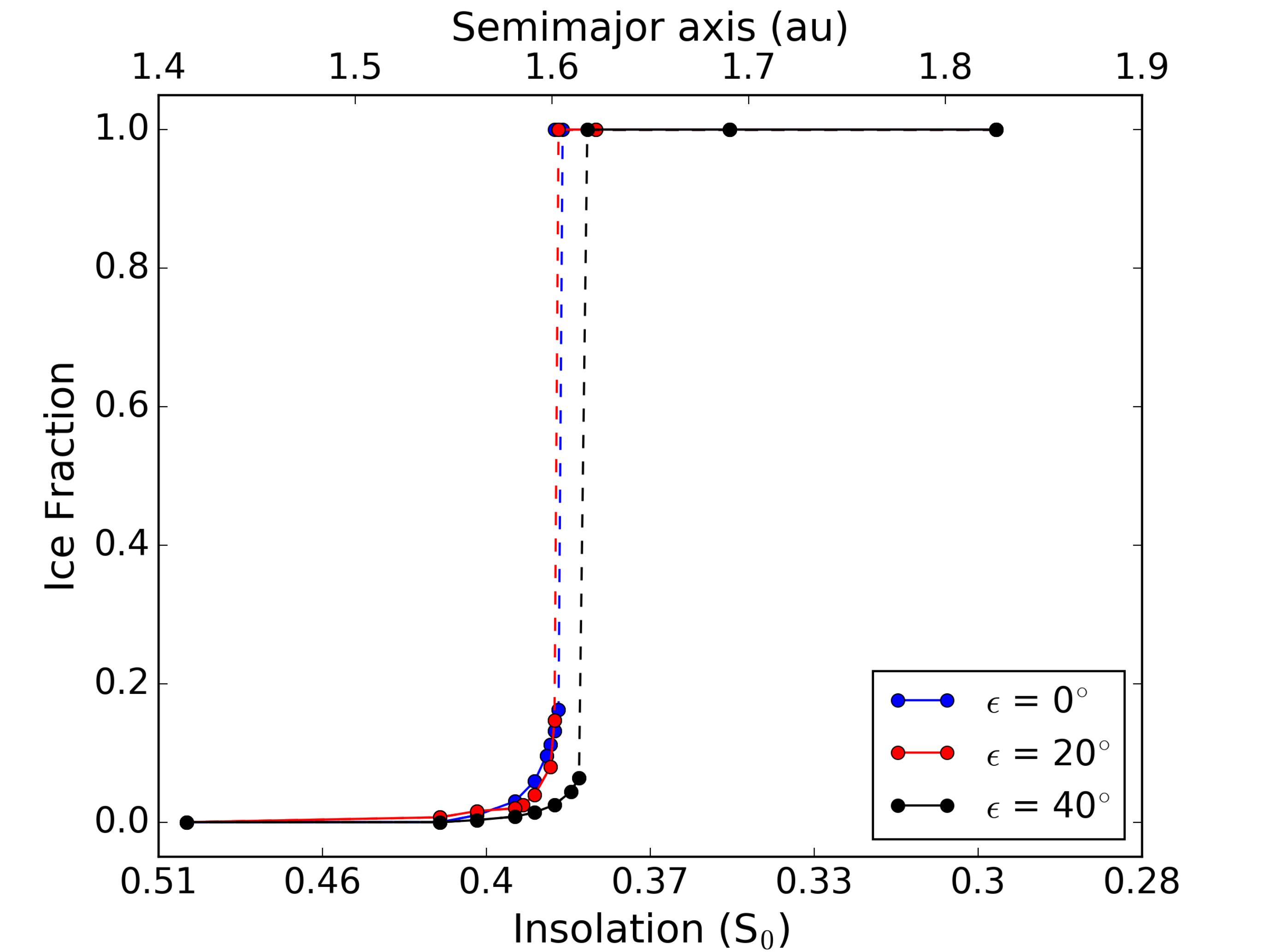}
    \includegraphics[width=80mm]{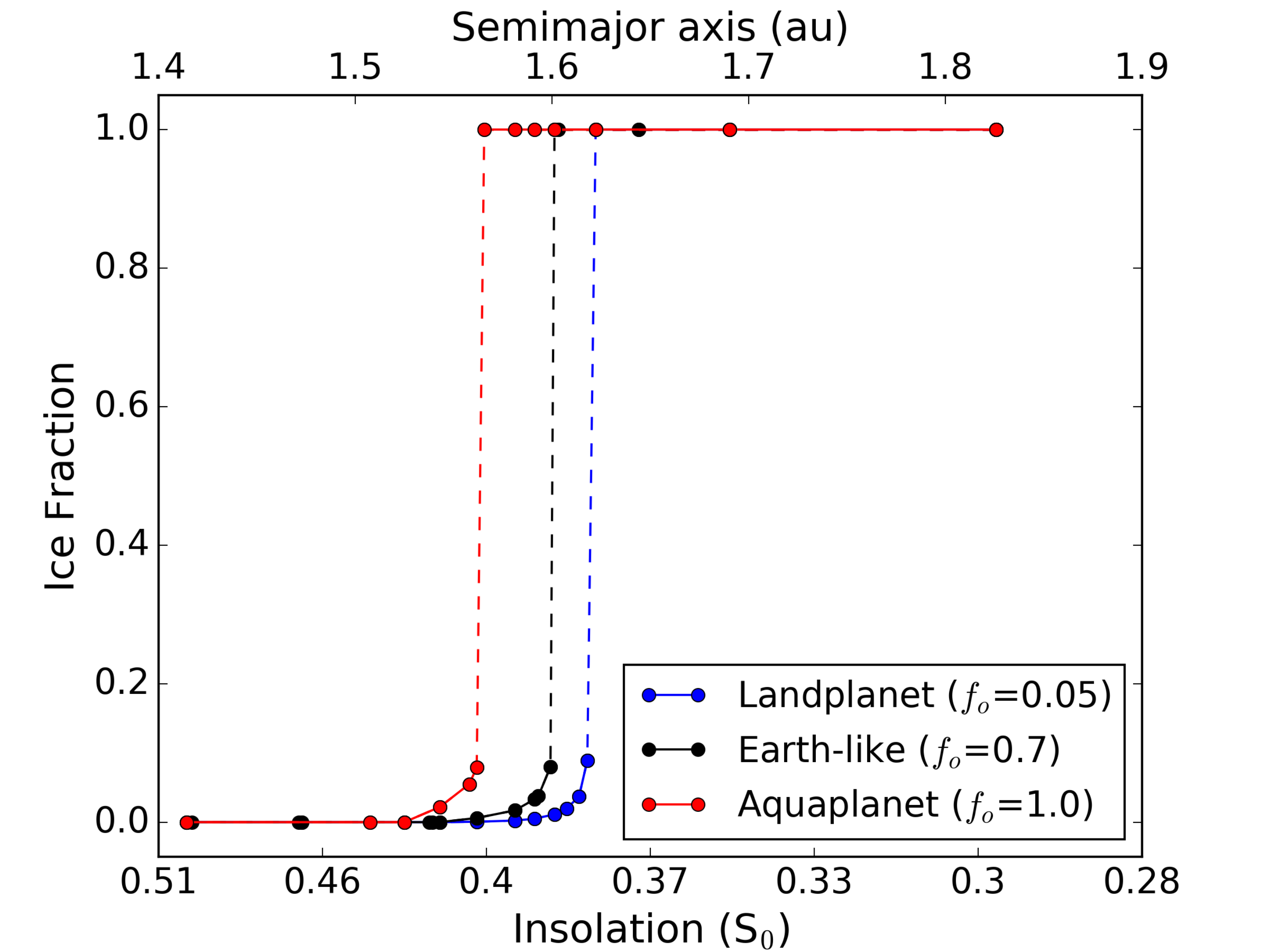}
    \caption{Dependence on planetary parameters of the fractional ice coverage calculated at the outer edge of the HZ. The results were obtained for a cloud-free Earth-like planet with a CO$_2$-dominated, maximum greenhouse atmosphere, the remaining parameters being fixed to Earth values; only the solutions obtained with warm initial conditions ($T_0=300$\,K) are shown (see Section \ref{outeredge}). Top left panel: rotation period, $P_\text{rot}$; top right panel: planet radius, $R_\text{p}$; bottom left panel: axis obliquity, $\epsilon$; bottom right panel: geography. 
     }
    \label{fig:OHZ}
\end{figure*}

The results of these experiments are shown in Fig. \ref{fig:OHZ},  where we plot the mean-annual global ice coverage as a function of insolation obtained for different values of planetary rotation, radius, axis tilt, and ocean/land distribution. 
In all cases we find that the transition to a snowball state is rather sharp,
taking place around $\simeq 1.6$\,AU.
This result is in general agreement with the maximum greenhouse limit for a solar-type star
found at 1.67\,AU by \citet[]{Kasting1993} and \citet{Kopparapu2013b} from single-column calculations.
The different treatment of the radiative transfer and of the climate recipes in our model 
can explain why we find the snowball transition at a location somewhat closer to the star than the classic outer edge. In particular, we used more recent spectral data (HITRAN2016) and a different H$_2$O continuum model with respect to \citet[]{Kasting1993} and \citet[]{Kopparapu2013a}. Our RT calculations employed a more coarse vertical pressure grid with respect of \citet[]{Kopparapu2013a}, which is known to slightly increase the OLR (thus increasing the lower insolation limit). Finally, we adopted less opaque CIA prescriptions for the CO$_2$ with respect to \citet[]{Kasting1993}, which are considered more in line with experimental results \citep[see][for a discussion on the subject]{wordsworth10}.

The extra dimension (latitude) and the ice-albedo feedback that are present in our model 
provide a detailed description of the climate changes that take place in the proximity of the outer edge.
A detailed analysis of Fig. \ref{fig:OHZ} highlights the role
played by different planetary properties in determining the onset of the snowball transition.
In the top left panel one can see that the transition occurs at increasing distance from
the star when the rotation period increases.
This effect is expected because
the heat transport from the equator to the poles becomes more efficient with increasing $P_\text{rot}$,
leading to a slower growth of the ice polar caps. 
In the top right panel one can see that the increase of planetary radius shifts the snowball boundary inwards. 
This is due to the fact that the heat transport to the poles is less effective in planets with larger $R_\text{p}$,
leading to a faster growth of the polar caps. 
The bottom left panel shows that an increase of the planetary axis tilt, $\epsilon$, shifts the snowball limit outwards.
The effect is negligible up to $\epsilon \simeq 20^\circ$, becoming evident above $\simeq 30^\circ$.
In this moderate range of obliquities the effect can be interpreted as follows.
The configuration at $\epsilon=0^\circ$ favours the formation of permanent ice caps in the
polar regions, where the zenith distance $Z$ is always large.
As the obliquity starts to increase, the polar regions undergo a period of higher insolation (lower $Z$) in some seasons, which tends to reduce the ice caps. 
At very high obliquities the behaviour is more complex (see Section 4.4.2 in V13) and can be properly investigated only using 3D models. 

In the bottom panel of Fig. \ref{fig:OHZ}, we show the impact of variations of ocean/land distribution. 
As one can see, the outer limit shifts outwards when the fraction of oceans, $f_o$, increases.
The land planet, with $f_o=0.05$, provides an extreme example of early snowball transition. 
These results can be understood in terms of the lower albedo and higher thermal capacity of the oceans 
compared to the continents. The snowball transition that we find is slightly sharper in ocean planets
than in desert planets owing to the slightly different temperature dependence of ice over oceans and lands (Section \ref{s:ice}).
Our results are in line with recent findings that planets covered largely by oceans have warmer average surface temperatures than land-covered planets \citet[]{Rushby2019}.

\section{Conclusions}
\label{sec:conclusions}

We have presented EOS-ESTM, a flexible climate model aimed at simulating the surface and atmospheric
conditions that characterize habitable planets. The model allows one to perform
a fast exploration of the parameter space representative
of planetary quantities, including those currently not measurable in rocky exoplanets.
EOS-ESTM has been built up starting from ESTM, a seasonal-latitudinal EBM featuring 
an advanced treatment of surface and cloud components and
a 2D (vertical and latitudinal) treatment of the energy transport.  
The main upgrades of EOS-ESTM can be summarized as follows:

\begin{itemize}

\item
The atmospheric radiative transfer is calculated using EOS \citep{Simonetti2021},
a procedure tailored for atmospheres of terrestrial-type planets, 
based on the opacity calculator HELIOS-K \citep[]{grimm15,grimm21} and the radiative transfer code HELIOS \citep[]{malik17,malik19}.
Thanks to EOS, the ESTM radiative transfer can  be now calculated
for a variety of atmospheres with different bulk and greenhouse compositions,
illuminated by stars with different SEDs. 

\item 
The parameterizations that describe the clouds properties have been largely upgraded.
New equations have been introduced for the albedo of the clouds and its dependence
on the albedo of the underlying surface. 
The clouds coverage over ice is now a function of the global planetary ice coverage.
A specific treatment for the transmittance and OLR forcing of clouds at very low temperature has been introduced.

\item
A generalized logistic function has been introduced to estimate the ice coverage as a function
of mean zonal surface temperature. Based on a detailed study of the ice distribution on Earth, 
the adopted algorithm discriminates between ice over lands and oceans.
The albedo and thermal capacity of transitional ice is now estimated using 
the fractional ice coverage. 

\end{itemize}

With the aim of providing a reference model for studies of habitable planets, we calibrated EOS-ESTM using a large set of Earth satellite and reanalysis data.  
The reference Earth model satisfies a variety of diagnostic tests, including mean global measurements (Table \ref{tab:eraceresdata}) and mean latitudinal profiles of surface temperature, TOA albedo, OLR and ice coverage (Fig. \ref{fig:earthannualzonal}). 
The positive results of the diagnostic tests were obtained by tuning the parameters 
within narrow ranges perfectly consistent with measurements of each climate component.  
All the Earth's data used in our analysis were selected for the same period (2005-2015)
and the atmospheric trace content of greenhouse gases was tuned accordingly (Section  \ref{earthatmodata}). 
Due to the lack of 3D treatment of clouds and atmospheric circulation, the model is not able to reproduce the detailed shape of the OLR latitudinal profile, even though it does reproduce correctly the mean global value. 

To test the consistency of EOS-ESTM with previous studies of non-terrestrial climate conditions
we performed a series of comparisons with a hierarchy of climate models (Section \ref{sec:model_validation}). 
The results of these tests  can be summarized as follows:

\begin{itemize}

\item
The latitudinal profiles of temperature and albedo of an Earth-like aquaplanet
are in agreement with predictions obtained using the 3D,
intermediate complexity model PlaSim. Differences that we find are due to the lack
of the 3D atmospheric circulation and the 3D
representation of clouds in our model.

\item
Comparisons performed at varying levels of insolation yield results which are in general agreement with other models.
However, critical differences appear at high insolation and temperature, when the resulting abundance of water vapour makes extremely model-dependent the radiative transfer calculations. 
Changing stellar spectrum at moderate and low levels of insolation yields consistent results. 

\item
Comparisons performed at varying planetary radius and rotation rate
yield consistent results, but suggest that the dependence of the meridional transport on these planetary quantities may be more moderate than estimated in V15. 
This test indicates that some parameters of our model can be recalibrated using a proper set of climate experiments carried out with state-of-the art GCMs.

\item
The application of EOS-ESTM to the case of a CO$_2$-dominated atmosphere in maximum greenhouse conditions
\citep{Kasting1993} yields a detailed description of the transition to a snowball state that takes place when the insolation decreases in the proximity of the outer edge of the HZ.  
Thanks to the flexibility of our model we can explore how this transition develops in different planetary conditions (e.g. rotation rate, radius, axis tilt, ocean coverage), taking also into account the presence of climate bistability. 

\end{itemize}

The possibility to easily adapt the input parameters to simulate a broad spectrum of planetary and atmospheric quantities allows one to apply EOS-ESTM to simulate a large variety of terrestrial-type exoplanets.
As in the case of the original ESTM, this flexibility can be used to explore in detail
the habitability conditions of individual exoplanets \citep{Silva2017B}
or to perform statistical studies of exoplanetary habitability \citep{Murante2020}.
With EOS-ESTM it will be possible to extend these types of studies
with a more accurate treatment of the climate effects of land, oceans, ice and clouds, 
and expanding the palette of atmospheres to non-terrestrial compositions
and the host stars to non-solar types. 

The flexibility of EOS-ESTM paves the road for building up multiparameter habitable zones, each parameter being representative a planetary property that affects the climate.
To achieve this ambitious goal it is important to assess the consistency with respect to a hierarchy of climate models, devising a dedicated series of experiments with the same set of initial conditions. 
Given the vastness of possibilities to be tested, a collaborative effort is required in order to establish proper protocols for a meaningful comparison of models developed by independent research groups, such as  the TRAPPIST-1 Habitable Atmosphere Intercomparison \citep[THAI,][]{Fauchez2020,Fauchez2021a}, and the future larger project Climates Using Interactive Suites of Intercomparisons Nested for Exoplanet Studies (CUISINES) NExSS\footnote{https://nexss.info/} Working Group \citep[][]{Fauchez2021b}.

\section*{Acknowledgements}

The Authors wish to thank the Italian Space Agency for co-funding the Life in Space project (ASI N. 2019-3-U.0).
The research reported in this work was supported by OGS and CINECA under HPC-TRES program award number 2022-02.
We thank the referee for his/her careful reading of the manuscript and helpful comments.

%and the National Institute of Oceanography and Applied Geophysics (OGS) for supporting the HPC-TRES (High Performance Computing Training and Research for Earth Sciences) program.

\section*{DATA AVAILABILITY}

The data used for this article will be shared on reasonable request to the corresponding author.



\bibliographystyle{mnras}
\bibliography{mnras} 

%%%%%%%%%%%%%%%%%%%%%%%%%%%%%%%%%%%%%%%%%%%%%%%%%%

%%%%%%%%%%%%%%%%% APPENDICES %%%%%%%%%%%%%%%%%%%%%

%%%%%%%%%%%%%%%%%%%%%%%%%%%%%%%%%%%%%%%%%%%%%%%%%%

% Don't change these lines
%\bsp	% typesetting comment
\label{lastpage}
\end{document}